\documentclass[12pt,a4paper]{report}
\usepackage{amsmath}
\usepackage{fancyheadings}          
\begin{document}
%%%%%%%%%%%%%%%%%%%%%%%%Abkuerzungen aus SBYM
\newcommand{\al}{\alpha}
\newcommand{\be}{\beta}
\newcommand{\de}{\delta}
\newcommand{\De}{\Delta}
\newcommand{\eps}{\varepsilon}
\newcommand{\ga}{\gamma}
\newcommand{\Ga}{\Gamma}
\newcommand{\tGa}{\tilde{\Gamma}}
\newcommand{\ka}{\kappa}
\newcommand{\la}{\lambda}
\newcommand{\La}{\Lambda}
\newcommand{\Lao}{\Lambda_0}
\newcommand{\Si}{\Sigma}
\newcommand{\om}{\omega}
\newcommand{\vep}{\varepsilon}
\newcommand{\vp}{\varphi}
\newcommand{\uvp}{\underline{\varphi}}
\newcommand{\uvph}{\underline{\varphi}}
\newcommand{\veph}{\vec{\varphi}}
\newcommand{\uveph}{\underline{\vec{\varphi}}}
\newcommand{\uP}{\underline{\Phi}}
\newcommand{\bc}{\bar{c}}
\newcommand{\uA}{\underline{A}}
\newcommand{\uB}{\underline{B}}
\newcommand{\uc}{\underline{c}}
\newcommand{\ubc}{\underline{\bar{c}}}
\newcommand{\uh}{\underline{h}}
\newcommand{\ulSi}{\underline{\Sigma}}
\newcommand{\uSi}{\underline{\dot{\Sigma}}}
\newcommand{\psib}{\overline{\psi}}
\newcommand{\etab}{\overline{\eta}}
\newcommand{\sib}{\overline{\sigma}}
\newcommand{\dsi}{\dot{\sigma}}
\newcommand{\pa}{\partial}
\newcommand{\ti}[1]{\tilde{#1}}
\newcommand{\qed}{\hfill \rule {1ex}{1ex}\\ }
\newcommand{\eq}{\begin{equation}}
\newcommand{\eqe}{\end{equation}}
\newcounter{saveeqn}
\newcommand{\alpheqn}{\setcounter{saveeqn}{\value{equation}}
\setcounter{equation}{0}
\addtocounter{saveeqn}{1}
\renewcommand{\theequation}{\mbox{\arabic{saveeqn}\alph{equation}}}}
\newcommand{\reseteqn}{\setcounter{equation}{\value{saveeqn}}%
\renewcommand{\theequation}{\arabic{equation}}}
%%%%%%Abkuerzungen aus RMP,alt %%%%%%%%%%%%%%%%%%%%%%%%
\newcommand{\Lb}{\mathcal{L}^{\Lambda _0,\Lambda _0}}
\newcommand{\Lf}{\mathcal{L}^{\Lambda ,\Lambda _0}}
\newcommand{\Lp}{\mathcal{L}^{0,\Lambda _0}}
\newcommand{\fit}{\varphi _{\tau }}
\newcommand{\fitx}{\varphi _{\tau }(x)}
\newcommand{\Jt}{J_{\tau }}
\newcommand{\Jtx}{J_{\tau }(x)}
\newcommand{\B}{B^{a}}
\newcommand{\Bx}{B^{a}(x)}
\newcommand{\jm}{j^{a}_{\mu}}
\newcommand{\jmx}{j^{a}_{\mu}(x)}
\newcommand{\jn}{j^{a}_{\nu}}
\newcommand{\ba}{b^{a}}
\newcommand{\bax}{b^{a}(x)}
\newcommand{\ca}{c^{a}}
\newcommand{\cax}{c^{a}(x)}
\newcommand{\cb}{{\bar c}^{a}}
\newcommand{\cbx}{{\bar c}^{a}(x)}
\newcommand{\Am}{A^{a}_{\mu}}
\newcommand{\Amx}{A^{a}_{\mu}(x)}
\newcommand{\An}{A^{a}_{\nu}}
\newcommand{\et}{\eta ^{a}}
\newcommand{\etx}{\eta ^{a}(x)}
\newcommand{\etb}{{\bar \eta}^{a}}
\newcommand{\etbx}{{\bar \eta}^{a}(x)}
\newcommand{\g}{\gamma ^{a}}
\newcommand{\gx}{\gamma ^{a}(x)}
\newcommand{\gm}{\gamma ^{a}_{\mu}}
\newcommand{\gmx}{\gamma ^{a}_{\mu}(x)}
\newcommand{\oma}{\omega ^{a}}%%%%%%%%geaendert
\newcommand{\omx}{\omega ^{a}(x)}
\newcommand{\LLz}{\Lambda, \Lambda_0}
\newcommand{\LLzz}{\Lambda_0, \Lambda_0}
\newcommand{\Lz}{\,0, \Lambda_0}
\newcommand{\si}{\sigma _{0, \Lambda_0}} %%%%gestrichen
\newcommand{\ufit}{\underline{\varphi }_{\tau}}
\newcommand{\uca}{\underline c^{a}}
\newcommand{\ucb}{\underline {\bar{c}}^{a}}
\newcommand{\uAn}{\underline A^{a}_{\nu}}
\newcommand{\uFi}{\underline \Phi }
\newcommand{\gt}{\gamma _{\tau}}
\newcommand{\Ri}{\mathcal{R}_{l,n}^{\lambda ,\Lambda _0}}
\newcommand{\Rf}{\mathcal{R}_{l,n}^{\Lambda ,\Lambda _0}}
\newcommand{\Rb}{\mathcal{R}_{l,n}^{\Lambda_0 ,\Lambda _0}}
\newcommand{\Db}{\mathcal{D}^{\Lambda _0,\Lambda _0}}
\newcommand{\Df}{\mathcal{D}^{\Lambda ,\Lambda _0}}
\newcommand{\ufi}{\underline{\varphi}}
\newcommand{\up}{\underline{p}}
\newcommand{\hufi}{\hat{\underline{\varphi}}}
\newcommand{\uk}{\underline{k}}
\newcommand{\Gf}{\Gamma ^{\Lambda, \Lambda_0}}
\newcommand{\Gtf}{{\tilde\Gamma}^{\Lambda, \Lambda_0}}

\begin{center}
{\bf \Large PERTURBATIVE RENORMALIZATION} \\  \vspace{1cm} 
         {\bf \Large BY FLOW EQUATIONS } \\
\vspace{4cm}
   {\large Volkhard F. M\"uller}  \\
Fachbereich Physik, Universit\"at Kaiserslautern,  \\
D-67653 Kaiserslautern, Germany   \\
E-mail: vfm@physik.uni-kl.de
\vspace{3cm}
\end{center}
{\small In this article a self-contained exposition of proving
 perturbative renormalizability of a quantum field theory based
 on an adaption of Wilson's differential renormalization group
 equation to  perturbation theory is given. The topics treated
 include the spontaneously broken SU(2) Yang-Mills theory.
 Although mainly a coherent but selective review, the article
 contains also some simplifications and extensions with respect  
 to the literature. \\
 In the original version of this review the spontaneously
 broken Yang-Mills theory, dealt with in Chapter 4,
 followed \cite{KM}. Recently, however, the authors of this
 article discovered a serious deficiency in their method to
 restore the Slavnov-Taylor identities (intermediately violated
 by the regularization), which invalidates their claim.
 Now in \cite{nKM} these authors have developed
 a new approach to accomplish the missing
 restoration of the Slavnov-Taylor identities.
 The present revised review concerns solely
 Chapter 4: the original Sections 4.1-4.3 are essentially 
 unaltered, whereas the former Section 4.4, now obsolete, has
 been replaced by the new Sections 4.4-4.6, following the
 recent article \cite{nKM}. }

\tableofcontents
\chapter{Introduction}

 Dyson's pioneer work \cite{Dy} opened the era of a systematic perturbative
renormalization theory long ago, and in the late sixties of the last century 
the rigorous BPHZ-version \cite{BPHZ} was 
accomplished. In place of the momentum space subtractions of BPHZ
 to circumvent UV-divergences
various intermediate regularization schemes were
 invented: Pauli-Villars regularization \cite{PaVi},
 analytical regularization \cite{Spe}, dimensional regularization \cite{dim}. 
These different methods, each with its proper merits, 
are equivalent up to finite counterterms, for
a review see e.g. \cite{VW}. All of them are based on the
 analysis of multiple integrals
corresponding to individual Feynman diagrams, the
 combinatorial complexity of which rapidly grows
with increasing order of the perturbative expansion.
 Zimmermann's famous forest formula \cite{Zi}
provides the clue to disentangle overlapping divergences,
 organizing the order of subintegrations to be followed. The
 BPHZ-renormalization, originally developed in case of massive theories, 
was extented by Lowenstein \cite{BPHZL} to cover also  zero mass particles.
From the point of view of elementary particle physics, renormalization theory
culminated in the work of 't Hooft and Veltman \cite{HV},
 demonstrating the renormalizability of non-Abelian gauge theories.

At the time of these achievements, Wilson's view \cite{Wil} 
of renormalization as a continuous evolution
of effective actions -- a primarily non-perturbative notion --
began to pervade the whole area of quantum field theory and
soon proved its fertility. In the domain of rigorous mathematical
analysis beyond formal perturbation expansion, the renormalizable
UV-asymptotically free Gross-Neveu model in two space-time
dimensions has been constructed, \cite{GaKu}, \cite{FMRS}, by
decomposing in the functional integral the full momentum
 range into a union of discrete, disjoint ``slices'' and 
integrating successively the corresponding quantum
fluctuations, thereby generating a sequence of effective
actions. This slicing can be seen as the equivalent of
introducing block-spins in lattices of Statistical Mechanics.
 Rigorous non-perturbative analysis of the 
renormalization flow is the general subject
of the lecture notes \cite{Bry, Riv1} and
of the monograph \cite{Riv2} which, among other topics,
also treats the problem of summing the formal perturbation 
series. In these lecture notes and in the monograph references
to the original work on non-perturbative renormalization
can be found.  \\
In the realm of perturbative renormalization Wilson's ideas
have proved beneficial, too.
Gallavotti and Nicolo, \cite{GN}, split the propagator
of a free scalar field in disjoint momentum slices, i.e.
decomposed the field into a sum of independent (generalized)
random variables, and developed a tree expansion to
perturbative renormalization. Here, due to the slicing, the
degrees of freedom are again integrated in finite steps.
This method has been applied in the monograph \cite{FHRW}
to present a proof of renormalizability of QED which only
involves gauge-invariant counterterms.
Polchinski \cite{Pol} realized, that considering renormalization in terms
of relevant and irrelevant operators as Wilson, is also effective
 in perturbation theory, and he gave in
the case of the $\Phi^4_4$-theory an inductive self-contained proof
 of perturbative renormalization,
based on Wilson's renormalization (semi-) group differential equation.
 His method avoids completely
the combinatoric complexity of generating Feynman diagrams and 
the following cumbersome
analysis of Feynman integrals with their overlapping divergences.
 It rather treats an $n$-point Green
function of a given perturbative order as a whole. 
Due to this fact, his method is particularly transparent.

Polchinski's approach has proved very stimulating in various directions. 
i) In mathematical physics it has been extended to present 
new proofs of general results
in perturbatively renormalized quantum field theory, which are
 simpler than those achieved before:
renormalization of the nonlinear $\sigma$-model \cite{MiRa},
a rigorous version of Polchinski's argument, together with 
physical renormalization conditions \cite {KKS},
renormalization of composite operators and Zimmermann 
identities \cite{KK3}, Wilson's operator product expansion \cite{KK4},
 Symanzik's improved actions \cite{Wiec,Impr}, large order bounds \cite{Ke2},
 renormalization of massless $\Phi ^4_4$ -theory
\cite{KK1}, renormalization of QED \cite{QED}, 
decoupling theorems \cite{Kim},
renormalization with flow equations in Minkowski space \cite{KKSch},
a renormalization of spontaneously broken Yang - Mills theory \cite{KM},
 temperature independent
 renormalization of finite temperature field theories \cite{KMR}.
The monograph \cite{Salm} contains a clear and detailed introduction 
to Polchinski's method formulated with Wick-ordered field products
\cite{Wiec}, and, in addition, the application of a similar
renormalization flow to the Fermi surface problem of condensed
matter physics. \\
ii) In the domain of theoretical physics there is a vast
amount of contributions with diverse applications of
Polchinski's approach.
Flow equations for  vertex functions have been introduced \cite{Vert}
 and also employed to investigate perturbative renormalizability
of gauge, chiral and supersymmetric theories \cite{Bon}.
In these articles also several explicit one-loop calculations  
are performed.  An effective quantum action principle has
 been formulated for 
the renormalization flow breaking gauge invariance \cite{D'AM}.
There are also interesting attempts to combine a
gauge invariant regularization with a flow equation \cite{AKMT}. 
 Besides aims within perturbation theory,
 there have been many activities to use truncated
versions of flow equations as appropriate non-perturbative
 approximations in strong interaction physics,
more in accord with Wilson's original goal.
In the physically distinguished case of (non-Abelian)
gauge theories \cite{EHW, ReWe} the effective action is
restricted in a local approximation to its relevant
part for all values of the flowing scale. As a consequence,
the flow equation for the effective action reduces to a
system of $r$ ordinary differential equations, $r$
being the number of relevant coefficients appearing.
This system is integrated from the UV-scale downward. 
In these non-perturbative approaches the problem 
arises to reconcile
the truncation with the gauge symmetry. 
This problem is discussed also in \cite{D'AM}.
In a very different field of interest
the question of the  non-perturbative renormalizability
of Quantum Einstein Gravity has been investigated,
based on truncated flow equations, \cite{LaRe}. 
These authors restrict the average effective action to the
Hilbert action together with a small number of
additional local terms. The flow of the coupling 
coefficients is studied numerically and the
existence of a non-Gaussian fixed point in the
ultraviolet found. This result is then
interpreted to support the conjecture, that
Quantum Einstein Gravity is ``asymptotically safe''
in Weinberg's sense. We  like to point out that  physical
applications of flow equations  are reviewed in \cite{BB},
 containing an extensive list of references.
 
The present article is intended to provide a self-contained exposition
 of perturbative renormalization based on Polchinski's inductive method,
 employing the differential renormalization group equation  of Wilson.
 Therefore, emphasis is laid on a coherent presentation of the topics
considered. A comprehensive overview of the literature on the
subject will not be pursued. The quantum field theories considered are 
treated in their Euclidean formulation on $d=4$
dimensional (Euclidean) space-time by means of functional integration.
Accordingly, their correlation functions are called \emph{Schwinger functions}
to distinguish them from the \emph{Green functions} on Minkowski space.
 In the intermediate steps of the derivations always 
regularized functionals are used, the controlled removal
 (within perturbation theory) of this regularization being our main concern. 
We avoid any manipulation of unregularized ``path integrals''.

The plan of this article is as follows. In Chapter 2 Polchinski's method
  to prove perturbative renormalizability is elaborated
 treating the nonsymmetric $\Phi ^4$-theory in detail. Besides
 the system of Schwinger functions of this theory, the Schwinger 
 functions with one composite field (operator) inserted are also dealt with. 
The presentation is mainly based on
 \cite{KKS,KK3} and on some simplifications \cite{Kopp}. Moreover,
considering the theory at finite temperature, its
 temperature independent renormalizability is reviewed,
 following closely \cite{KMR}. In chapter 3 two simple cases of the
quantum action principle are demonstrated, again treating the
 nonsymmetric $\Phi ^4$-theory: the field equation and the variation
 of a coupling constant. These applications of the method 
seem not to have been treated in the literature.
 Hereafter, somewhat disconnected, flow equations
for proper vertex functions are dealt with, \cite{Vert,KKSch}.
 Chapter 4 is devoted to the proof of
renormalizability of the physically most important spontaneously
 broken Yang-Mills theory.
Because of the necessity to implement nonlinear field variations,
 this problem can be regarded as a further instance of
 the quantum action principle. The initial presentation followed
the line of \cite{KM}. The authors of this article, however,
recently found, that the restoration of the violated Slavnov-Taylor
identities claimed there in fact has not been acchieved, since
Lemma 2 used does not take into account irrelevant boundary
terms which are inevitably present in the bare action, thus
rendering this Lemma obsolete. Keeping to the general line
of the earlier paper the authors have developed in \cite{nKM}
a new approach to cope with the appearance of irrelevant
contributions in the violated Slavnov-Taylor identities,
which is based on tracing the super-renormalizable couplings
in the perturbative expansion. Hence, the revised presentation
here keeps with a few adaptions the former Sections 4.1 - 4.3,
but replaces the former Section 4.4 by the new Sections 4.4 - 4.6,
following \cite{nKM}.

\chapter{The Method}
\section{Properties of Gaussian measures}
Our point of departure is a Gaussian probability measure $ d\mu $
 on the space $ \mathcal{C}(\Omega ) $ of continuous real-valued
 functions on a $d$-dimensional torus $ \Omega. $ Such a
 function we identify with a periodic function on $ \mathbf{R}^d,$ i.e.
$ \phi (x) = \phi (x + nl ),$ 
where $ x \in \mathbf{R}^d ,\ n \in \mathbf{Z}^d ,\
 l=(l_1, \cdots ,l_d) \in {\mathbf{R}}^d_{+} $ and
$ nl = n_1l_1 + \cdots + n_d l_ d .$ A Gaussian measure with mean 
zero is uniquely defined by its covariance $ C(x,y) $, 
\begin{equation}
\int d\mu_C (\phi ) \phi (x) \phi (y) = C(x,y) = C(y,x) . 
\end{equation}
The covariance is a positive non-degenerate bilinear form
 on $ \mathcal{C}^\infty (\Omega ) \times \mathcal{C}^\infty (\Omega ) $,
 we assume it to be translation invariant, $ C(x,y) = C(x-y) $ , too.
 Moreover, the function $ C(x) $ is assumed to have 
 a given number $ N \in \mathbf{N} $ of derivatives
continuous everywhere on $ \Omega $. We list some properties of this
 Gaussian measure employed in the sequel,
 proofs can be found e.g. in \cite{GJ}.
\begin{itemize}
\item Using the notation 
$$ \langle \phi , J \rangle = \int_{\Omega } dx \phi (x) J(x) ,\qquad 
     \langle J, C J \rangle =
 \int_{\Omega } dx \int_{\Omega } dy J(x) C(x-y) J(y) $$
where $ J \in \mathcal{C}^\infty (\Omega ) $ is a test function,
 the generating functional of the correlation functions is given explicitly as
\begin{equation}\label{g1}
\int d\mu_C ( \phi  ) \ e^{\langle \phi , J \rangle } =
 e^{\ \frac{1}{2} \langle J, C J \rangle } . 
\end{equation}
\item The translation of the Gaussian measure by a function
 $ \varphi \in \mathcal{C}^\infty (\Omega ) $ results in
\begin{equation}\label{g2}
d\mu_C (\phi  - \varphi ) =
 e^{- \frac{1}{2}\langle \varphi , C^{-1} \varphi \rangle } \ d\mu_C (\phi )
   \ e^{\langle \phi , C^{-1} \varphi \rangle} .
\end{equation}\
\item Let $ A(\phi ) $ denote a polynomial formed of local  powers
of the field, $ \phi (x)^n , n \in \mathbf{N}, $ and of its derivatives
 $(\partial_{\mu} \phi (x))^m , m  \in \mathbf{N}, 2m < N,$
 at various points $ x $. If the covariance $ C $ is the sum of
 two covariances, $ C = C_1 + C_2 $ , then 
\begin{equation}\label{g3}
\int d\mu _C (\phi ) A( \phi ) = \int d\mu _{C_1}(\phi _1) 
\int d\mu _{C_2}(\phi _2) A( \phi _1 + \phi _2 ).
\end{equation}
\item Integration by parts of a function $ A(\phi ) $
 as considered in (\ref{g3}) yields
\begin{equation}\label{g4}
 \int d\mu_C (\phi ) \phi (x) A(\phi ) = \int d\mu_C (\phi )
 \int_{\Omega } dy \ C(x-y) \frac{\delta }
 {\delta \phi (y) } A(\phi ) .
\end{equation} 
\item Finally, let the covariance of the Gaussian measure
 depend differentiably on a parameter,
 $$ C(x-y) = C_t (x-y) \ , \qquad \dot C_t \equiv  \frac{d}{dt} C_t (x-y) \ .$$
 Given again a function $ A(\phi ) $ as in (\ref{g3}), then
\begin{equation}\label{g5}
\frac{d}{dt} \int d\mu_{C_t} (\phi ) A(\phi ) =
 \frac{1}{2} \int d\mu_{C_t} (\phi ) 
 \langle \frac{\delta }{\delta \phi }\ , \dot C_t \ 
\frac{\delta }{\delta \phi } \rangle A(\phi ). 
\end{equation}
\end{itemize}
  As an example of the class of covariances considered,
 we present already here the particular covariance
which will be mainly used in the flow equations envisaged.
 The torus $ \Omega $ has volume 
$ \mid \Omega \mid = l_1 l_2 \cdots l_d $ and a
 point $ x\in \Omega $ has coordinates 
$ -\frac{1}{2} l_i \leq x_i < \frac{1}{2} l_i \ , i= 1,\dots ,d . $
 Hence, the dual Fourier variables ( momentum vectors) $ k $ form
 a discrete set: $$ k = k(n) = \left ( \frac{2\pi n_1}{l_1} , \cdots ,
 \frac{2\pi n_d}{l_d} \right) \ , n\in \mathbf{Z}^d . $$ 
Let $ m, \Lambda _0 $ be positive constants, $ 0 < m \ll \Lambda _0 $,
 and the nonnegative parameter $ \Lambda $ satisfy
 $ 0 \leq \Lambda \leq \Lambda _0 $ , we define the covariance 
\begin{equation}\label{g6}
C^{\Lambda ,\Lambda _0}(x-y) = \frac{1}{\mid \Omega \mid }
 \sum_{n \in \mathbf{Z}^d}
    \frac{ e^{ik(x-y)}}{k^2 + m^2 } \left( e^{- \frac{k^2 + m^2}
{ {\Lambda_0}^2}} - e^{- \frac{k^2 + m^2}{ {\Lambda}^2}} \right) .
\end{equation}
This covariance obviously has the well-defined infinite volume limit
 $ \Omega \rightarrow \mathbf{R}^d $,
with $ x,y,k \in \mathbf{R}^d $,
\begin{equation}\label{g7}
C^{\Lambda ,\Lambda _0}(x-y) = \frac{1}{{(2\pi)}^d} \int_{ \mathbf{R}^d} dk
    \frac{ e^{ik(x-y)}}{k^2 + m^2 } \left( e^{- \frac{k^2 + m^2}
{ {\Lambda_0}^2}} - e^{- \frac{k^2 + m^2}{ {\Lambda}^2}} \right) .
\end{equation}
Abusing slightly the notation we did not choose a
 different symbol for the limit. 
Later on, however, the case referred to will be clearly stated.
 Choosing the values $ \Lambda _0 = \infty , \Lambda = 0 $ the
 covariances (\ref{g6}), (\ref{g7}) become the Euclidean propagator
 of a free real scalar field with mass $m$ on $\Omega $
 and $ \mathbf{R}^d $, respectively. A finite value of
 $ \Lambda _0 $ generates an UV-cutoff thus regularizing
 the covariances: they now satisfy the regularity condition
 assumed for all $N$. This property is kept introducing the
 additional term governed by the ``flowing'' parameter
 $ 0 \leq \Lambda \leq \Lambda _0 $. Its role is to interpolate
 differentiably between a vanishing covariance at 
$\Lambda =\Lambda _0 $, corresponding to a $\delta $-measure
 on the function space, and the free UV-regularized covariance
 at $ \Lambda =0 $. As a consequence we remark 
 that the Gaussian measure with covariance (\ref{g6})
 is supported with probability one on (the nuclear space) 
  $\mathcal{C}^\infty (\Omega) $, \cite{HiYa}.
 Clearly, a modification of the Euclidean
 propagator showing these properties can be accomplished
 with a large variety of cutoff functions.
 In (\ref {g6}),(\ref {g7}) a factor of the form
\begin{equation}\label{g8}
R^{\Lambda ,\Lambda _0}(k²) = \sigma _{\Lambda _0} (k^2) -
 \sigma _{\Lambda} (k^2)  
\end{equation}
has been introduced, with the particular function 
\begin{equation}\label{g9}
\sigma _\Lambda (k^2) = e^{- \frac{k^2 + m^2}{\Lambda ^2}} \, .
\end {equation}
 We observe, that regularization and interpolation is caused by 
 any positive function $  \sigma _{\Lambda} (k^2) $  satisfying:
 i) For fixed $ \Lambda $ it decreases as a function of $ k^2 $,
 vanishing rapidly for $ k^2 > \Lambda ^2 $ .
 ii) For fixed $ k^2 $ it increases with $\Lambda $ from the
 value zero at $ \Lambda = 0 $ to the value one at $ \Lambda = \infty $.
 Later on, our particular choice will prove advantageous.
%%%%%%%%%%%%%%%%%%%%%%%%%%%%%%%%%%%%%%%%%%%%%%%%%%%%%%%%%%%%%%%%%
\section{The flow equation }
Perturbative renormalizability of a quantum field theory is based
 on locality of its action and on
power counting. The qualification ``perturbative" means expansion
 of the theory's Green (or Schwinger) functions
as  formal power series in the loop parameter
 \footnote{ If one considers Feynman diagrams, 
the power of $\hbar$ counts the number of loops formed by
 such a diagram.} $\hbar$, and treating them order-by-order,
 i.e. disregarding questions of convergence. The notions of
 locality and power counting
 can be introduced looking at the classical precurser of the
 quantum field theory to be constructed.
There, a local action in $d$ space-time dimensions is the
 space-time integral of a Lagrangian (density),
 having the form of a polynomial in the fields entering the theory
 and their derivatives. The propagators
are determined by the free part, which is bilinear in the fields.
 For a scalar field
 and a spin-$\frac{1}{2}$-field
this free part is of second and first order in the derivatives,
 respectively. Defining the canonical (mass)
dimension of the corresponding fields to be $\frac{1}{2}(d-2)$ and
  $\frac{1}{2}(d-1)$, respectively,
and attributing the mass dimension $1$ to each partial derivative,
 the free part of the Lagrangian 
has the dimension $d$, the action thus is dimensionless.
 Vector fields, especially (non-Abelian) gauge
fields, pose particular problems to be considered later.
 Still looking at the classical theory, local interaction terms
 in the Lagrangian involve by definition more than two fields
. Their respective coupling constant
 has a mass dimension, derived from the mass dimension of
 the interaction term, the coupling constant
of an interaction term of mass dimension $d$ being dimensionless.
 Any local term entering the Lagrangian
is called a \emph{relevant} operator \cite{Wil} if it has 
 a mass dimension $\leq d$, but 
\emph{irrelevant}, if its mass dimension is greater than $d$.
 In the physically distinguished case $d=4$ 
 the central result of pertubative renormalization theory is
  that UV-finite Green (or Schwinger) 
functions can be obtained in any order,  if the interaction terms
 have mass dimensions $\leq 4$,
 by prescribing a finite number of renormalization conditions.
This number equals the number of relevant operators forming the
 full classical Lagrangian. 

We consider the quantum field theory of a real scalar field $\phi $
 with mass $m$ on four-dimensional Euclidean space-time within
 the framework of functional integration. The emerging vacuum
 effects require a finite space-time volume. Therefore we start
 with a real-valued field $ \phi \in \mathcal{C}^1(\Omega ) $
 on a four-dimensional torus $\Omega $. Its bare interaction,
 labeled by an UV-cutoff $\Lambda _0 \in \mathbf{R}_{+} $, is chosen as
\begin{eqnarray}\label{f1}
L^{\Lambda _0,\Lambda _0}(\phi ) &=& 
\int_{\Omega } dx \Bigl( \frac{f}{3!}\phi ^3(x) +
\frac{g}{4!}\phi ^4(x) \Bigr)  \nonumber\\
&+&\int_{\Omega } dx \Bigl( v(\Lambda _0)\phi (x)+
\frac{1}{2}a(\Lambda_0)\phi ^2(x)
+\frac{1}{2}z(\Lambda_0)\bigl(\partial _\mu \phi \bigr)^2(x) \nonumber \\
 & \quad & \qquad \quad + \frac{1}{3!}b(\Lambda_0)\phi^3 (x)
+\frac{1}{4!}c(\Lambda_0)\phi^4 (x) \Bigr) .
\end{eqnarray}
The first integral has classical roots: its integrand is
 formed of the field's self-interaction with real coupling
 constants $f$ and $g$ having mass dimension equal to one and zero,
 respectively.\footnote{Stability requires $g$ to be positive,
 but this property is not felt in a perturbative treatment.}
 The second integral contains the related counterterms,
 determined according to the following rule. The canonical mass
 dimension of the field $\phi $ is equal to one. As counterterms
 in the integrand of the bare interaction have
to appear all local terms of mass dimension $\leq 4$ that can
 be formed of the field and of its derivatives
but respecting the (Euclidean) $O(4)$-symmetry. This symmetry is not
 violated by the intermediate UV-regularization procedure
 and can thus be maintained. In contradistinction to the
 coupling constants \,$ f,g $ the five coefficients
 $ v(\Lambda _0), a(\Lambda _0), z(\Lambda _0), b(\Lambda _0),
 c(\Lambda _0) $  of the counterterms cannot be chosen freely
 but have to depend on the UV-cutoff $\Lambda _0$. This dependence
 is dictated by the aim that after functional integration the
 UV-regularization can be removed, i.e. the limit
$\Lambda _0 \longrightarrow \infty $ can be performed keeping 
 the physical content of the theory  finite. 
As a consequence the coefficients stated above, however, turn out 
to diverge with $ \Lambda_0 \longrightarrow \infty $.
 If we restrict the bare interaction (\ref{f1}) to the case
 $ f=0, v(\Lambda _0) = b(\Lambda _0) = 0 $, it is also invariant
 under the mirror transformation $ \phi (x) \rightarrow  - \phi (x) $
 implying an additional symmetry of the theory.

 The regularized quantum field theory on finite volume is defined
 by the generating functional of its Schwinger functions 
\begin{equation}\label{f2}
Z^{\Lambda ,\Lambda _0} (J)
       = \int d\mu _{\Lambda ,\Lambda _0}(\phi )
 e^{- \frac{1}{\hbar}L^{\Lambda _0,\Lambda _0}(\phi )
         +\frac{1}{\hbar}\langle \phi , J \rangle }
\end{equation}
with a real source $ J \in \mathcal{C}^\infty (\Omega ) $,
 bare interaction (\ref{f1}) and a Gaussian measure 
$ d\mu _{\Lambda ,\Lambda _0} $ with mean zero and covariance
 $\hbar C^{\Lambda ,\Lambda _0}$, (\ref{g6}).
 The positive parameter $\hbar$ has been introduced with
 regard to a systematic loop expansion considered later.
For fixed $\Omega$ and $\Lambda_0$, and assuming
$g+c(\Lambda_0) > 0 ,\, z(\Lambda_0)\geq 0 $ in the bare
interaction, (\ref{f1}), the functional integral (\ref{f2})
is well-defined. As a functional on $\mathcal{C}^\infty (\Omega ) $,
the support of the Gaussian measure 
$ d\mu _{\Lambda ,\Lambda _0}(\phi )$, the bare interaction is
continuous in any Sobolev norm of order $ \, n \geq 1$,
and, furthermore, bounded below, 
 $ L^{\Lambda _0,\Lambda _0}(\phi ) > \kappa $.
Hence, with $\Lambda_0 $ fixed, we have the uniform bound
for $ 0 \leq \Lambda \leq \Lambda_0,$
\begin{equation} \label{f2a}
|Z^{\Lambda ,\Lambda _0} (J)| < e^{- \kappa}
 \int d\mu _{\Lambda ,\Lambda _0}(\phi )\,
 e^{ \, \frac{1}{\hbar}\langle \phi , J \rangle }
 \leq e^{- \kappa + \frac{1}{2\hbar} \langle J,
     C^{\Lz} J \rangle } .
\end{equation}
 From (\ref{f2}) one obtains the generating functional
 $ W^{\Lambda ,\Lambda _0}(J) $ of the truncated  Schwinger functions
\footnote{In a representation of these functions in terms of
(Feynman) diagrams only \emph{ connected} diagrams appear.}
\begin{equation}\label{f3}
e^{\frac{1}{\hbar} W^{\Lambda ,\Lambda _0}(J) }
  = \frac{Z^{\Lambda ,\Lambda _0}(J)}{Z^{\Lambda ,\Lambda _0}(0) }
\end{equation}
which provides the $ n$-point functions, $ n \in \mathbf{N}$,
 upon functional derivation:
\begin{equation}\label{f4}
 W^{\Lambda ,\Lambda _0}_n ( x_1, \cdots ,x_n )
  = \frac{\delta ^n}{\delta J(x_1) \cdots \delta J(x_n) }
 W^{\Lambda ,\Lambda _0}(J) |_{J=0 } \ .
\end{equation}  
 Besides the UV-regularization determined by the cutoff
 $\Lambda _0 $, imperative to have a well-defined
functional integral (\ref{f2}) , an additional flowing cutoff
 $\Lambda $ has been built in, suppressing smaller momenta.
 It is a merely technical device, introduced by Polchinski \cite{Pol}
 and inspired by Wilson's view of renormalization \cite{Wil}.
 Decreasing $\Lambda $ from its maximal value $\Lambda = \Lambda _0 $
 to its physical value $\Lambda =0 $  gradually takes into
account the momentum domain, starting at high momenta -- in
mathematical terms: the parameter $\Lambda $ interpolates
 continuously between a $\delta$-measure
(i.e. absence of quantum effects) at $ \Lambda =\Lambda _0 $ and
 the Gaussian measure $ d\mu_{0,\Lambda _0} $ on a UV-regularized field,
 at $\Lambda =0$. Of course, as stressed after eq.(\ref{g9}),
 such an interpolation can also be realized by other
 cutoff functions than (\ref{g6}) used here. In order to make use
 of the flow parameter $\Lambda $ it is advantageous to consider
 the (free propagator-) amputated truncated Schwinger functions
 with generating functional $ L^{\Lambda ,\Lambda _0}(\varphi ) ,
 \varphi \in \mathcal{C}^\infty(\Omega ) $, defined as 
 \begin{eqnarray}
     \label{f5} 
     e^{-\frac{1}{\hbar}\left( L^{\Lambda ,\Lambda _0}(\varphi ) +
 I^{\Lambda ,\Lambda _0} \right)}
  & = & \int d\mu _{\Lambda ,\Lambda _0}(\phi )
 e^{-\frac{1}{\hbar} L^{\Lambda_0 ,\Lambda _0}( \phi  +\varphi ) } ,  \\ 
       \label{f6}
    L^{\Lambda ,\Lambda _0}(0) & = & 0 .
 \end{eqnarray}
The constant $ I^{\Lambda ,\Lambda _0} $ is the vacuum part of the
 theory. Translating on the r.h.s. in (\ref{f5}) the source function
 $\varphi $ to the measure  and using (\ref{g2}) leads to
\begin{equation}\label{f7}
e^{- \frac{1}{\hbar}\left( L^{\Lambda ,\Lambda _0}(\varphi )
 + I^{\Lambda ,\Lambda _0} \right)}
   = e^{-\frac{1}{2}\langle \varphi ,
 (\hbar C^{\Lambda ,\Lambda _0})^{-1} \varphi \rangle }
  Z^{\Lambda ,\Lambda _0}( ( C^{\Lambda ,\Lambda _0})^{-1} \varphi )
\end{equation}
relating  the generating functionals $Z$ and $L$. Hereupon,
 together with  the definition (\ref{f3}) follows finally
\begin{equation}\label{f8}
 L^{\Lambda ,\Lambda _0} (\varphi ) = 
 \frac{1}{2}\langle \varphi , ( C^{\Lambda ,\Lambda _0})^{-1} \varphi \rangle 
  - W^{\Lambda ,\Lambda _0}( ( C^{\Lambda ,\Lambda _0})^{-1} \varphi ) .
\end{equation}
Denoting by $\dot  C ^{\Lambda ,\Lambda _0} $ the derivative of
 the covariance   $ C ^{\Lambda ,\Lambda _0} $
 with respect to the flow parameter  $\Lambda $ we observe,
with $\Lambda_0 $ kept fixed,
\begin{eqnarray}
 \frac{d}{d\Lambda }\int d\mu _{\Lambda ,\Lambda _0}(\phi )
 e^{-\frac{1}{\hbar} L^{\Lambda_0 ,\Lambda _0}( \phi +\varphi )}
 &= &\frac{\hbar}{2} \int d\mu _{\Lambda ,\Lambda _0}(\phi )
 \langle \frac{\delta }{\delta \phi } \, ,\dot  C ^{\Lambda ,\Lambda _0}
 \frac{\delta }{\delta \phi }\rangle 
e^{-\frac{1}{\hbar} L^{\Lambda_0 ,\Lambda _0}( \phi  +\varphi ) }
 \nonumber  \\  
  &= &\frac{\hbar}{2}\langle \frac{\delta }{\delta \varphi }\,,
\dot C ^{\Lambda ,\Lambda _0} \frac{\delta }{\delta \varphi }\rangle 
  \int d\mu _{\Lambda ,\Lambda _0}(\phi )
 e^{-\frac{1}{\hbar} L^{\Lambda_0 ,\Lambda _0}( \phi  +\varphi ) } \nonumber 
\end{eqnarray} 
where in the first step (\ref{g5}) has been used, whereas the second
 step follows from the integrand's particular dependence on
 the field $\phi $. Hence, because of eq. (\ref{f5}) we 
obtain the differential equation
\begin{equation}\label{f9} 
   \frac{d}{d\Lambda }    
  e^{-\frac{1}{\hbar}\left( L^{\Lambda ,\Lambda _0}(\varphi ) +
 I^{\Lambda ,\Lambda _0} \right)} =
  \frac{\hbar}{2}\langle \frac{\delta }{\delta \varphi }\,,
\dot C ^{\Lambda ,\Lambda _0} \frac{\delta } {\delta \varphi }\rangle \,
 e^{-\frac{1}{\hbar}\left( L^{\Lambda ,\Lambda _0}(\varphi ) +
 I^{\Lambda ,\Lambda _0} \right) } .
 \end{equation}
The reader notices that the relation (\ref{g5}) has been used in
 the case of a nonpolynomial function.
Therefore this extension has to be understood in terms of a
 formal power series expansion, i.e. disregarding the question
 of convergence.
 Upon explicit differentiation  in (\ref{f9}) follows
 the Wilson flow equation 
\begin{eqnarray}\label{f10}
 \frac{d}{d\Lambda }\left( L^{\Lambda ,\Lambda _0}(\varphi ) +
 I^{\Lambda ,\Lambda _0} \right) &=& 
 \frac{\hbar}{2} \langle \frac{\delta }{\delta \varphi }\,,
\dot C ^{\Lambda ,\Lambda _0} \frac{\delta } {\delta \varphi }\rangle
 L^{\Lambda ,\Lambda _0}(\varphi ) \nonumber\\  
&{}& - \frac{1}{2} \langle \frac{\delta }{\delta \varphi }
 L^{\Lambda ,\Lambda _0}(\varphi ) ,\dot C ^{\Lambda ,\Lambda _0}
 \frac{\delta }{\delta \varphi } L^{\Lambda ,\Lambda _0}(\varphi ) \rangle .
\end{eqnarray}
The form of eq.(\ref{f9}) strongly resembles the heat equation.
 Defining the functional Laplace operator 
\begin{equation}
\Delta _{\Lambda ,\Lambda _0} = \frac{1}{2}\langle
 \frac{\delta }{\delta \varphi }\,, C ^{\Lambda ,\Lambda _0}
 \frac{\delta }{\delta \varphi } \rangle \, ,
\end{equation} 
the unique solution of the differential equation (\ref{f9}) ,
 already given in the form (\ref{f5}), can also be written as
\begin{equation}\label{f11}
e^{-\frac{1}{\hbar}\left( L^{\Lambda ,\Lambda _0}(\varphi ) +
 I^{\Lambda ,\Lambda _0} \right)}
= e^{\hbar \Delta _{\Lambda ,\Lambda _0} }\,
 e^{-\frac{1}{\hbar} L^{\Lambda _0,\Lambda _0}(\varphi )} .
\end{equation}
Since $\Delta _{\Lambda ,\Lambda _0} $ commutes with its
 derivative $ \dot \Delta _{\Lambda ,\Lambda _0} $ with respect
 to $\Lambda $, the r.h.s. of (\ref{f11}) satisfies the
 differential equation. Moreover, the
initial condition holds because of $ \Delta _{\Lambda_0 ,\Lambda _0} = 0 $
 and $ I^{\Lambda _0 , \Lambda _0 } = 0 $.\\
At this point, several remarks concerning the mathematical aspect
of the steps performed are in order:\\
i) Our aim with these preparatory steps is to generate the
system of flow equations satisfied by the regularized
Schwinger functions of the theory, when considered in 
the perturbative sense of formal power series. This system then
is taken as the starting point for a proof of perturbative
renormalizability. As basic ``root'' acts the UV-regularized
finite-volume generating functional (\ref{f2}) or one of its
direct descendants (\ref{f3}),(\ref{f5}). Expanding in their
respective integrands the exponential function in a power
series would provide the standard perturbation expansion in terms
of (regularized) Feynman integrals. Bearing in mind our
goal stated, we could already view the steps performed
in the retricted sense as formal power series. \\
ii) We mention, that in \cite{Salm} to begin on safe ground 
the generating functional of the theory has first been
formulated on a finite space-time lattice in order to
derive the (perturbative) flow equation - implying a
finite-dimensional Gaussian integral -, and the limit
to continuous infinite space-time taken afterwards.\\
iii) Rigorous analysis beyond perturbation theory 
 of flow equations of the Wilson type (\ref{f10}) 
is the subject dealt with in \cite{Bry}, using convergent
expansion techniques. Such techniques are
developed in the monograph \cite{Riv2}. \\ 
The flow equation (\ref{f10}) for the generating functional
 $ L^{\Lambda, \Lambda_0} (\varphi) $
 encodes a system of flow equations for the corresponding 
$n$-point functions, $ n \in \mathbf{N} $, and for the vacuum part
 $ I^{\Lambda ,\Lambda _0} $. The flow of the latter is determined
 considering eq. (\ref{f10}) at $ \varphi =0 $. From the
 translation invariance of the theory it follows that the 2-point function
 is a distribution depending on the difference variable $x-y$ only,
 (suppressing momentarily the superscript $\Lambda ,\Lambda _0 $)
$$ \frac{\delta }{\delta \varphi (x)}\frac{\delta }{\delta \varphi (y)}
L(\varphi )\mid _{\varphi =0}\,  =: \mathcal{L}_2(x-y) ,$$
and thus
$$ \langle \frac{\delta }{\delta \varphi }\,, 
\dot C \frac{\delta }{\delta \varphi} \rangle L(\varphi )\mid _{\varphi =0}
 = \int_{\Omega } dx \int_{\Omega }dy \dot C(x-y) \mathcal{L}_2 (x-y)
    = \mid \Omega \mid \int_\Omega  dz \dot C (z) \mathcal{L}_2 (z) . $$
Because of the emerging dependence on the volume
 $ \mid \Omega \mid $ the flow equation of the vacuum part cannot
be treated in the infinite volume limit. However, due to the
 covariance (\ref{g7}) which corresponds to a massive particle and
 thus decays exponentially, the flow equation for the $n$-point functions 
can and in the sequel will be treated in this limit. Hence, at least
 one functional derivative has to act on the flow equation (\ref{f10}).

Due to the translation invariance of the theory it is convenient
 to consider the generating functional
$ L^{\Lambda ,\Lambda _0}(\varphi ) , 
\varphi \in {\mathcal S}(\mathbf{R}^4 ) $, in terms of the Fourier
 transformed source field $ \hat \varphi $, the conventions used are
\begin{equation}\label{f12}
\varphi (x) = \int_{p} e^{ipx} \hat \varphi (p) \ ,
 \quad \int_{p} := \int_{\mathbf{R}^4} \frac{d^{\,4} p}{(2\pi)^4}  \ ,
\end{equation}
implying for the functional derivative 
$ \delta _{\varphi (x)} := \frac{\delta }{\delta \varphi (x)} $ the
 transformation
$$  \delta _{\varphi (x)} = (2\pi )^4 
\int_{p} e^{-ipx} \ \delta _{ \hat \varphi (p)} \ . $$ 
From the generating functional
 $ L^{\Lambda ,\Lambda _0} (\varphi) $ the correlation
functions are obtained by functional derivation , $ n \in \mathbf{N} $,
\begin{equation}\label{f14}
(2\pi)^{4(n-1)} \delta _{\hat \varphi (p_n)} \cdots 
\delta _{\hat \varphi (p_1)}L^{\Lambda ,\Lambda _0}(\varphi )
 \mid _{\varphi = 0} 
 \, =  \delta ( p_1 + \cdots + p_n)
 \mathcal{L}^{\Lambda ,\Lambda _0}_n (p_1 , \cdots , p_n) .
\end{equation}
 The  amputated truncated $n$-point function
 $ \mathcal{L}^{\Lambda ,\Lambda _0}_n (p_1, \cdots ,p_n) $
 is a totally symmetric function of the momenta $ p_1 , \cdots ,p_n $ and,
 moreover, due to the $ \delta $ -function, $ p_n := - p_1 - \cdots - p_{n-1}$.
 (In the case where the bare interaction
(\ref{f1}) shows the mirror symmetry $ L^{\Lambda_0, \Lambda _0}(- \phi  )
 = L^{\Lambda_0, \Lambda _0}(\phi ) $ all $n$-point functions
 with $n$ odd vanish.) Observing the definition (\ref{f14})
we obtain from (\ref{f10}) the system of flow equations for
 the $n$-point functions, $n \in \mathbf{N} $,   
\begin{eqnarray}\label{f15}
\partial _{\Lambda }\mathcal{L}^{\Lambda ,\Lambda _0}_n (p_1, \cdots , p_n)
 = \frac{\hbar}{2}
  \int \limits_{k} \partial _{\Lambda } C^{\Lambda ,\Lambda _0}(k)
 \cdot \mathcal{L}^{\Lambda ,\Lambda _0}_{n+2} (k, p_1, \cdots , p_n, -k)
    {}  \nonumber\\
    - \frac{1}{2} \sum_{r=0}^{n} \sum_{i_1<\cdots <i_r}
    \mathcal{L}^{\Lambda ,\Lambda _0}_{r+1} (p_{i_{1}}, \cdots , p_{i_{r}}, p) 
\partial _{\Lambda } C^{\Lambda ,\Lambda _0}(p) \cdot
 \mathcal{L}^{\Lambda ,\Lambda_0}_{n-r+1}(-p, p_{j_{1}}, \cdots, p_{j_{n-r}})
          {} \nonumber\\
       p_1 + \cdots + p_n = 0, \qquad -p = p_{i_{1}} + \cdots + p_{i_{r}}.
    \qquad \qquad \qquad
\end{eqnarray} 
In the quadratic term a given set of momenta
 $ (p_{i_1}, \cdots , p_{i_r} ) , \, i_1 < \cdots < i_r \, $,
determines (uniquely) the corresponding set
 $ (p_{j_1} , \cdots , p_{j_{n-r}}), \, j_1 < \cdots <j_{n-r} \, $, such
 that the union of this pair is the set of momenta $ ( p_1, \cdots , p_n) $.
 Furthermore, the Fourier transform of the covariance (\ref{g7}), 
\begin{equation}\label{f16}
\hat C^{\Lambda ,\Lambda _0}(k) = \frac{1}{k^2 + m^2}
 \left( e^{-\frac{k^2 + m^2}{{\Lambda _0}^2}} -  
 e^{-\frac{k^2 + m^2}{\Lambda ^2}} \right) \, ,
\end{equation} 
is written with a slight abuse of notation  omitting the ``hat''.
 In the sequel we shall write the quadratic term appearing
 in (\ref{f15}) more compactly as
\begin{eqnarray}\label{f17}
 - \frac{1}{2} \sum_{r=0}^{n}  \sum_{i_1<\cdots <i_r} \cdots
 \qquad  \qquad \qquad \qquad  \qquad \qquad \qquad  \qquad  {} \nonumber\\
= - \frac{1}{2}\sum_{n_1, n_2}' \Bigl [
 \mathcal{L}^{\Lambda ,\Lambda _0}_{n_1 + 1} (p_1, \cdots,  p_{n_1}, p)
 \partial _\Lambda C^{\Lambda ,\Lambda _0}(p) \cdot
 \mathcal{L}^{\Lambda ,\Lambda _0} _{n_2 + 1} ( -p, p_{n_1 + 1}, \cdots , p_n)
 \Bigr]_{rsym} {} \nonumber\\
    p:= - p_1 - \cdots - p_{n_1} = p_{n_1 + 1} + \cdots + p_n \, ,
 \qquad  \qquad \qquad 
\end{eqnarray} 
where the prime on top of the summation symbol imposes the
 restriction to $ n_1 + n_2 = n $. Moreover, the symbol
 ``rsym" means summation over those permutations of the
 momenta $ p_1 , \cdots , p_n $, which do not leave invariant
 the (unordered)  subsets $ (p_1, \cdots , p_{n_1} ) $ and
 $ ( p_{n_1 + 1}, \cdots , p_n ) $, and, in addition, produce
mutually different pairs of (unordered ) image subsets. 
 The system of flow equations (\ref{f15}) will be treated
 perturbatively employing a loop expansion of the $ n $-point functions
 as  formal power series, $ n \in \mathbf{N} $,
\begin{equation}\label{f18}
 \mathcal{L}^{\Lambda ,\Lambda _0}_n (p_1, \cdots ,p_n) =
 \sum_{l=0}^{\infty} \hbar^l
       \mathcal{L}^{\Lambda ,\Lambda _0}_{l,n} (p_1, \cdots ,p_n) \,  .
\end{equation}    
 Since also flow equations for momentum derivatives of 
 $ n $-point functions have to be considered, we introduce
 the shorthand notation
\begin{eqnarray}\label{f19}
 w = (w_{1,1} , \cdots , w_{n-1,4} ) ,
 \quad w_{i,\mu } \in \mathbf{N}_0  ,   
 \quad \mid w\mid = \sum_{i,\mu } w_{i,\mu }\, \nonumber \\
\partial ^w :\, = \prod_{i=1}^{n-1} \prod_{\mu =1}^{4}
 \Bigl(\frac{\partial }{\partial p_{i,\mu }} \Bigr)^{w_{i,\mu }}, 
 \qquad w ! = \, \prod_{i=1}^{n-1} \prod_{\mu =1}^{4} w_{i, \mu} ! \,.
\end{eqnarray} 
From (\ref{f15}) then follows the system of flow equations,
 $\, n \in \mathbf{N} , l \in \mathbf{N}_0 \,$ : 
\begin{eqnarray}\label{f20}
\partial _\Lambda  \partial^w
 \mathcal{L}^{\Lambda ,\Lambda _0}_{l,n}(p_1,\cdots, p_n) \, =
 \, \frac{1}{2}\int \limits_{k} \partial _\Lambda C^{\Lambda ,\Lambda _0}(k)
 \cdot \partial ^w \mathcal{L}^{\Lambda ,\Lambda _0}_{l-1,n+2}
(k,p_1,\cdots,p_n,-k)  \nonumber\\
 - \frac{1}{2} \sum_{n_1,n_2}' \sum_{l_1,l_2}'\sum_{w_1,w_2,w_3}' c_{\{w_i\}}
 \Bigl[ \partial ^{w_1}\mathcal{L}^{\Lambda ,\Lambda _0}_{l_1,n_1 +1} 
(p_1, \cdots, p_{n_1}, p )\cdot 
\partial ^{w_3}\partial _{\Lambda}C^{\Lambda ,\Lambda _0}(p) 
 \nonumber\\ \cdot \, \partial ^{w_2}
\mathcal{L}^{\Lambda ,\Lambda _0}_{l_2,n_2 +1}(-p,p_{n_1 +1}, \cdots, p_n)
 \Bigr]_{rsym}  \quad  \nonumber\\
p = -p_1 - \cdots -p_{n_1} \, = \, p_{n_1 +1} + \cdots +p_n  .\quad
\end{eqnarray} 
 One should not overlook that the residual symmetrization rsym
 acts on the momentum $ p $, too.The primes restrict the summations to
 $ n_1+n_2 = n, l_1+l_2=l, w_1+w_2+w_3=w,$ respectively.
Moreover, the combinatorial factor 
$ c_{\{w_i\}} = w! (w_1 ! w_2 ! w_3 !)^{-1} $ comes from Leibniz's rule.
 In the loop
 order $l=0 $,\, obviously, the first term on the r.h.s. is absent. 
%%%%%%%%%%%%%%%%%%%%%%%%%%%%%%%%%%%%%%%%%%%%%%%%%%%%%%%%%%
\section{Proof of perturbative renormalizability}
Perturbative renormalizability of the regularized field theory
 (\ref{f5}) amounts to the following: For given coupling constants $ f,g $ 
in the bare interaction (\ref{f1}) the coefficients
 $ v(\Lambda _0),a(\Lambda_0), z(\Lambda_0), b(\Lambda_0), c(\Lambda_0) $ of
 the counterterms can be adjusted within a loop expansion of the theory, i.e. 
\begin{equation} \label{p1}
  v(\Lambda_0) = \sum_{l=1}^{\infty} {\hbar}^l v_l(\Lambda_0) \, ,
 \cdots \cdots \, , c(\Lambda_0) = \cdots
\end{equation}
in such a way, that all (infinite volume) $n$-point functions (\ref{f18})
 in every loop order $l$ have finite
limits
\begin{equation} \label{p2}
\lim_{\Lambda_0 \to \infty} \lim_{\Lambda \to 0} \Lf_{l,n}(p_1, \cdots , p_n)
 \, , \quad n\in \mathbf{N}   , \, l\in \mathbf{N_0} \, .
\end{equation}
These limits emerge directly in the tree order $ l=0 $, of course.
 The counterterms are adjusted by requiring that the corresponding
 $n$-point functions at the physical value $\Lambda = 0 $ of the flow
parameter have prescribed values for a chosen set of momenta.
 Since the theory is massive it is convenient to prescribe
 these \emph{ renormalization conditions} at vanishing momenta. Hence, taking
into account the Euclidean symmetry we require for all $ l \in \mathbf{N} $ :
\begin{eqnarray} 
   \Lp_{l,1} & = & v_l^R ,  \label{p3} \\
  \Lp_{l,2}(p,-p) & = & a_l^R + z_l^R p^2 + \mathcal{O}\bigl((p^2)^2 \bigr) ,
 \label{p4}   \\
  \Lp_{l,3}(p_1,p_2,p_3) & = & b_l^R + \mathcal{O}(p_{1}^2, p_{2}^2, p_{3}^2 )
  ,  \label{p5}   \\
  \Lp_{l,4}(0,0,0,0) & = & c_l^R . \label{p6}
\end{eqnarray}
In each loop order $ l \in \mathbf{N} $ these five real
 \emph{renormalization constants} $ v_l^R, \cdots , c_l^R $ 
can be chosen freely, not depending \footnote{ A weak dependence of
 these constants  on $\Lambda_0 $ with finite limits when
 $ \Lambda_0 \to \infty $ could be permitted.} on $\Lambda_0 $ . Together
 with the corresponding constants of the tree order $l=0$ they
 fix the relevant part of the theory
 completely. A particular (simple) choice would be to set
 $ v_l^R = a_l^R = z_l^R = b_l^R = c_l^R = 0 $. 

 The tree order has to be treated first. It is fully determined
 by the classical part appearing in the bare interaction (\ref{f1}).
 This classical interaction acts as initial condition at
 $ \Lambda = \Lambda_0 $ when integrating the flow equations (\ref{f20})
 for $ l=0 $ downwards to smaller values of $ \Lambda $, ascending
 successively in the number of fields $ n $. The classical interaction
contains no terms linear or quadratic in the fields. To bring the
 system of flow equations to bear, however,
at first the crucial properties, $ 0 \le \Lambda \le \Lambda_0 $,
\begin{equation} \label{p7}
\Lf_{0,1} = 0 \, , \quad \Lf_{0,2}( p, -p) = 0  .
\end{equation}
have to be inferred directly from the representation (\ref{f5}). Hereupon
 and with the initial condition for $ n = 3 $ follows from (\ref{f20})
\begin{equation} \label{p8}
   \Lf_{0,3}( p_1, p_2, p_3 ) = f ,
\end{equation}
and then for $ n=4 $ :
\begin{eqnarray}\label{p9}
\Lf_{0,4}(p_1, p_2, p_3, p_4) = g  \qquad \qquad \qquad \qquad \qquad
 \qquad \qquad \qquad \qquad \nonumber  \\
- f^2 \left ( C^{\Lambda,\Lambda_0}(p_1 + p_2)
 + C^{\Lambda,\Lambda_0}(p_1 + p_3)
 + C^{\Lambda,\Lambda_0}(p_1 + p_4) \right ) .
\end{eqnarray}
Ascending further in the number of fields yields the whole tree order.
 (For $ n > 4 $ all initial conditions 
 at $ \Lambda = \Lambda_0 $ are equal to zero.)

The first step in proving renormalizability is to establish the

\textbf{Proposition 2.1} (Boundedness) \\
 \emph{ For all} $ l \in \mathbf{N}_0 , n \in \mathbf{N} $ ,
  $ w $ \emph{from (\ref{f19}) and for}
 $ 0 \le \Lambda \le \Lambda_0 $ \emph{holds} 
\begin{equation} \label {p10}
| \partial ^w  \Lf_{l,n}(p_1, \cdots ,p_n) | \le (\Lambda + m )^{4-n-|w|}\,
  \mathcal{P}_1 (log \frac{\Lambda + m}{m})
 \mathcal{P}_2 (\{ \frac{|p_i|}{\Lambda +m} \} ) \, ,
\end{equation}
 \emph{where $ \mathcal{P}$ denotes polynomials having 
nonnegative coefficients. These coefficients, as
 well as the degree of the polynomials, depend on $ l,n,w $ but
 not on $ \{p_i \}, \Lambda, \Lambda_0 $.
For $l=0 $ all polynomials $ \mathcal{P}_1 $ reduce to positive constants.} 
\\ \emph{Remark}: In the following the symbol $\mathcal{P} $ always
 denotes a polynomial of this type, possibly
 a different one each time it appears.

\emph{Proof}: Using again the shorthand (\ref{f19}) in the
 case of one momentum the covariance
(\ref{f16}) satisfies the bounds, $ 0 \le \Lambda $,
\begin{equation} \label{p11}
|\partial ^w \partial _{\Lambda} C^{\Lambda,\Lambda_0}(k)| \le \Lambda^{-3-|w|}
  \mathcal{P}\bigl(\frac{|k|}{\Lambda}\bigr ) e^{-\frac{k^2+m^2}{\Lambda^2}} .
\end{equation}
Here, the polynomials $\mathcal{P} $ are of respective degree $|w| $
 (and obviously do not depend on $ n,l $ ). A weaker version, used too, is
\begin{equation} \label{p12}
  |\partial ^w \partial _{\Lambda} C^{\Lambda,\Lambda_0}(k)|
 \le (\Lambda +m)^{-3-|w|}
    \mathcal{P} \bigl( \frac{|k|}{\Lambda+m} \bigr)  .
\end{equation}
We first consider the tree order $l=0$ . Due to (\ref{p7}),
 (\ref{p8}) the bounds (\ref{p10}) evidently
hold for $ n \le 3 $. From (\ref{p9}) and the very crude bound
 $ |C^{\Lambda,\Lambda_0}(k)| < 2m^{-2} $ follows the claim for 
$ (n=4,w=0) $. Now in all remaining cases we have $ n+|w|>4$ .
 Due to the crucial
properties (\ref{p7}) they can be treated successively ascending
 in $n$, and for given $n$ the various
$ w $ dealt with in arbitrary order,
 by integrating the respective flow
equation (\ref{f20}) downwards from the initial point
 $ \Lambda = \Lambda_0 $. In each such case
the initial condition is equal to zero. Then bounds already
 established together with (\ref{p12}) yield 
\begin{eqnarray} \label{p13}
|\partial ^w \Lf_{0,n}(p_1, \cdots ,p_n)| & 
\le & \int_{\Lambda}^{\Lambda_0} d\lambda
|\partial _{\lambda}\partial ^w 
\mathcal{L}^{\lambda,\Lambda_0}_{0,n}(p_1, \cdots ,p_n)|
   \qquad  \qquad \qquad \quad \nonumber \\
 & \le & \mathcal{P}\bigl(\{\frac{|p_i|}{\Lambda+m}\}\bigr)
 \int_{\Lambda}^{\Lambda_0} d\lambda
     (\lambda+m)^{4-n-|w|-1}    \nonumber  \\
 & < & \mathcal{P}\bigl(\{\frac{|p_i|}{\Lambda+m}\}\bigr) \frac{1}{n+|w|-4}
                       (\Lambda+m)^{4-n-|w|} \, .    
\end{eqnarray}
Thus the assertion (\ref{p10}) is shown for the tree order.

Given the bounds for $l=0 $, those of the higher loop orders can
 be generated inductively by successive
integration of the system of flow equations (\ref{f20}):
 i) Ascending in the loop order $l$, ii) for fixed $l$
ascending in $n$, iii) for fixed $l,n$ descending with
 $w$ down to $ w=0$.
 We observe that in the inductive
order adopted the terms on the r.h.s. of a flow equation are
 always prior to that on the l.h.s. since the
linear term has lower loop order and to the quadratic 
term - because of the key properties (\ref{p7}) - 
 only terms of the same loop order contribute which have a smaller
 value $ n $.  To comply with the growth properties of the bounds
(\ref{p10}) the integrations are performed as follows:

$A_1$) If $ n+|w|>4 $, the bound decreases with increasing
 $\Lambda $. Hence, the flow equation is integrated from the
 initial point $\Lambda = \Lambda_0 $ downwards to smaller
 values of $\Lambda $ with the initial condition
\begin{equation} \label{p14}
  \partial ^w \Lb_{l,n}(p_1, \cdots,p_n) \, = \, 0 \, ,\quad  n+|w|>4 \, ,
\end{equation}
as a consequence of the bare interaction (\ref{f1}) chosen.

$A_2$) In the cases $ n+|w| \le 4 $ the bounds increase with
 increasing $ \Lambda $. Therefore, the
corresponding flow equations are integrated for a
prescribed set of momenta (the \emph{renormalization point})
 with the physical value $ \Lambda = 0 $
as initial point. The respective initial values can be freely chosen
 order by order, but in accordance
with the (Euclidean) symmetry of the theory.
 As already stated before we choose vanishing momenta
as renormalization point, together with the renormalization
 conditions (\ref{p3} - \ref{p6}) as initial values.
 Thus, for $ n+|w| \le 4  $ ,
\begin{equation} \label{p15}
\partial ^w \Lf_{l,n}(0,\cdots,0) = \partial ^w \Lp_{l,n}(0,\cdots,0 ) \, +\,
 \int_{0}^{\Lambda} d\lambda  \partial _{\lambda}
 \partial ^w \mathcal{L}_{l,n}^{\lambda,\Lambda_0} (0, \cdots ,0) \, .
\end{equation}
(For $ n=1 $ there is no momentum dependence and $ w=0 $ .) Once a bound
has been obtained at the renormalization point,
 it is extended to general momenta using the Taylor formula
\begin{equation} \label{p16}
f(p) = f(0) + \sum_{i=1}^{n} p^{i} \int_{0}^{1} dt (\partial _i f) (tp) 
\end{equation}
for a differentiable function on $ \mathbf{R}^n $ . Applying this formula, 
 the bound of the integrand ( due to the
derivative ) yields an additional factor $ (\Lambda +m)^{-1} $
 which combines  with the momentum factor in front to give
 a new momentum bound of the type considered. 

To generate inductively the assertion (\ref{p10}) we use it
 in bounding the r.h.s. of the flow equation 
(\ref{f20}), together with the bounds (\ref{p11}) and (\ref{p12})
 in the linear and in the quadratic term, respectively,
\begin{eqnarray}
 &|&\partial_{\Lambda} \partial ^w \Lf_{l,n}(p_1, \cdots, p_n) \, |
  \nonumber   \\
& \le & \int_{k} \frac{e^{- \frac{k^2+m^2}{\Lambda^2}}}{\Lambda^3}
 (\Lambda +m)^{4-n-2-|w|}
  \mathcal{P}_1 (\log\frac{\Lambda+m}{m})
 \mathcal{P}_2 \bigl( \frac{|k|}{\Lambda+m}, 
   \{\frac{|p_i|}{\Lambda+m} \} \bigr )    \nonumber  \\
 & +  &  (\Lambda +m)^{4-n-|w|-1}
  \mathcal{P}_3 (\log\frac{\Lambda+m}{m}) 
\mathcal{P}_4 \bigl(  \{\frac{|p_i|}{\Lambda+m} \} \bigr ) \, . \nonumber
\end{eqnarray}
The second term on the r.h.s. results from combining a sum of 
such terms into a single one with new 
polynomials. In the first term the $k$-integration is performed
 substituting $ k \rightarrow \Lambda k$ .
 The result, easily majorized  and combined with the second term 
 yields the bound
\begin{equation} \label{p17}
 | \partial_{\Lambda} \partial ^w \Lf_{l,n}(p_1, \cdots, p_n) | \, \le \,
  (\Lambda +m)^{4-n-|w|-1}
  \mathcal{P}_5 (\log\frac{\Lambda+m}{m})
    \mathcal{P}_6 \bigl( \{\frac{|p_i|}{\Lambda+m} \} \bigr ) .
\end{equation}
$\, \quad a_1)$ Following the order of the induction stated before,
 the (irrelevant) cases $ n+|w|> 4 $ have always
to be considered first (for fixed $ l,n $ ). In these cases 
the bound (\ref{p17}) is integrated downwards, observing
 (\ref{p14}) , similarly as in the tree order, (\ref{p13}).
 In place of the pure power behaviour, however, we now 
 have 
$$ \int_{\Lambda}^{\Lambda_0} d \lambda (\lambda+m)^{4-n-|w|-1}
  \mathcal{P} \bigl( \log \frac{\lambda+m}{m} \bigr) < (\Lambda+m)^{4-n-|w|}
  \mathcal{P}_1 \bigl( \log \frac{\Lambda+m}{m} \bigr)   $$
with a new polynomial on the r.h.s. , see the end
   of this chapter, section 2.6. 
Thus, the assertion is established in the cases  $ n+|w| > 4 $ . 

$ a_2) $ In the cases $n+|w| \le 4 $ the claim (\ref{p10}) has to
 be deduced from the respective integrated
flow equation (\ref{p15}) at the renormalization point 
followed by an extension to general momenta 
by way of (\ref{p16}), proceeding in the order of induction.
 That is, to start with the particular
(momentum independent) case $n=1$ and continue successively with
 the cases $(n=2,|w|=2), (n=2,|w|=1)$, and so on.
 Converting in the obvious way equation (\ref{p15}) into an inequality
for absolute values, a bound on the integral is gained
 using the bound (\ref{p17}) at vanishing momenta:
\begin{eqnarray}
{\bigg |} \int_{0}^{\Lambda}d\lambda \partial _{\lambda} \partial ^w
 \mathcal{L}_{l,n}^{\lambda,\Lambda_0} (0,\cdots,0){\bigg |}
 & \le & \int_{0}^{\Lambda}d \lambda (\lambda+m)^{4-n-|w|-1}
       \mathcal{P} \bigl( \log \frac{\lambda+m}{m} \bigr )    \nonumber  \\
  & \le & (\Lambda +m )^{4-n-|w|} \mathcal{P}_1
 \bigl( \log \frac{\Lambda+m}{m} \bigr ) , \nonumber
\end{eqnarray}         
where $\mathcal{P}_1 $ is a new polynomial, see section 2.6.
 Hence, the assertion (\ref{p10}) is
established at the renormalization point. In each case extension
 to general momenta  via (\ref{p16})
is guaranteed by bounds established before.
 This concludes the proof of Proposition 2.1 .

The boundedness due to Prop.2.1 would still allow an oscillatory dependence
 on $\Lambda_0 $. Such a (implausible) behaviour is excluded by the \\
\textbf{Proposition 2.2 } ( Convergence) \\
\emph{ For all $ l \in \mathbf{N}_0 , n \in \mathbf{N} , 
 w $ from (\ref{f19})
 and for $ 0 \le \Lambda \le \Lambda_0 $ holds } 
\begin{equation} \label{p18}
|\partial _{\Lambda_0}\partial ^w \Lf_{l,n}(p_1, \cdots,p_n) |
 \le \frac{(\Lambda+m)^{5-n-|w|}}
  {(\Lambda_0+m)^2} \mathcal{P}_3 \bigl(\log \frac{\Lambda_0+m}{m} \bigr )
  \mathcal{P}_4 \bigl( \{ \frac{|p_i|}{ \Lambda+m} \} \bigr ) .
\end{equation}
Since we need this Proposition for large values of $\Lambda_0 $ only,
 we then obviously can write
\begin{equation} \label{p18a}
 |\partial _{\Lambda_0}\partial ^w \Lf_{l,n}(p_1, \cdots,p_n) | \le 
 \frac{(\Lambda+m)^{5-n-|w|}}{(\Lambda_0)^2}
 \Bigl(\log \frac{\Lambda_0}{m} \Bigr)^\nu 
 \mathcal{P}_4 \bigl( \{ \frac{|p_i|}{ \Lambda+m} \} \bigr ) 
\end{equation}
with a positive integer $\nu $ depending on $ l,n,w $.
Integration of these bounds with respect to $\Lambda_0 $ finally
 shows that for fixed $\Lambda$ all
$ \Lf_{l,n}(p_1, \cdots,p_n) $ converge to finite limits with
 $\Lambda_0 \rightarrow \infty $. In particular, one obtains
for all $\Lambda_0' > \Lambda_0$ :
$$ | \Lp_{l,\, n}(p_1, \cdots, p_n) - 
              {\mathcal{L}}_{l,\, n}^{0,\Lambda_0'}(p_1, \cdots, p_n)| 
< \frac{m^{5-n}}{\Lambda_0}\Big( log\frac{\Lambda_0}{m}\Big)^{\nu} 
  \mathcal{P}_{\,5}(\{\frac{|p_i|}{m}\})\, .$$ 
 Thus, due to the Cauchy criterion, finite limits (\ref{p2}) exist, i.e.
 perturbative renormalizability of
 the theory considered is demonstrated.  \\
\emph{Proof of Proposition 2.2 }: We integrate the system of
 flow equations (\ref{f20}) according to the 
induction scheme employed before and derive the
 individual $n$-point functions with respect to
$\Lambda_0 $. The r.h.s. of (\ref{f20}) will be denoted by the
 shorthand $ \partial ^w \Rf (p_1, \cdots , p_n) $ .
 Due to ( \ref{p7}-\ref{p9}) the cases $ (l=0, n+|w| \le 4 ) $
 evidently satisfy the claim 
(\ref{p18}) .\\
$ b_1) \quad n+|w| > 4 $: In these cases, because of the
 initial condition (\ref{p14}), we have
$$ - \, \partial ^w \Lf_{l,n}(p_1,\cdots,p_n) \, = \,
 \int_{\Lambda}^{\Lambda_0} d \lambda \partial^w
     \Ri (p_1,\cdots,p_n)  $$
and hence
\begin{eqnarray} \label{p19}
- \, \partial _{\Lambda_0} \partial ^w \Lf_{l,n}(p_1,\cdots,p_n)  & = &
 \partial ^w \Rb(p_1,\cdots,p_n)  \nonumber   \\
 & {} & + \int_{\Lambda}^{\Lambda_0} d \lambda \partial _{\Lambda_0}
 \partial ^w \Ri(p_1,\cdots,p_n) .
\end{eqnarray} 
To the first term on the r.h.s. only the quadratic part of (\ref{f20})
contributes, cf. (\ref{p14}). It is bounded
 using Proposition 2.1 and the bound (\ref{p12}) :
\begin{eqnarray} \label{p20}
|\partial ^w \Rb (p_1,\cdots,p_n) |& \le & 
(\Lambda_0+m)^{3-n-|w|} \mathcal{P}_1 \bigl( \log
              \frac{\Lambda_0+m}{m} \bigr) \mathcal{P}_2
           \bigl (\{ \frac{|p_i|}{\Lambda_0+m} \}\bigr)    \nonumber   \\
&\le & \frac{(\Lambda+m)^{5-n-|w|}}{(\Lambda_0+m)^2} \mathcal{P}_1 \bigl( \log
              \frac{\Lambda_0+m}{m} \bigr) \mathcal{P}_2
    \bigl (\{ \frac{|p_i|}{\Lambda+m} \}\bigr) ,  \nonumber  \\
\end{eqnarray}
valid for $ 0 \le \Lambda \le \Lambda_0 $, since $ n+|w| > 4 $.
 The integrand of the second term on the r.h.s. of (\ref{p19}) is
 the derivative with respect to $\Lambda_0 $ of the r.h.s.
 of (\ref{f20}). Observing
\begin{equation} \label{p21}
  \partial _{\Lambda_0} \partial _\Lambda C^{\Lambda , \Lambda_0} (k) \,
      = \, 0 ,
\end{equation}
we bound the $\Lambda_0 $ - derivative considered using Propositions 2.1 and
- in accord with the induction hypothesis -  2.2  together with 
the bounds (\ref{p11}) and (\ref{p12}) which are employed in the
 linear and in the quadratic part, respectively. 
Proceeding then similarly as in deducing (\ref{p17}) yields
\begin{equation} \label{p22}
|\partial _{\Lambda_0} \partial ^w \Rf (p_1,\cdots,p_n) | \le
 \frac{(\Lambda+m)^{5-n-|w|-1}}{(\Lambda_0+m)^2}
        \mathcal{P}_7 \bigl( \log \frac{\Lambda_0+m}{m} \bigr)
 \mathcal{P}_8 \bigl (\{ \frac{|p_i|}{\Lambda+m} \}\bigr) .
\end{equation}
From this follows upon integration, with the bound on the momenta
 majorized, a bound on the second term on the r.h.s. of (\ref{p19})
 that has the form (\ref{p20}). Therefore,
the assertion (\ref{p18}) is deduced if $ n+|w| > 4 $. \\
$ b_2) \quad  n+|w| \le 4 $. Here, the respective flow equations
 integrated at the renormalization point 
 (\ref{p15}) are derived with respect to $\Lambda_0 $. They imply,
 observing that the initial conditions, i.e. the renormalization constants
 (\ref{p3}-\ref{p6}) do not depend on $\Lambda_0 $, the bound
\begin{equation} \label{p23}
|\partial _{\Lambda_0} \partial ^w \Lf_{l,n}(0,\cdots,0) | \, \le \,
 \int_{0}^{\Lambda} d \lambda
     | \partial _{\Lambda_0} \partial ^w \Ri (0,\cdots,0) | ,
\end{equation}
where on the r.h.s. the shorthand introduced before has been used.
 In deducing the bound (\ref{p22})
 no restriction on $ n,w $ entered. Therefore, we can use it in
 (\ref{p23}) also and obtain upon integration
\begin{equation} \label{p24}
|\partial _{\Lambda_0} \partial ^w \Lf_{l,n}(0,\cdots,0) | \, \le \,
 \frac{(\Lambda+m)^{5-n-|w|}}{(\Lambda_0+m)^2}
 \mathcal{P}_3 \bigl( \log \frac{\Lambda_0+m}{m} \bigr) .
\end{equation}
Extension of these bounds  to general momenta is again achieved
 via the Taylor formula (\ref{p16}) as in the proof
 of Proposition 2.1. Thus, Proposition 2.2 is proven.

\emph{Remarks}: Renormalizability is a consequence of
 Proposition 2.2 at the value $\Lambda = 0 $.
 From this
point of view Proposition 2.1 is of preparatory, technical nature.
 The bounds established in both Propositions are
not optimal but sufficient. Their virtue is to allow a concise
 and complete proof of renormalizability. 
These bounds can be refined in various ways. We mention that
 Kopper and Meunier \cite{Ko+Me},  
by sharpening the induction hypothesis with respect to momentum
 derivatives of $n$-point functions, obtained optimal bounds on the
 momentum behaviour related to Weinberg's theorem \cite{Wei}.

The Propositions 2.1 and 2.2 established, it is physically
 important to notice that they even remain
valid, when the original bare interaction (\ref{f1}) is extended
 by appropriately chosen irrelevant terms: It is
sufficient to replace the condition (\ref{p14}) by requiring for $ n+|w|>4 $ :
\begin{eqnarray}  \label{p25}
|\partial ^w\Lb_{l,n}(p_1,\cdots,p_n)| & \le &
 (\Lambda_0 +m)^{4-n-|w|} \mathcal{P}_1
   \bigl(\log \frac{\Lambda_0+m}{m} \bigr)
 \mathcal{P}_2 \bigl(\{ \frac{|p_i|}{\Lambda_0+m}\} \bigr) , \nonumber   \\
|\partial _{\Lambda_0}\partial ^w\Lb_{l,n}(p_1,\cdots,p_n)| & \le &
 (\Lambda_0 +m)^{3-n-|w|} \mathcal{P}_3
   \bigl(\log \frac{\Lambda_0+m}{m} \bigr)
 \mathcal{P}_4 \bigl(\{ \frac{|p_i|}{\Lambda_0+m}\} \bigr) ;
           \nonumber    \\
\end{eqnarray}
evidently, we can also write $\Lambda_0 $ instead of $ \Lambda_0+m $
 everywhere. One first observes
that these bounds agree with Propositions 2.1 and 2.2 considered
 at $ \Lambda = \Lambda_0 $. 
Moreover, as bounds on the initial conditions to be
 added in (\ref{p13}), $a_1) $ and (\ref{p19}), 
respectively, they can be absorbed in the corresponding bounds
 on the integrals appearing.
%%%%%%%%%%%%%%%%%%%%%%%%%%%%%%%%%%%%%%%%%%%%%%%%%%%%%%%%%%%%%%%%
\section{Insertion of a composite field}  
 Besides the system of $n$-point functions dealt with up to now,
 $n$-point functions with one or more additional 
composite fields  inserted are of considerable physical interest.
 In particular, the generators 
of symmetry transformations of a theory appear generally in
 the form of composite fields. But there are further instances where
 inserted composite fields -- sometimes also
 called inserted operators -- occur. 
In the sequel we treat the perturbative renormalization of \emph{one}
 composite field inserted. Since, by definition, a composite field
 depends nonlinearly on the basic field (or fields) of the theory considered,
new divergences have to be circumvented and hence additional 
renormalization conditions are required.

As before, we examine the quantum field theory of a real 
scalar field $\phi(x) $ with mass $m$ in 
four-dimensional Euclidean space-time. Then, a composite 
field $ Q(x) $ is a local polynomial formed in
general of the field $\phi(x) $ and of its space-time derivatives.
 It is determined by its classical version
$ Q_{class}(x) $. If we restrict to achieve a renormalized theory
 with \emph{one} insertion, $Q(x)$ has to be chosen as follows:
 Let $ Q_{class}(x) $ be a monomial having the canonical mass
 dimension $D$, then
\begin{equation} \label{i1}
  Q(x) \, = \, Q_{class}(x) + Q_{c.t.}(x) ,
\end{equation}
where $Q_{c.t.}(x)$ is a polynomial which is formed of all local terms
 of canonical mass dimension $\le D$.
 This polynomial  $Q_{c.t.}(x)$ acts as  counterterm.
 If $Q_{class}(x)$ shows a symmetry not violated in the 
intermediate process of regularization this symmetry can be imposed
 on $Q_{c.t.}(x)$ , too. Since the regularization (\ref{g7})
keeps the Euclidean symmetry the counterterms $Q_{c.t.}(x)$ can be 
restricted to those showing the
same tensor type as  $ Q_{class}(x) $. We illustrate the notion
 introduced with the example of a scalar composite field having $D=3$:
\begin{eqnarray}
Q_{class}(x) & = & \frac{1}{3!}\phi (x)^3 ,   \label{i2}  \\
Q_{c.t.}(x) & = &  \frac{1}{3!} r_1(\Lambda_0) \phi (x)^3 -
      r_2(\Lambda_0) \Delta  \phi (x) 
      +\frac{1}{2!} r_3(\Lambda_0) \phi (x)^2   \nonumber  \\
  & {}& +  r_4(\Lambda_0) \phi (x) + r_5(\Lambda_0) ,  \label{i3}
\end{eqnarray}
with coefficients $ r_i(\Lambda_0) = \mathcal{O}(\hbar) , i=1,\cdots,5$.
 Our aim here is to show the renormalizability of the theory considered
 in section 2.3 with one insertion of a scalar composite field
$Q(x)$ of mass dimension $D$. It will turn out that we can
 essentially proceed as before, taking
minor modifications into account. Hence, we can refrain
 from repeating definitions and arguments already
introduced. In place of the bare interaction (\ref{f1})
 one starts with a modified one:
\begin{eqnarray} \label{i4}
\tilde{L} ^{\Lambda_0,\Lambda_0}(\varrho \, ;\phi ) +
 \tilde I^{\Lambda_0, \Lambda_0}(\varrho )
& = & L^{\Lambda_0,\Lambda_0}(\phi )
     + \int dx \varrho (x) Q(x) ,   \\
 \tilde{L} ^{\Lambda_0,\Lambda_0}(\varrho \, ; 0 ) & = & 0 ,   \nonumber
\end{eqnarray}
where the composite field (\ref{i1}), coupled to an external
 source $\varrho \in C^{\infty}(\Omega ) $,
has been added.\footnote{$ \tilde I^{\Lambda_0, \Lambda_0}(\varrho ) $
 is the field independent part 
that possibly has to enter the modified bare action, as e.g. in (\ref{i3}). }
  Then, as in (\ref{f2}), the generating functional 
of regularized Schwinger functions
with insertions $Q$ is obtained upon functional integration:
\begin{equation} \label{i5}
\tilde{Z}^{\Lambda ,\Lambda _0} (\varrho \, ;J)
       = \int d\mu _{\Lambda ,\Lambda _0}(\phi ) e^{- \frac{1}{\hbar}
     \bigl (\tilde{L}^{\Lambda _0,\Lambda _0}(\varrho \, ;\phi ) +
       \tilde {I}^{\Lambda_0, \Lambda_0}(\varrho ) \bigr ) 
             +\frac{1}{\hbar}\langle \phi , J \rangle } .
\end{equation}
Moreover, passing similarly as before to regularized 
amputated truncated Schwinger functions with insertions,
 the equations (\ref{f5}), (\ref{f6}) are replaced by
\begin{eqnarray} \label{i6} 
 e^{-\frac{1}{\hbar}\left(\tilde{L}^{\Lambda ,\Lambda _0}(\varrho\,;\varphi )
 +\tilde {I}^{\Lambda ,\Lambda _0} (\varrho ) \right)}
  & = &\int d\mu _{\Lambda ,\Lambda _0}(\phi )
 e^{-\frac{1}{\hbar}\bigl(\tilde{L}^{\Lambda_0 ,\Lambda_0}( \varrho \,;\phi
 +\varphi ) +\tilde {I}^{\Lambda_0, \Lambda_0}(\varrho ) \bigr ) },
   \qquad \\ \label{i7}
         \tilde{L}^{\Lambda ,\Lambda _0}(\varrho \, ;0) & = & 0 \, .
     \end{eqnarray}
In view of the implicit notation (\ref{i1}) we stress that the
 shift of the field to $\phi +\varphi $
involves the field dependent insertion $ Q(x) $ in (\ref{i4}), too.
In exactly the same way as (\ref{f7}) was obtained, we find 
the relation between the generating functionals
$ \tilde L $ and $ \tilde Z $ :
\begin{equation} \label{i8}
 e^{-\frac{1}{\hbar}\left(\tilde{L}^{\Lambda ,\Lambda _0}(\varrho \,;\varphi )
 +\tilde {I}^{\Lambda ,\Lambda _0} (\varrho ) \right)} = 
e^{-\frac{1}{2}
  \langle \varphi , (\hbar C^{\Lambda ,\Lambda _0})^{-1} \varphi \rangle }
          \tilde Z^{\Lambda ,\Lambda _0}( \varrho \, ;
       ( C^{\Lambda ,\Lambda _0})^{-1} \varphi ) .
\end{equation}
Since (\ref{i6}) and (\ref{f5}) have the same form we obtain the
 flow equation of the functional
$ \tilde{L}^{\Lambda ,\Lambda _0}(\varrho \, ;\varphi ) $ by
 substituting in the flow equation (\ref{f10}):
$$  L^{\Lambda ,\Lambda _0}(\varphi ) \rightarrow 
 \tilde{L}^{\Lambda ,\Lambda _0}(\varrho \, ;\varphi ) 
  \quad , \quad  I^{\Lambda, \Lambda_0} \rightarrow 
 \tilde{I}^{\Lambda, \Lambda_0}(\varrho ) .$$
As a consequence the generating functional of the amputated
 truncated Schwinger functions with one 
insertion $Q$ ,
\begin{equation} \label{i9}
L_{(1)}^{\Lambda, \Lambda_0}(x ;\varphi ) := \frac{\delta }{\delta \varrho (x)}
 \tilde{L}^{\Lambda ,\Lambda _0}(\varrho \, ;\varphi ) |_{\varrho (x) = 0 }
   \,  ,
\end{equation}
then satisfies the flow equation
\begin{eqnarray}\label{i10}
 \frac{d}{d\Lambda }\left( L_{(1)}^{\Lambda ,\Lambda _0}( x ;\varphi )
 + I_{(1)}^{\Lambda ,\Lambda _0}(x) \right) &=& 
 \frac{\hbar}{2} \langle \frac{\delta }{\delta \varphi }\,,
\dot C ^{\Lambda ,\Lambda _0} \frac{\delta } {\delta \varphi }\rangle
 L_{(1)}^{\Lambda ,\Lambda _0}(x ;\varphi ) \\  
& \, & - \,  \langle \frac{\delta }{\delta \varphi }
 L^{\Lambda ,\Lambda _0}(\varphi ) ,\dot C ^{\Lambda ,\Lambda _0}
   \frac{\delta }{\delta \varphi }                     
 L_{(1)}^{\Lambda ,\Lambda _0}( x ;\varphi ) \rangle , \nonumber
\end{eqnarray}
involving the vacuum part with one insertion
\begin{equation} \label{i11}
I_{(1)}^{\Lambda, \Lambda_0}(x) := \frac{\delta }{\delta \varrho (x)}
       \tilde{I}^{\Lambda ,\Lambda _0}(\varrho) |_{\varrho (x) = 0 } \,  .
\end{equation}
In deriving the r.h.s. of (\ref{i10}) use of the symmetry
 $ \dot {C}^{\Lambda,\Lambda_0}(x-y) =
  \dot {C}^{\Lambda,\Lambda_0}(y-x) $ has been made. We note
 that the functional $ L_{(1)} $
satisfies a \emph{linear} equation. Because of the insertion the full
flow equation (\ref{i10}) can be studied  in the infinite 
volume limit $ \Omega \rightarrow \mathbf{R}^4, 
\varphi \in \mathcal{S}(\mathbf{R}^4) $. The 
Fourier transform with respect to the insertion is defined as
\begin{equation} \label{i12}
\hat{L}_{(1)}^{\Lambda, \Lambda_0}(q \, ;\varphi ) := 
\int dx \, e^{iqx} \, L_{(1)}^{\Lambda, \Lambda_0}(x ;\varphi ) ,
\end{equation} 
\begin{equation} \label{ii12}
\hat {I}^{\Lambda,\Lambda_0}_{(1)}(q) : = \int dx \, e^{iqx}
    I^{\Lambda,\Lambda_0}_{(1)} (x) =
  (2\pi )^4 \, i^{\Lambda,\Lambda_0}\, \delta (q) .
\end{equation}
Furthermore, the generating functional is decomposed, 
observing the conventions (\ref{f12}), $ n \in \mathbf{N} $,
\begin{eqnarray}\label{i13}
(2\pi)^{4(n-1)} \delta _{\hat \varphi (p_n)} \cdots 
\delta _{\hat \varphi (p_1)}\, \hat{L}_{(1)}^{\Lambda, \Lambda_0}(q \,;\varphi)
 \mid _{\varphi\, =\, 0} \qquad \qquad \qquad \nonumber \\ 
 \qquad \qquad \qquad  =  \delta (q+ p_1 + \cdots + p_n)\,
    \Lf_{(1) \, n}(q \,;p_1, \cdots , p_n) .
\end{eqnarray}
 The amputated truncated $n$-point function with one insertion carrying
 the momentum $q$,
$$ \Lf_{(1) \, n}(q \,;p_1, \cdots , p_n) ,$$  
is at fixed $q$ totally symmetric in the momenta $p_1, \cdots ,p_n $.
 Furthermore, the sum of all momenta has to vanish because of the
 $\delta $-constraint in (\ref{i13}).  From (\ref{i10}) and by proceeding
exactly as before from (\ref{f14}) to (\ref{f20}), after
 a loop expansion of the $n$-point functions,
$ n \in \mathbf{N} $, and of the vacuum part $i^{\Lambda,\Lambda_0} $ ,
\begin{equation} \label{i14}
\Lf_{(1) \, n}(q \, ; p_1, \cdots , p_n) \, = \,
 \sum_{l=0}^{\infty} {\hbar}^l \Lf_{(1) \, l, n}(q \,; p_1, \cdots , p_n) \ ,
\end{equation}
we arrive at the system of flow equations with one insertion:
\begin{eqnarray} \label{ii15}
\partial _{\Lambda} i^{\Lambda, \Lambda_0}_l & = &
 \frac{1}{2}\int \limits_{k} \partial _\Lambda
C^{\Lambda ,\Lambda _0}(k) \cdot \Lf_{(1)\, l-1, 2}(0 \, ; k, -k) \nonumber \\ 
& \, & - \sum_{l_1,l_2}' \mathcal{L}^{\Lambda ,\Lambda _0}_{l_1, 1} (0)
 \partial _\Lambda C^{\Lambda ,\Lambda _0}(0) \cdot
 \mathcal{L}^{\Lambda ,\Lambda _0}_{(1) \, l_2, 1} (0 \, ;0) ,  
\end{eqnarray}
\begin{eqnarray}\label{i15}
\partial _\Lambda  \partial^w
 \mathcal{L}^{\Lambda ,\Lambda _0}_{(1) \, l,n}(q; p_1,\cdots, p_n)  =
\qquad \qquad \qquad \qquad \qquad\\ 
\frac{1}{2}\int \limits_{k} \partial _\Lambda C^{\Lambda ,\Lambda _0}(k)
 \cdot \partial ^w \mathcal{L}^{\Lambda ,\Lambda _0}_{(1) \, l-1,n+2}
  (q;k,-k,p_1,\cdots,p_n) \qquad \qquad \qquad \nonumber\\
 -  \sum_{n_1,n_2}' \sum_{l_1,l_2}'\sum_{w_1,w_2,w_3}' c_{\{w_i\}}
    \Bigl[ \partial ^{w_1}\mathcal{L}^{\Lambda ,\Lambda _0}_{l_1, 
n_1 +1} (p_1, \cdots, p_{n_1}, p )\cdot \partial ^{w_3}
  \partial _{\Lambda}C^{\Lambda ,\Lambda _0}(p)  \qquad \qquad \nonumber \\
 \cdot \,
 \partial ^{w_2}\mathcal{L}^{\Lambda ,\Lambda _0}_{(1) \, l_2,n_2 +1}
   (q \, ;-p,p_{n_1 +1}, \cdots, p_n) \Bigr]_{rsym}  \quad  \qquad \nonumber\\
p = -p_1 - \cdots -p_{n_1} \, = \, q+p_{n_1 +1} + \cdots +p_n  .
     \qquad \nonumber 
\end{eqnarray}  
The notation used above has been introduced in (\ref{f17} - \ref{f20}).
 Furthermore, the derivations $ \partial ^w $ can be restricted
 to the momenta $p_1, \cdots ,p_n$, because of $q = -p_1 - \cdots - p_n $.
The vacuum part does not act back on the functional $ L_{(1)} $.
 We therefore disregard its flow 
(\ref{ii15}) and just state that in each loop order the
 bare parameter $ i^{\Lambda_0, \Lambda_0}_l $ is
determined by a renormalization constant $ i^{\, 0, \Lambda_0}_l $
 prescribed at $ \Lambda=0 $.

The task  is to show that finite limits 
\begin{equation} \label{i16}
\lim_{\Lambda_0 \to \infty} \lim_{\Lambda \to 0} \Lf_{(1) \, l,n}
(q \, ;p_1, \cdots , p_n) \, ,\quad n\in \mathbf{N} , \, l\in \mathbf{N_0}\, ,
\end{equation}
can be obtained, given the $n$-point functions without insertion
 which satisfy the Propositions 2.1 and 2.2.
This can be achieved following closely the corresponding 
steps performed before without insertion in section 2.3; the
 demonstration here can thus be presented in a concise way.
 Due to (\ref{i12} - \ref{ii12}), (\ref{i9}),(\ref{i4}),
 the bare functional is given by
\begin{equation} \label{i17}
\hat{L}_{(1)}^{\Lambda_0, \Lambda_0}(q \, ;\varphi ) +
     \hat{I}^{\Lambda_0, \Lambda_0}_{(1)} (q)
 : \, = \int dx e^{iqx} \, Q(x) \, . 
\end{equation} 
The tree order is completely determined by the classical part of 
the insertion (\ref{i1}) . This classical part
$ Q_{class} $ yields the initial condition in integrating
 the system of flow equations (\ref{i15}) for 
$l=0$ and general momenta from the initial point $\Lambda = \Lambda_0 $ 
downwards to smaller values 
of $\Lambda$ ascending successively in $n$. It is not necessary
 to prescribe the order in which  the derivations
$\partial ^w$ are treated. Since $Q_{class}$ is assumed to be a
 monomial of canonical mass dimension
$D$ and containing $n_0$ field factors, $2 \le n_0 \le D$ ,
 the key properties (\ref{p7}) imply for all $w$:
\begin{equation} \label{i18}
 \partial ^w \Lf_{(1) \, 0, 1}(q \,;p) = 0 \quad , \quad
 \partial _{\Lambda } \partial ^w \Lf_{(1) \, 0, 2}   (q \,;p_1, p_2)  = 0 \, .
\end{equation} 
The nonvanishing tree order starts at $n=n_0$ with
\begin{equation} \label{i19}
 \partial ^w \Lf_{(1) \, 0, n_0}(q \, ; \{p_i\}) \, = \,
  \partial ^w \Lb_{(1) \, 0, n_0}(q \, ; \{p_i\}) \, ,    
\end{equation}   
i.e. the initial condition. Of course the limits (\ref{i16}) exist
 in the tree order (since no integrations occur). In view of the mass
dimension $D$ of the insertion, the  bounds 
\begin{equation} \label{i20}
| \partial ^w \Lf_{(1) \, 0, n}( q \, ; p_1, \cdots , p_n) | \le
        (\Lambda +m)^{D-n-|w|} \,
    \mathcal{P} \bigl ( \{ \frac{|p_i|}{\Lambda+m} \} \bigr) 
\end{equation} 
are established for later use. If $n_0 < D$, apart from (\ref{i19}) 
there are other relevant instances $n+|w| \leq D$; they
should be evaluated explicitly  without using the bounds in
 the flow equation. (It is instructive to compare
the examples $Q_{class}=\phi \Delta \phi\, ,\, \phi ^4 $ both having $D=4$.) 
The irrelevant cases
$n+|w|>D $ in (\ref{i20}) are established similarly as (\ref{p13}).

For $l>0$ the coefficients of the counterterms inherent in (\ref{i17}),
extracted via (\ref{i13}-\ref{i14}), have to depend on $\Lambda_0$,      
 \begin{equation} \label{i21}
 \partial ^w \Lb_{(1) \, l,n}(0 \,;0,\cdots,0) =
         r_{l,n,w}(\Lambda_0) ,\quad n+|w|\leq D, \, l>0 .
\end{equation} 
They are determined by prescribing related renormalization conditions
 for vanishing flow parameter $\Lambda$ at a renormalization point
 which is again chosen at vanishing momenta. Hence, order-by-order,
the real constants, $l>0$,
 \begin{equation} \label{i22}
 \partial ^w \Lp_{(1) \, l,n}(0 \,;0,\cdots,0) = : r_{l,n,w}^R  \, ,
       \quad n+|w|\leq D ,
 \end{equation} 
can be freely chosen, provided they respect the symmetry of
 $ L_{(1)}^{\Lambda,\Lambda_0}$.\\
 \textbf{Proposition 2.3} (Boundedness) \\
 \emph{ Let} $ l \in \mathbf{N}_0 ,\, n \in
 \mathbf{N},\, w $ 
  \emph{from (\ref{f19}) and } $ 0 \le \Lambda \le \Lambda_0 $ , \emph{then} 
\begin{equation} \label {i23}
| \partial ^w  \Lf_{(1) \, l,n}(q \, ;p_1, \cdots ,p_n) |
 \le (\Lambda + m )^{D-n-|w|}\,
  \mathcal{P}_1 (log \frac{\Lambda + m}{m})
 \mathcal{P}_2 (\{ \frac{|p_i|}{\Lambda +m} \} ).
 \end{equation} 
 \emph{The symbol $ \mathcal{P}$ denotes polynomials with nonnegative
  coefficients which depend 
 on $ l,n,w $ but not on $ \{p_i \}, \Lambda, \Lambda_0 $.
For $l=0 $ all polynomials $ \mathcal{P}_1 $ reduce to positive constants.}\\
\emph{Proof}: Due to (\ref{i20}), the assertion (\ref{i23}) is
 already shown in the tree order. Given the
set of $n$-point functions without insertions satisfying
 Proposition 2.1, one proceeds for $l>0$ inductively
as in the proof of the latter proposition, i.e. i) ascending in $l$, ii)
 at fixed $l$ ascending in $n$, iii) at
fixed $l,n$ descending in $w$. 
 Inspecting the flow equations (\ref{i15}), it is
easily seen that for any given $l,n,w$ on the l.h.s. the
 contributions to the r.h.s. - because of the key
properties (\ref{p7}) - always precede those on the l.h.s.
 in the order of induction adopted. Imitating the steps leading to
 (\ref{p17}) provides the bound for $l,n \in \mathbf{N}$, 
 \begin{equation} \label{i24}
 | \partial_{\Lambda} \partial ^w 
    \Lf_{(1) \, l,n}(q \, ;p_1, \cdots, p_n) | \, \le \,  
  (\Lambda +m)^{D-n-|w|-1}
  \mathcal{P}_5 (\log\frac{\Lambda+m}{m})
   \mathcal{P}_6 \bigl( \{\frac{|p_i|}{\Lambda+m} \} \bigr ) .
\end{equation} 
$a_1)$ In the  cases $ n+|w| > D $ this bound is
 integrated downwards from the initial point $\Lambda = \Lambda_0 $ with
 vanishing initial conditions, see (\ref{i17}). From this
 follows easily (\ref{i23}) for $n+|w|>D$ . \\
$a_2)$ If $n+|w| \le D$ , however, the respective flow equation
 (\ref{i15}) has to be integrated upwards at the 
renormalization point, as in (\ref{p15}), employing now the
 renormalization conditions (\ref{i22}).
The bound (\ref{i24}) then implies the claim (\ref{i23}) at 
  vanishing momenta. Again as in section 2.3,
 extension to general momenta is accomplished appealing to 
 the Taylor formula (\ref{p16}). Thus, 
Proposition 2.3 is proven.\\
In complete analogy with the steps taken in section 2.3,
 the proposition just proven prepares the decisive\\
 \textbf{Proposition 2.4} (Convergence)  \\
\emph{Let $ l \in \mathbf{N}_0,\, n \in \mathbf{N},\, w $ from (\ref{f19}),
 \, $  0 \leq \Lambda \leq \Lambda_0$, 
  and $ \Lambda_0 > \underline {\Lambda_0}$, sufficiently large, then   
\begin{equation} \label{i25}
 |\partial _{\Lambda_0}\partial ^w 
\Lf_{(1) \, l,n}(q \, ;p_1, \cdots,p_n) | \le 
 \frac{(\Lambda+m)^{D+1-n-|w|}}{(\Lambda_0)^2}
 \Bigl(\log \frac{\Lambda_0}{m} \Bigr)^\nu 
 \mathcal{P}_4 \bigl( \{ \frac{|p_i|}{ \Lambda+m} \} \bigr ) 
\end{equation}
with a positive integer $\nu $ depending on $l,n,w$ only.}  \\
\emph{Proof}: One first verifies the assertion directly in
 the tree order for the relevant cases 
 $ n+|w| \leq D $. Herewith, the further course of the proof
 is just a replica of the proof given for
Proposition 2.2. As a consequence, in each place the
 exponent $4+1-n-|w|$ appears there, this exponent is changed
 here into $D+1-n-|w|$, thus proving the assertion (\ref{i25}). \\
Finally, integration of the bound (\ref{i25}) of Proposition 2.4
 demonstrates, that  the renormalized regularized $n$-point functions with
 one insertion have finite limits (\ref{i16}).
%%%%%%%%%%%%%%%%%%%%%%%%%%%%%%%%%%%%%%%%%%%%%%%%%%%%
\section{Finite temperature field theory}
 There are essentially two formulations of quantum fields
at finite temperature: a real-time approach to treat
 dynamical effects, and an imaginary-time approach
to describe equilibrium properties \cite{LvW}.In this section the
 problem of renormalization in a
temperature independent way is considered. Such a renormalization
 is required studying the $T$- dependence
of observables, since then the relation between bare and renormalized
 coupling constants must not
depend on the temperature. Our aim here is to show within
 the imaginary-time formalism, that a
quantum field theory renormalized at $T=0$ stays
 also renormalized at any $T> 0$. For the sake of a 
succinct presentation the symmetric $\Phi ^4$-theory is
 treated and the generalization to the
 nonsymmetric  theory is stated at the end.

The first steps to be taken do not differ from the zero temperature
 case: Starting from a finite domain, given by a $4$-dimensional 
torus $\Omega $, and the Gaussian measure with the regularized covariance
(\ref{g6}), we obtain Wilson's flow equation (\ref{f10}). Here,
 the bare interaction (\ref{f1}) is restricted  to the symmetric theory,
 as already mentioned, i.e. putting
\begin{equation} \label{t1}
f = v(\Lambda_0) = b(\Lambda_0) \equiv 0 .
\end{equation}
Disregarding as before the flow of the vacuum part
 $I^{\Lambda, \Lambda_0}$, 
we imagine at least one functional derivative acting on the
 flow equation (\ref{f10}). Then we can pass
 to the spatial infinite volume limit, but keeping the periodicity
 in the imaginary time $x_4$ and choosing  
the period equal to the inverse temperature: $l_4 = \beta  \equiv 1/T $.
 Hence, in this limit the space-time domain is $ \mathbf{R}^3 \times S^1 $ 
 and the theory shows the reduced symmetry
$ O(3) \times  Z_2$, as compared to the $O(4)$- symmetry at $T=0$.
 Correspondingly, the dual Fourier variables (momentum vectors) are
\begin{equation} \label{t2}
\up := (\vec p, \up_{ \, 4}) ,\,\, \vec p \in \mathbf{R}^3, \,\,
 \up_{\, 4} = 2\pi nT,\,\, n \in \mathbf{Z}\, ,
\end{equation}
and hence we define
\begin{equation} \label{t3}
\int_{\up} \, := \, T\, \sum_{n \in \mathbf{Z}}^{}\, \int_{\mathbf{R}^3}^{} 
\frac{d^3 \vec p}{(2\pi)^3} \, .
 \end{equation}
In the sequel we underline a symbol denoting a quantity at finite
 temperature or write the T-dependence
explicitly. In place of (\ref{f12}) the Fourier transform takes the form
\begin{equation} \label{t4}
\ufi (x) = \int_{\up} e^{i \up x} \, \hat {\ufi}(\up) \, ,  \qquad 
    \hufi(\up) =  \int_{\mathbf{R}^3} d^3 x \int_{0}^{\beta }
    dx_4 \, e^{- i \up x}\, \ufi (x) \, ,
 \end{equation}  
implying for a functional derivation:
 \begin{equation}\label{t5}
\delta _{ \ufi (x)} = \frac{(2 \pi )^3}{T} \int_{\up}
  e^{- i \up x}\, \delta _{\hufi (\up)} \, , \quad
\delta _{\hufi (\up)} = \frac{T}{(2\pi)^3}
 \int_{\mathbf{R}^3} d^3 x \int_{0}^{\beta } dx_4 \, 
        e^{i \up x} \,\delta _{\ufi (x) } \, .             
 \end{equation}
Furthermore, the regularized covariance (\ref{f16}) is restricted to
 momenta (\ref{t2}),
\begin{equation}\label{t6}
 C^{\Lambda ,\Lambda _0}(\up) = \frac{1}{{\up}^2 + m^2}
 \left( e^{-\frac{{\up}^2 + m^2}
{{\Lambda _0}^2}} -   e^{-\frac{{\up}^2 + m^2}{\Lambda ^2}} \right) \, .
\end{equation} 
Denoting by $L^{\LLz} (\ufi \, ; T)$ the generating functional
 of the amputated truncated Schwinger functions
 at finite temperature $T$, we define the $n$-point functions,
 $n \in \mathbf{N}$, similar to (\ref{f14}) 
as \footnote{In the symmetric theory the $n$-point functions
 with odd $n$ vanish.}
\begin{eqnarray} \label{t7}
\Bigl ( \frac{(2\pi)^3}{T} \Bigr )^{n-1} \delta _{\hufi (\up_1)}
 \dots \delta _{\hufi (\up_n)}\,
  L^{\LLz} ( \ufi \, ; T) |_{ \ufi \, \equiv \, 0}  \nonumber \\
 =  \delta ( \vec p_1+ \cdots +\vec p_n) \,
\delta _{0, ( \up_1 + \cdots + \up_n)_{,4}} \,
 \Lf _n ( \up_1, \cdots, \up_n ; T) \, .
\end{eqnarray} 
These $n$-point functions, after a respective loop expansion
 in complete analogy to (\ref{f18}), then
satisfy a system of flow equations obtained from (\ref{f20}) by 
 replacing  every momentum
vector appearing by its underlined analogue, and moreover,
 restricting the momentum derivatives
$\partial ^w$ to spatial momentum components.
 Employing this system of flow equations, we could prove renormalizability 
of the theory at finite temperature proceeding similarly as in the case
of zero temperature. However, because of the reduced spacetime symmetry,
 the renormalization conditions (\ref{p3}-\ref{p6})
 for $l\geq 1$ would have to be extended by an additional constant:
 \begin{equation} \label{tt1}
 \Lp _{l,2}(\up, -\up \, ;T) = a^R_l (T) +
 z^{R,1}_l(T)\, {\vec p}^{\, 2} + z^{R,2}_l (T) \, 
{\up_{\,4}}^2+ \mathcal{O}(\up^4) \, , 
 \end{equation} 
 \begin{equation} \label{tt2}
 \Lp _{l,4}(0, 0, 0, 0 \,; T) = c^R_l (T) \, . 
 \end{equation} 
 The constants for $n=1,3$ are set equal to zero
 in the symmetric theory (\ref{t1}). Only at
$T=0$, the emerging $O(4)$-symmetry implies
the equality $ z^{R,1}_l (0) = z^{R,2}_l (0) $. Our aim is
 to prove renormalizability in a temperature independent way,
 i.e. with counterterms that do not depend on the temperature.
 In this case, the renormalization constants  $a^R_l (T), z^{R,1}_l (T),
 z^{R,2}_l (T), c^R_l (T) $ cannot be prescribed arbitrarily,
 since they are related dynamically to the
three renormalization constants at $T=0$. Therefore, we follow a
 different course and study the respective difference
of a $n$-point function at $T>0$ and at $T=0,\, n \in \mathbf{N}$,
 with momenta $ \{\up \} \equiv (\up_1, \cdots, \up_n) $
 of the form (\ref{t2}): 
\begin{equation} \label{t8}
\Df_{l,n} ( \{\up\}) := \Lf_{l,n} ( \{\up\}\, ; T) - \Lf_{l,n} (\{ \up \}) .
\end{equation}
These functions are well-defined. From the system of flow equations
 (\ref{f20}) and from its analogue
at finite temperature  follows the system of
 flow equations satisfied by the difference 
functions (\ref{t8}), with $l,n \in \mathbf{N}$: 
\begin{equation} \label{t9}
 \partial _\Lambda \Df_{l,n}( \{\up \}) \, = \, \frac{1}{2}\int_{\uk}
 \partial _\Lambda C^{\Lambda ,\Lambda _0} (\uk) \cdot
 \Df_{l-1,n+2}(\uk, -\uk, \{\up \}) \qquad  
 \end{equation}  $$ 
+\frac{1}{2} \Bigl (\int_{\uk}\partial _\Lambda C^{\Lambda ,\Lambda _0} (\uk)
 \cdot  \Lf_{l-1,n+2}(\uk, -\uk, \{\up \}) 
  - \int_k \partial _\Lambda C^{\Lambda ,\Lambda _0} (k)\cdot
 \Lf_{l-1,n+2}(k, -k, \{\up \}) \Bigr )  $$  
 $$ - \frac{1}{2} \sum_{n_1,n_2}' \sum_{l_1,l_2}'
 \Bigl[ \Lf_{l_1,n_1 +1} (\up_1,\cdot \cdot,\up_{n_1},\up\, ;T)
    \partial _{\Lambda}C^{\Lambda ,\Lambda _0}(\up) \cdot 
 \Df_{l_2,n_2 +1}(-\up, \up_{n_1 +1},\cdot \cdot,\up_n) \Bigr]_{rsym}  $$
$$- \frac{1}{2} \sum_{n_1,n_2}' \sum_{l_1,l_2}'
   \Bigl[ \Df_{l_1,n_1 +1} (\up_1,\cdot \cdot,\up_{n_1},\up)
  \partial _{\Lambda}C^{\Lambda ,\Lambda _0}(\up) \cdot 
 \Lf_{l_2,n_2 +1}(-\up, \up_{n_1 +1},\cdot \cdot,\up_n) \Bigr]_{rsym}  $$
$$ \up = - \up_1 - \cdots -\up_{n_1} \, = \, \up_{n_1 +1} + \cdots +\up_n .$$ 
In the (tree) order $l=0$ we infer directly
\begin{equation} \label{t10}
\Df_{0,n} (\up_1, \cdots , \up_n) \, = \, 0 , \, n\in \mathbf{N} ,
 \end{equation}
since on the set of momenta considered the $n$-point function
 at $T=0$ is equal to the $n$-point function
at $T>0$ in this order. The assertion of a temperature
 independent renormalization now requires
the bare difference functions to vanish for $l\geq 1$:
\begin{equation} \label{t11}
\Db_{l,n} (\up_1, \cdots , \up_n) \, = \, 0 , \quad l, n\in \mathbf{N} ,
\end{equation}
Given the bounds (\ref{p10})  and (\ref{p18}) satisfied by the
 $n$-point functions at zero temperature, then follows the \\
\textbf{Theorem} \\
\emph{ For $l,n \in \mathbf{N}$ and for $0\leq \Lambda \leq \Lambda_0$ holds }
\begin{eqnarray}
|\Df_{l,n}(\up_1, \cdots, \up_n )|  \leq  (\Lambda +m)^{-s-n} \,
       \mathcal{P}_1 (log \frac{\Lambda + m}{m})
 \mathcal{P}_2 (\{ \frac{|\up_i|}{\Lambda +m} \} ) ,  \label{t12}  \\
|\partial_{\Lambda_0} \Df_{l,n}(\up_1, \cdots, \up_n )| 
 \leq  \frac{(\Lambda +m)^{-s-n}} {(\Lambda_0 )^2} 
       \mathcal{P}_3 (log \frac{\Lambda_0}{m}) 
    \mathcal{P}_4 (\{ \frac{|\up_i|}{\Lambda +m} \} ) . \label{t13}
\end{eqnarray} 
\emph{The polynomials $\mathcal{P}$ have positive coefficients,
 which depend on $l,n,s,m$ and 
(smoothly) on $T$, but not on $\{\up\}, \Lambda, \Lambda_0$.
 The positive integer $s$ may be
 chosen arbitrarily. \\ 
The $n$-point functions at finite temperature
 $T, \, \Lf_{l,n} (\up_1, \cdots, \up_n \, ;T) $, when
renormalized with the same counterterms as the 
zero temperature functions, (\ref{t11}),  
satisfy the bounds (\ref{p10}) and (\ref{p18})
 restricted to the case $w=0$ and to momenta (\ref{t2}).
 The coefficients in the polynomials
$\mathcal{P}$ may now depend also (smoothly) on $T$.} \\
For the proof we refer to \cite{KMR}, pp.396-399, and just indicate 
that the system of flow equations (\ref{t9})
 is integrated inductively from the initial
 point $\Lambda = \Lambda_0$ downwards, observing 
(\ref{t11}). The difference of the two terms
 not involving any function $ \Df_{l',n'}$, which
 appears in (\ref{t9}), however, is not accessible by
 induction. It is bounded
 separately, matching the sharp bound on $\Lambda$ asserted,
 by use of the Euler - MacLaurin formula, see e.g. \cite{Bour}. 

 Due to the Theorem, the $n$-point functions of the theory at $T>0$,
 renormalized at zero temperature, satisfy the bound (\ref{p18})
 for $w=0$ and momenta (\ref{t2}). Hence, they have finite limits
$$ \lim_{\Lambda_0 \rightarrow \infty}
 \Lp_{l,n} (\up_1, \cdots, \up_n \, ; T) , \quad l,n \in \mathbf{N}, $$
 upon removing the UV-cutoff $\Lambda_0$ . 

As already indicated before, a finite theory at given
 temperature $ T_0 > 0$ could also be generated 
imposing renormalization conditions at this temperature.
 The price to be paid ( in the symmetric theory
considered ) are in each loop order  the four constants
 $a^R_l (T_0), z^{R,1}_l (T_0), z^{R,2}_l (T_0), c^R_l (T_0) $,
 (\ref{tt1}-\ref{tt2}) , instead of the three constants
 $ a^R_l, z^R_l, c^R_l $ at zero temperature.
 However, an arbitrary choice of $z^{R,1}_l (T_0), z^{R,2}_l (T_0) $
 would not correspond to a theory at zero temperature, which
 shows the $O(4)$- symmetry of Euclidean
space-time. Starting from  an $O(4)$-invariant theory
 at zero temperature, the functional
\begin{equation} \label{t25}
L^{\LLz} (\ufi \, ; T) \, - \, L^{\LLz} (\ufi ) 
\end{equation}  
with initial condition (\ref{t11}) has been proven to
 satisfy the bound (\ref{t13}). Hence, the function
$\mathcal{D}^{0, \Lambda_0}_{l,2}(\up, - \up)$, converging for all $l$ 
with $\Lambda_0 \rightarrow \infty $
to a finite limit, produces a dynamical relation between the
 renormalization constants $ z^{R,1}_l (T_0) $
and $z^{R,2}_l (T_0) $, i.e. fixing one of them determines
 the other. Thus, a renormalization does
not depend on temperature, if this relation is satisfied.
 It becomes manifest in the equality $ z^1_l (\Lambda_0) = z^2_l (\Lambda_0) $
 of the corresponding \emph{bare} parameters.

Concluding we remark that the proof can be easily extended 
to the nonsymmetric $\Phi^4 $-theory. In this case, the $n$-point 
functions with $n$ odd no longer vanish, since the $Z_2\,$-symmetry 
is now lacking. Hence, the bare interaction will be of the general form 
(\ref{f1}).
Correspondingly, the theory at zero temperature is renormalized 
by the conditions (\ref{p3}-\ref{p6}),
involving five renormalization constants. Proceeding inductively 
as before, considering now odd and even
values of $n$, establishes the Theorem for the nonsymmetric theory, too.
%%%%%%%%%%%%%%%%%%%%%%%%%%%%%%%%%%%%%%%%%%%%%%%%%
\section{ Elementary Estimates }
Here, rather obvious estimates on some elementary integrals are listed,
 which we used repeatedly in generating inductive bounds
 on Schwinger functions. \\
$a_1)$ In the irrelevant cases, the integrals have the form
$$ \int_{a}^{b} dx \, x^{-r-1} ( \log x)^s \, ,
 \quad {\rm with} \,\, 1\leq a \leq b  \,\,{\rm and}\, \,
 r \in \mathbf{N}, \,s \in \mathbf{N}_0 \, .  $$
Defining correspondingly the function
$$ f_{r,s}(x) := \frac{1}{r} x^{-r} \bigg ( ( \log x)^s 
    + \frac{s}{r} ( \log x)^{s-1} + \frac{s(s-1)}{r^2}
  (\log x)^{s-2} + \cdots +\frac{1\cdot 2 \cdots s}{r^s} \bigg ) $$
we observe $f_{r,s}'(x) = - \, x^{-r-1}( \log x)^s < 0$ and
    $ f_{r,s}(x) > 0 $ for $x>1$, hence 
$$ \int_{a}^{b} dx \, x^{-r-1} ( \log x)^s \, = \, f_{r,s}(a) -
 f_{r,s}(b) < f_{r,s}(a) \, .$$
$a_2)$ The integrals to be bounded in the relevant cases have the form
$$ \int_{1}^{b} dx \, x^{\, r-1} ( \log x)^s \, ,
    \quad {\rm with} \,\, 1 \leq b 
\,\,{\rm and}\, \,r,s \in \mathbf{N}_0 .  $$
If $r=0$, we just integrate. For $r>0$ , defining
$$ g_{r,s}(x) := \frac{1}{r} x^r \bigg ( ( \log x)^s 
  - \frac{s}{r} ( \log x)^{s-1} + \frac{s(s-1)}{r^2}
  (\log x)^{s-2} + \cdots +(-)^s \frac{1\cdot 2 \cdots s}{r^s} \bigg ) $$
we notice $ g_{r,s}' (x) \, = \,x^{\, r-1} (\log x)^s $ and hence
$$ \int_{1}^{b} dx \, x^{\, r-1} ( \log x)^s \,
 = g_{r,s}(b) - g_{r,s}(1) \, < \, 
        \frac{1\cdot 2 \cdots s}{r^s} + |g_{r,s}(b) | \, .$$

\chapter{The Quantum Action Principle}
The Green functions of a relativistic quantum field theory
 depend on the adjustable parameters of
this theory and are in general related according to the
 inherent symmetries of the theory. Clearly, all
types of Green functions, whether truncated, amputated, 
or one-particle-irreducible, show these properties.
The \emph{ quantum action principle } deals with the 
variation of Green functions caused by diverse operations performed: \\
i) applying the differential operator appearing
 in the (classical) field equation, \\
ii) (nonlinear) variations of the fields, \\
iii) variation of an adjustable parameter of the theory. \\
The quantum action principle relates each of these 
different operations on Green functions to the
insertion of a corresponding composite field into the 
Green functions: as a local operator in the first
two cases, whereas integrated over space-time in the third.
 Moreover, in general the local operation
has a precursor within classical field theory (e.g. the field equation,
 the Noether theorem). Then the
local composite field to be inserted in the case of a quantum field 
theory is a sum formed of its
classical  precursor and of assigned local counterterms,
 whose canonical mass dimensions are equal
to or smaller than the canonical mass dimension ascribed to 
the term of classical descent. \\
The quantum action principle has been established first by Lam \cite{Lam}
 and Lowenstein \cite{Low} using the
BPHZ-formulation of perturbation theory. This principle is 
extensively used in the method of algebraic renormalization \cite{PiSo}. 

Our aim is to demonstrate the parts i) and iii) of the
 quantum action principle by means of flow
equations in the case of the scalar field theory. The particularly 
interesting part ii) is deferred to a
later section, where nonlinear BRST-transformations
 have to be implemented in showing the
renormalizability of a non-Abelian gauge theory.
%%%%%%%%%%%% 
\section{Field equation}
We consider again the quantum field theory of a real scalar field 
 on four-dimensional Euclidean space-time,
which has been treated in the preceding sections. To derive a
 field equation, we act on the generating
functional of its regularized Schwinger functions (\ref{f2}) as follows:
$$ \hbar \int dy \bigl ( C^{\Lambda, \Lambda_0} \bigr )^{-1} (x-y)
     \frac{\delta }{\delta J(y)} Z^{\Lambda, \Lambda_0}(J) $$
$$ = \int dy \bigl ( C^{\Lambda, \Lambda_0} \bigr )^{-1} (x-y)
        \int d\mu _{\Lambda, \Lambda_0}(\phi )
    \, \phi (y) \, e^{- \frac{1}{\hbar}L^{\Lambda _0,\Lambda _0}(\phi )
         +\frac{1}{\hbar}\langle \phi , J \rangle } . $$
In presence  of the regularization the inverse of the regularized
 covariance (\ref{g7}) replaces the differential
operator $ -\Delta +m^2 $. Integration by parts (\ref{g4}) on the r.h.s.
 and recalling that the covariance  
 of the Gaussian measure $  d\mu _{\Lambda, \Lambda_0}(\phi ) $
is  $ \hbar  C^{\Lambda, \Lambda_0} $
yields the field equation of the regularized generating functional (\ref{f2}) 
\begin{eqnarray} \label{a1}
\Bigl ( J(x) - \hbar \int dy \bigl ( C^{\Lambda, \Lambda_0} \bigr )^{-1} (x-y)
 \frac{\delta }{\delta J(y)}  \Bigr ) Z^{\Lambda, \Lambda_0}(J)
  \qquad \qquad \nonumber  \\ 
\qquad \qquad  =  \int d\mu _{\Lambda, \Lambda_0}(\phi )
    \, Q(x) \, e^{- \frac{1}{\hbar}L^{\Lambda _0,\Lambda _0}(\phi )
         +\frac{1}{\hbar}\langle \phi , J \rangle } . 
\end{eqnarray}                    
On the r.h.s. the inserted  composite field $ Q(x) $  is given by  
\begin{equation} \label{a3}
Q(x) = \frac{\delta }{\delta \phi (x)} L^{\Lambda_0, \Lambda_0}(\phi ) .
\end{equation}
If we employ the generating functional of regularized Schwinger functions
 with insertions (\ref{i4} - \ref{i5}) we
can rewrite the field equation (\ref{a1}) in the form
 \begin{equation} \label{a4}
\Bigl ( J(x) - \hbar \int dy \bigl ( C^{\Lambda, \Lambda_0} \bigr )^{-1}(x-y)
 \frac{\delta }{\delta J(y)}
    \Bigr ) Z^{\Lambda, \Lambda_0}(J)  =
   - \hbar \frac{\delta }{\delta \varrho  (x)} 
 \tilde { Z}^{\Lambda, \Lambda_0}(\varrho \, ; J) |_{\varrho (x)=0} .
 \end{equation}    
Taking into account the relations (\ref{f7}) and (\ref{i8}) on the l.h.s.
 and on the r.h.s. of this equation,
respectively, provides the field equation for the generating functional
 of the amputated truncated
Schwinger functions,
 \begin{equation} \label{a5}
\frac{\delta }{\delta \varphi (x)} L^{\Lambda, \Lambda_0}(\varphi ) \, = \,
 L^{\Lambda, \Lambda_0}_{(1)}(x ; \varphi) +I ^{\Lambda, \Lambda_0}_{(1)}(x ) .
 \end{equation}
 Hence, in momentum space we have
 \begin{equation} \label{a6}
 (2\pi )^4 \frac{\delta }{\delta \hat {\varphi} (q)}
 L^{\Lambda, \Lambda_0}(\varphi ) \, = \,
\hat { L}^{\Lambda, \Lambda_0}_{(1)}(q \, ; \varphi) +
  \hat {I} ^{\Lambda, \Lambda_0}_{(1)}(q ) ,
 \end{equation} 
using the conventions (\ref{f12}) and (\ref{i12}). Our goal is to
 show within perturbation theory, i.e. in a formal loop expansion,
 that the field equation (\ref{a6}) remains valid taking 
 the limit $ \Lambda = 0,
\Lambda_0 \rightarrow \infty $. To this end we proceed as follows:  \\
$\alpha )$ In section 2.3 it has been shown that the
 generating functional $ L^{\Lambda, \Lambda_0} (\varphi ) $ of the theory 
 considered (perturbatively) converges to a finite limit 
with $ \Lambda_0 \rightarrow \infty $. The limit theory is 
determined by the choice of renormalization conditions (\ref{p3} - \ref{p6})
 at the renormalization point (chosen at vanishing momenta). \\
$\beta ) $ The generating functional
 $ L^{\Lambda, \Lambda_0}_{(1)}(q \, ; \varphi ) $ with insertion
of one composite field of mass dimension $D$ and momentum $q$ has
 been shown in section 2.4 to have a finite limit with
 $ \Lambda_0 \rightarrow \infty $, too, provided the 
counterterms of the insertion
(\ref{i1}) are introduced at first as indeterminate functions
 of $\Lambda_0 $ and then determined by
a choice of renormalization conditions at the renormalization point. 
In the case entering the field equation,
 however, the dependence on  $\Lambda_0 $ of the
 insertion (\ref{a3}) is already given by
(\ref{f1}), the counterterms of the theory without insertion.
 In order to maintain in an intermediate stage the 
freedom in choosing renormalization
 conditions, we use instead of (\ref{a3}) indeterminate 
counterterms to begin with:
\begin{eqnarray}\label{a7}
Q(x)  &=&  \frac{f}{2!}\phi ^2(x) +\frac{g}{3!}\phi ^3(x)  \nonumber\\
& \, & + v_1(\Lambda _0)+a_1(\Lambda_0)\phi(x)
   - z_1(\Lambda_0) \Delta \phi (x) \nonumber \\  & \, & 
 + \frac{1}{2!}b_1(\Lambda_0)\phi^2 (x)+\frac{1}{3!}c_1(\Lambda_0)\phi^3 (x) .
\end{eqnarray}
Then, as has been demonstrated in section 2.4, any choice of admissible
 renormalization conditions leads
to a finite limit of the generating functional
 $ L^{\Lambda, \Lambda_0}_{(1)}(q \, ; \varphi ) $  in
sending $ \Lambda_0  \rightarrow \infty $, and thereby a related
 dependence of the coefficients $ v_1 (\Lambda_0), \cdots, c_1(\Lambda_0) $
 on $\Lambda_0 $ arises. \\
$\gamma )$ We define the functional
\begin{equation} \label{a8}
\hat {D}^{\Lambda, \Lambda_0}(q \, ; \varphi ) :=
 \hat{L}^{\Lambda, \Lambda_0}_{(1)}(q \, ; \varphi)
       +  \hat {I}^{\Lambda, \Lambda_0}_{(1)}(q ) - (2\pi)^4 
 \frac{\delta }{\delta \hat {\varphi} (q)} L^{\Lambda, \Lambda_0}(\varphi ) .
\end{equation}
If this functional can be forced to vanish at $\Lambda = 0 $ and for
 all $\Lambda_0, \Lambda_0 >
\underline{\Lambda_0} $, by an appropriate fixed choice of
 renormalization conditions in $\beta )$,
then (\ref{a6}) converges to a finite renormalized field equation
 for $(\Lambda=0, \Lambda_0 \rightarrow \infty ) $ .

The functional (\ref{a8}) obeys the linear flow equation
\begin{eqnarray} \label{a9}
  \frac{d}{d \Lambda} \hat D^{\Lambda, \Lambda_0} (q \, ; \varphi ) & = & 
 \frac{\hbar}{2} \langle \frac{\delta }{\delta \varphi }\,
  ,\dot C ^{\Lambda ,\Lambda _0} \frac{\delta } {\delta \varphi } \rangle 
 \hat{D}^{\Lambda ,\Lambda _0}(q \, ;\varphi )  \nonumber  \\   
 & \, & \, - \, \langle \frac{\delta }{\delta \varphi }L^{\Lambda ,\Lambda _0}
 (\varphi ) ,\dot C ^{\Lambda ,\Lambda _0} \frac{\delta }{\delta \varphi }  
           \hat{D}^{\Lambda ,\Lambda _0}( q \, ;\varphi ) \rangle , 
 \end{eqnarray}
which follows directly from the flow equations (\ref{f10}) and (\ref{i10}),
 performing a functional
derivation of the first and Fourier transforming the latter.
 To make use of it we decompose
\begin{eqnarray} \label{a10}
 (2\pi)^{-4} \hat {D}^{\Lambda ,\Lambda _0}(q \, ;0 )& = & 
 \delta (q) \, \mathcal{D}^{\Lambda ,\Lambda _0}_0 
  \qquad \qquad \qquad \nonumber  \\
(2\pi)^{4(n-1)} \delta _{\hat \varphi (p_n)} \cdots 
\delta _{\hat \varphi (p_1)} \hat{D}^{\Lambda , \Lambda_0}
    ( q \, ; \varphi)|_{\varphi\, = \,0}& = &  \\
 {} \qquad \delta (q +p_1 +& \cdots & +p_n)\,
 \mathcal{D}^{\Lambda ,\Lambda _0}_ {n }(q \,; p_1 ,\cdots , p_n) \,.
 \nonumber
 \end{eqnarray}
From (\ref{f14}), (\ref{i12} - \ref{ii12}) then results
\begin{equation} \label{a11} 
\mathcal{D}^{\Lambda, \Lambda_0}_0  =  i^{\Lambda, \Lambda_0} 
  - \mathcal{L}^{\Lambda, \Lambda_0}_1 (0) \, , \qquad \qquad \qquad \quad \, 
\end{equation}
 \begin{equation} \label{a12} 
\mathcal{D}^{\Lambda, \Lambda_0}_n (q \, ; p_1, \cdots, p_n)  = 
\mathcal{L}^{\Lambda, \Lambda_0}_{(1) \,n }(q \, ; p_1, \cdots, p_n) 
   - \mathcal{L}^{\Lambda, \Lambda_0}_{n +1}(q , p_1, \cdots, p_n) .  
 \end{equation}
We notice that the  flow equations (\ref{a9}) and (\ref{i10}) have
 the same form. Thus, after a loop expansion, the strict analogue of
 the system of flow equations
 (\ref{ii15} - \ref{i15}) is obtained, $l \in \mathbf{N}_0 , n \in \mathbf{N}$:
\begin{eqnarray} \label{a13}
\partial _{\Lambda} \mathcal{D}^{\Lambda, \Lambda_0}_{l,0} \, & = &
 \frac{1}{2}\int \limits_{k} \partial _\Lambda C^{\Lambda ,\Lambda _0}(k)
 \cdot \mathcal{D}^{\Lambda, \Lambda_0}_{l-1, 2} (0 \, ; k, -k )  \nonumber \\ 
& \, & - \sum_{l_1,l_2}' \mathcal{L}^{\Lambda ,\Lambda _0}_{l_1, 1} (0)
 \partial _\Lambda C^{\Lambda ,\Lambda _0}(0)
 \cdot \mathcal{D}^{\Lambda ,\Lambda _0}_{l_2, 1} (0 \, ;0) ,  
 \end{eqnarray}
 \begin{eqnarray} \label{a14}
 \partial _\Lambda  \partial^w \mathcal{D}^{\Lambda ,\Lambda _0}_{ l,n}
 (q \, ; p_1,\cdots, p_n)  = 
     \qquad \qquad  \qquad \qquad \qquad \quad \\
 \frac{1}{2}\int \limits_{k} \partial _\Lambda C^{\Lambda ,\Lambda _0}(k)
 \cdot \partial ^w \mathcal{D}^{\Lambda ,\Lambda _0}_{l-1,n+2}
 (q \,;k,-k,p_1,\cdots,p_n)  \qquad \qquad \qquad \nonumber\\ 
-  \sum_{n_1,n_2}' \sum_{l_1,l_2}'\sum_{w_1,w_2,w_3}' c_{\{w_i\}}
   \Bigl[ \partial ^{w_1}\mathcal{L}^{\Lambda ,\Lambda _0}_{l_1, n_1 +1}
 (p_1, \cdots, p_{n_1}, p )\cdot
 \partial ^{w_3}\partial _{\Lambda}C^{\Lambda ,\Lambda _0}(p) 
 \qquad \qquad \nonumber \\ \cdot \,
 \partial ^{w_2}\mathcal{D}^{\Lambda ,\Lambda _0}_{ l_2,n_2 +1}
 (q \, ;-p,p_{n_1 +1}, \cdots, p_n) \Bigr]_{rsym}  \quad  \qquad \nonumber\\
p = -p_1 - \cdots -p_{n_1} \, = \, q+p_{n_1 +1} + \cdots +p_n .
 \qquad  \nonumber
  \end{eqnarray}
We first treat the tree order. From (\ref{f1}), (\ref{i17}) follows
 $ \hat{D}^{\Lambda_0, \Lambda_0} (q \, ; \varphi ) |_{l=0} = 0 $.
 Integrating the flow equations with $l=0 $ and general momenta
 from the initial point $ \Lambda = \Lambda_0 $ downwards to smaller
 values of $ \Lambda $, ascending successively in $n$ , we find due to
 the properties (\ref{p7}) for  $ n \in \mathbf{N}, 0 \leq \Lambda
\leq \Lambda_0 $,
\begin{equation} \label{a15}
\mathcal{D}^{\Lambda, \Lambda_0}_{0,0} = 0 \, ,
 \quad \mathcal{D}^{\Lambda, \Lambda_0}_{0,n} 
  (q \, ;p_1, \cdots, p_n) = 0 .
\end{equation} 
The extension to all loop orders $l$ is achieved by the  \\
\textbf{Proposition 3.1} \\
\emph{ For all $l \in \mathbf{N} $ and $ n+|w| \leq 3 $ let }
\begin{equation} \label {a16}
\mathcal{D}^{0, \Lambda_0}_{l,0} = 0 \, ,
 \quad \partial ^w \mathcal{D}^{0, \Lambda_0}_{l,n} 
  (0 \, ;0, \cdots, 0) = 0 ,
\end{equation} 
\emph{then for} $l \in \mathbf{N}_0 , n \in \mathbf{N}, |w| \leq 3,$ \emph{and}
 $ 0\leq \Lambda \leq \Lambda_0$\,:
\begin{equation} \label{a17}
\mathcal{D}^{\Lambda, \Lambda_0}_{l,0} = 0 \, ,
 \quad \partial ^w \mathcal{D}^{\Lambda, \Lambda_0}_{l,n}
 (q \, ;p_1, \cdots, p_n) = 0 .
\end{equation} 
\emph{Proof}: In the order $l=0$ the assertion is already established
 because of (\ref{a15}). We now
assume (\ref{a17}) to hold for all orders smaller than a fixed
 order $l$. As a consequence, on the 
respective r.h.s. of the flow equations (\ref{a13}) and (\ref{a14})
 the first term vanishes and and in 
the second term only the pair $ (l_1=0, l_2=l ) $ has to be
 taken into account. Looking first at the
vacuum part, we observe that the r.h.s. of (\ref{a13}) vanishes
 due to (\ref{p7}). Thus, integration from the initial point
 $ \Lambda=0 $ yields $  \mathcal{D}^{\Lambda, \Lambda_0}_{l,0} = 0 $. To
demonstrate the assertion for general $n$ we proceed inductively:
 ascending in $n$, and for fixed  $n$
descending with $w$ from $|w|=3 $. Thus, for each $n$ the
 irrelevant cases $ n+|w| > 3 $
always precede the relevant ones, $n+|w| \leq 3 $, if present at all. Since
\begin{equation} \label{a18}
\quad \partial ^w \mathcal{D}^{\Lambda_0, \Lambda_0}_{l,n} 
        (q \, ;p_1, \cdots, p_n) = 0 \, , \quad  n+|w| > 3,
 \end{equation} 
the flow equations (\ref{a14}) of these cases are integrated
 from the initial point $ \Lambda = \Lambda_0 $ downwards.
 On the other hand, the respective flow equation of the cases
 $ n+|w| \leq 3 $ is first integrated at zero momentum from the
 initial point $ \Lambda = 0 $ with vanishing initial condition
(\ref{a16}) and in a now familiar second step the result is 
 extended to general momenta via the Taylor
formula (\ref{p16}). Following the inductive order stated one notices
 that for each pair $ (n,w) $ 
occuring, the r.h.s. of (\ref{a14}) vanishes due the key properties 
  (\ref{p7}) and preceding instances
(\ref{a17}). Hence, (\ref{a17}) also holds for all $n$ in the
 order $l$ and the proposition is proven.  \\
From $\alpha ), \beta ) $ we know, that letting
 $ \Lambda_0 \rightarrow \infty $, each term on the r.h.s.
of equation (\ref{a8}) converges to a finite limit. Hence, if
 the renormalization conditions chosen for
$ L^{\Lambda, \Lambda_0}_{(1)} (q \, ; \varphi ) $ are inferred
 from those of  $ L^{\Lambda, \Lambda_0}
( \varphi ) $ to satisfy (\ref{a16}), the l.h.s. of (\ref{a8})
 vanishes for $ 0 \leq \Lambda \leq  \Lambda_0 $.
 Thus, the field equation (\ref{a6}) remains valid after removing the 
cutoffs $ \Lambda, \Lambda_0 $, written suggestively as
 \begin{equation} \label{a19}
 (2\pi )^4 \frac{\delta }{\delta \hat {\varphi} (q)}
  L^{0, \, \infty}(\varphi ) \, = \,
\hat { L}^{0, \, \infty}_{(1)}(q \, ; \varphi)
   +\hat {I} ^{\, 0, \, \infty}_{(1)}(q ) \, ;
 \end{equation} 
in the realm of a formal loop expansion, of course.
 Considering the relations (\ref{a17}) at $\Lambda = \Lambda_0 $ reveals
 that the counterterms entering the insertion
(\ref{a7}) have to be chosen identical to those of 
the bare interaction (\ref{f1}), $ l \in \mathbf{N} $:
 \begin{equation} \label{a20}
v_{1,l} (\Lambda_0) = v_{l} (\Lambda_0), \, \cdots \, ,
 c_{1,l} (\Lambda_0) = c_{l} (\Lambda_0) .
  \end{equation} 
%%%%%%%%%%%%%%%%%%%%%%%%%%%%%%%%%%%%%%%%%%%%%%%%%%%%%%%%%%%%   
 \section{Variation of a coupling constant}
The renormalized amputated truncated Schwinger functions (\ref{p2})
 depend on the coupling constants
$f$ and $g$, which can be freely chosen in the bare
 interaction (\ref{f1}). Our aim is to find a
representation for the derivative of these Schwinger
 functions with respect to $f$ or $g$. To this end
we start from the defining equation (\ref{f5}) of the
 regularized generating functional. Denoting by
$\kappa $ either $f$ or $g$, and defining
 \begin{equation} \label{c1j}
W_{\kappa }(\phi ) := \frac{\partial }{\partial \kappa }
     L^{\Lambda_0,\Lambda_0} (\phi )
   = : \int_{\Omega } dx Q_{\kappa }
\end{equation} 
where the integrand $ Q_{\kappa }(x) $ is a composite field and
 $ W_{\kappa }(\phi ) $ the space-time  integral of it,
we obtain from deriving (\ref{f5}): 
\begin{eqnarray} \label{c2j} 
    \partial _{\kappa}\left( L^{\Lambda ,\Lambda _0}(\varphi ) +
 I^{\Lambda ,\Lambda _0} \right) \cdot
e^{-\frac{1}{\hbar}\left( L^{\Lambda ,\Lambda _0}(\varphi ) +
 I^{\Lambda ,\Lambda _0} \right)}
   \qquad \quad \nonumber \\
    \qquad \quad =  \int d\mu _{\Lambda ,\Lambda _0}(\phi )
 e^{-\frac{1}{\hbar} L^{\Lambda_0 ,\Lambda _0}( \phi +\varphi ) }
 W_{\kappa }(\phi +\varphi  ) . 
\end{eqnarray}
On the other hand , the functional derivation of eq. (\ref{i6})
 with respect to $\varrho (x) $ at
$\varrho (x) =0 $ yields, observing the shift $ \phi \rightarrow \phi
 +\varphi $ to be performed in
(\ref{i4}) and employing the notations (\ref{i9}),(\ref{i11}):
 \begin{eqnarray} \label{c3j} 
  \left( L^{\Lambda ,\Lambda _0}_{(1)}(x \, ;\varphi ) +
 I^{\Lambda ,\Lambda _0}_{(1)}(x) \right) 
e^{-\frac{1}{\hbar}\left( L^{\Lambda ,\Lambda _0}(\varphi ) +
 I^{\Lambda ,\Lambda _0} \right)}
    \qquad \qquad \nonumber \\
   \qquad \qquad =  \int d\mu _{\Lambda ,\Lambda _0}(\phi )
 e^{-\frac{1}{\hbar} L^{\Lambda_0 ,\Lambda _0}( \phi +\varphi ) }
 Q(x)|_{\phi \rightarrow \phi +\varphi } .
 \end{eqnarray}
In writing this equation we have already taken account of the identities
$$ L^{\Lambda, \Lambda_0}(0 \, ;\varphi ) =
 L^{\Lambda, \Lambda_0} (\varphi ) \quad ,
 \quad I^{\Lambda, \Lambda_0} (0) = I^{\Lambda, \Lambda_0} . $$
Choosing in (\ref{c3j}) the particular composite field
 $ Q(x) = Q_{\kappa }(x) $ introduced in (\ref{c1j}),
and integrating over the finite space-time  $\Omega $, implies
 by comparison with (\ref{c2j}),
$$\partial _{\kappa } L^{\Lambda,\Lambda_0} (\varphi ) \,=
 \,  \int_{\Omega } dx L^{\Lambda, \Lambda_0}_{(1)} (x \, ;\varphi ) .$$
 We can now pass to the infinite volume limit $\Omega \rightarrow \mathbf{R}^4,
\varphi \in \mathcal{S}(\mathbf{R}^4) $. Hence,
\begin{equation} \label{c4j}
\partial _{\kappa } L^{\Lambda, \Lambda_0} (\varphi )  \, =
 \, \hat{L}^{\Lambda, \Lambda_0}_{(1)} (0 \, ;\varphi ) ,
\end{equation}
with the Fourier transform (\ref{i12}) at vanishing momentum.
 In the sequel, $ \hat{L}^{\Lambda, \Lambda_0}_{(1)} (0 \, ;\varphi ) $
 is always understood as the generating functional with the insertion
 (\ref{c1j}). \\
The task posed is to produce a finite limit of the eq. (\ref{c4j})
 upon removing the cutoffs, i.e. letting
$ \Lambda = 0, \Lambda_0 \rightarrow \infty $. A finite limit of
 $ L^{\Lambda,\Lambda_0} (\varphi ) $ has been established
in section 2.3. Furthermore, the insertion appearing in (\ref{c4j})
 is a particular instance of the insertion
of a composite field $Q(x)$ dealt with in section 2.4. The
 composite field $ Q_{\kappa}(x) $ involved here 
follows from (\ref{c1j}) and (\ref{f1}) as
\begin{eqnarray}\label{c5j}
Q_{\kappa}(x ) &=& 
 \frac{1}{3!} \delta _{\kappa f} \,\phi ^3(x)
 +\frac{1}{4!}\delta _{\kappa g} \,\phi ^4(x)   \nonumber\\
& \, & + v_{\kappa}(\Lambda _0)\phi (x)+
 \frac{1}{2}a_{\kappa}(\Lambda_0)\phi ^2(x)
+\frac{1}{2}z_{\kappa}(\Lambda_0)\bigl(\partial _\mu \phi \bigr)^2(x)
 \nonumber \\ & \quad &  + \frac{1}{3!}b_{\kappa}(\Lambda_0)\phi^3 (x)
+\frac{1}{4!}c_{\kappa}(\Lambda_0)\phi^4 (x)  ,
\end{eqnarray}
where $\delta _{\kappa f} $ is the Kronecker symbol:
 $\delta _{\kappa f} = 1$, if $\kappa = f$, and
 $\delta _{\kappa f} = 0$, if $\kappa \neq f $. One should note
 that this composite field in both cases
$\kappa =f$ or $\kappa =g$ has the canonical mass dimension
 $D=4$, in contrast to its classical part.
The coefficients of the counterterms appearing in (\ref{c5j}) are the
coefficients entering the bare interaction (\ref{f1}) derived with
respect to $ \kappa $. However, since in the process of renormalization
the counterterms are determined by the renormalization conditions chosen,
we at first treat the counterterms in (\ref{c5j}) as free functions of
$ \Lambda_0$, which are then determined by the renormalization conditions
prescribed in the case of the insertion.
 We do not assume the renormalization
conditions (\ref{p3}-\ref{p6}) of $ L^{\Lambda, \Lambda_0} (\varphi ) $
 to depend on $f$ and $g$,
 hence, their derivative with respect to $\kappa$ vanishes.
 Requiring (\ref{c4j}) to be valid at
the renormalization point for all values of $\Lambda_0 $ then implies,
 that in all loop orders $ l \geq 1 $ an $n$-point function of
$ \hat{L}^{\Lambda, \Lambda_0}_{(1)} (0 \, ;\varphi ) $ and
 its momentum derivatives vanish for
$\Lambda=0$ at zero momenta, if $n+|w| \leq 4$. The
 renormalization conditions fixed, we know
from Proposition 2.4, that the functional 
 $ \hat{L}^{\Lambda, \Lambda_0}_{(1)} (0 \, ;\varphi )$
has a finite limit $ \Lambda=0, \Lambda_0 \rightarrow \infty$.
 To control the renormalization of eq. (\ref{c4}) we define
\begin{equation} \label{c6j}  
D^{\Lambda, \Lambda_0}(\varphi ) := 
 \hat{L}^{\Lambda, \Lambda_0}_{(1)} (0 \, ;\varphi )
     - \partial _{\kappa} L^{\Lambda, \Lambda_0} (\varphi ) .
\end{equation}
This functional satisfies, as easily seen, a linear flow equation
 of the form (\ref{a9}); hence,
after decomposition, a system of flow equations of the
 form (\ref{a14}) results. (Here, no vacuum part appears.) Comparing
the bare interaction (\ref{f1}) with the insertion (\ref{c5j})
 we observe that $D^{\Lambda_0, \Lambda_0}$ vanishes in the tree order
     $$D^{\Lambda_0, \Lambda_0}|_{l\,=\,0} =0 \, ,$$
and its irrelevant part vanishes for $l > 0 $,
$$\partial ^w \mathcal{D}^{\Lambda_0, \Lambda_0}_{l,n}( p_1, \cdots, p_n)
\, = 0 , \quad  n+|w| > 4\, .$$
Given these initial conditions we have the \\ 
\textbf{Proposition 3.2} \\
\emph{ Assume for all } $ l \in \mathbf{N}, n+|w| \leq 4 $:
 \begin{equation} \label{c7j}
\partial ^w \mathcal{D}^{0, \Lambda_0}_{l,n}(0, \cdots, 0) \, = \, 0,
 \end{equation}
\emph{then follows for} $ l \in \mathbf{N}_0, n \in \mathbf{N},
 |w|\leq  4$, \emph{and} $ 0 \leq \Lambda \leq \Lambda_0 $:
 \begin{equation} \label{c8j}
\partial ^w \mathcal{D}^{\Lambda, \Lambda_0}_{l,n}(p_1, \cdots, p_n) \,
 = \, 0  \,. 
 \end{equation}
The proof by induction proceeds exactly as the proof of
 Proposition 3.1 and is omitted.
 Proposition 3.2 implies, that equation
 (\ref{c4j}) has a finite limit for
$ \Lambda=0, \Lambda_0 \rightarrow \infty $,
\begin{equation} \label{c9j}
    \partial _{\kappa} L^{0,\, \infty} (\varphi ) \,
 = \, \hat{L}^{0,\, \infty}_{(1)}( 0 \, ; \varphi ) ,
  \end{equation}
again to be read in terms of a formal loop expansion. Furthermore, from
(\ref{c8}) at $ \Lambda = \Lambda_0 $ follows the relation of the
counterterms
 \begin{equation} \label{c10j}
v_{\kappa, \, l}(\Lambda_0) = \partial_{\kappa} v_{\,l} (\Lambda_0),
\cdots, c_{\kappa, \, l}(\Lambda_0) = \partial_{\kappa} c_{\,l} (\Lambda_0). 
 \end{equation} 
%%%%%%%%%%%%%%%%%%%%%%%%%%%%%%%%%%%%%%%%%%%%%%%%%
\section{Flow equations for \\ proper vertex functions}
In perturbative renormalization based on the analysis of
 Feynman integrals, (proper) vertex functions form the building blocks.
 They are represented by one - particle - irreducible
 (1PI) Feynman diagrams, see
e.g. \cite{ZJ}. Although their generating functional has no
 representation as a functional
integral, flow equations for vertex functions can be
 derived \cite{Vert}. Our goal in this section
is to deduce in the case of the symmetric $\Phi ^4$-theory
 from Wilson's differential flow equation
(\ref{f10}) for the $L$-functional the system of flow equations
 satisfied by the regularized
$n$-point vertex functions, $n \in \mathbf{N}$. After that,
 an inductive proof of renormalizability
based on them is outlined.

We start from the regularized generating functional $W^{\LLz} (J)$
 of the truncated Schwinger functions, (\ref{f3}), decomposed as
  \begin{equation} \label{s1}
W^{\LLz} (J) = \sum_{n=1}^{\infty} \frac{1}{(2n) !} \int d x_1 \cdots
 \int d x_{2n}  W^{\LLz}_{2n}(x_1, \cdots, x_{2n}) J(x_1) \cdots J(x_{2n}) ,
 \end{equation} 
according to (\ref{f4}). Due to the symmetry $\phi \rightarrow - \phi $
 of the theory, all $n$-point
functions with $n$ odd vanish identically. Defining the ``classical field "
 \begin{equation} \label{s2}
\ufi (x) := \frac{\delta W^{\LLz} (J)}{\delta J(x)} \, ,
 \end{equation} 
we then notice, that
$$ \ufi (x) |_{J \equiv 0} \, = 0 \, ,$$
and, moreover, that $\ufi$ depends on the flow parameter $\Lambda$
 ( and on $\Lambda_0$). Since 
the 2-point function $W^{\LLz}_2$ is different from zero,
 (\ref{s2}) can be inverted iteratively
 as a formal series in $\ufi (x)$ to yield the source $J(x)$ in the form
  \begin{equation} \label{s3}
J(x)  = \mathcal{J} (\ufi (x) ) \equiv 
 \end{equation}   
 $$   \sum_{n=0}^{\infty} \frac{1}{(2n+1) !}
   \int d x_1 \cdots \int d x_{2n+1}
  F_{2n+2} (x, x_1, \cdots, x_{2n+1}) \ufi (x_1) \cdots \ufi (x_{2n+1}) ,$$
      where 
$$ \int dy F_2 (x,y )\,  W^{\LLz}_2  (y,z) = \delta (x-z)\ . $$
The generating functional $\Gf (\ufi)$ of the regularized vertex
 functions results from the Legendre transformation
 \begin{equation} \label{s4}
\Gf (\ufi) := \Big [ - W^{\LLz} (J) +
 \int d y J(y) \ufi(y)\, \Big]_{ J = \mathcal{J} (\ufi) } 
 \end{equation} 
implying, due to (\ref{s2}), 
 \begin{equation} \label{s5}
\frac{\delta \Gf (\ufi)}{\delta \ufi (x)} \, = \, J(x) \, .
 \end{equation} 
The functional $\Gf (\ufi)$ is even under $\ufi \rightarrow - \ufi$
 and vanishes at $ \ufi = 0$.
Finally, performing the functional derivation of (\ref{s5})
 with respect to $J(y)$ and using again (\ref{s2}),
provides the crucial functional relation
 \begin{equation} \label{s7}
\int d z \, \frac{\delta ^2\, W^{\LLz} (J)}{\delta J(y) \, \delta J(z)}\cdot
     \frac{\delta ^2 \,\Gf (\ufi)}{\delta \ufi (z) \,\delta \ufi (x)} \,
     = \, \delta (y-x)  .
 \end{equation} 
As an immediate consequence follows
$$ \int d z \, W^{\LLz}_2 (y, z)\, \Gf_2 (z, x) \, = \, \delta (y-x) ,$$
considering (\ref{s7}) at $\ufi = 0$, and thus also at $J=0$.
 In order to obtain the relation between the functionals
 $L^{\LLz}(\varphi )$ and $\Gf (\ufi) $, we write (\ref{f8}) in the form
  \begin{equation} \label{s8}
W^{\LLz} (J) = - \, L^{\LLz} (\varphi ) + \frac{1}{2} 
\langle J, C^{\LLz} J \,\rangle \, ,
 \end{equation} 
 \begin{equation} \label{s9}
\varphi (x) \, = \, \int d y \, C^{\LLz} (x-y) J(y) \, .
 \end{equation} 
Deriving (\ref{s8}) twice with respect to $J$ as required in (\ref{s7})  
one obtains after operating on this equation with $(C^{\LLz})^{-1}$ ,
\begin{eqnarray} \label{s10}
( C^{\LLz})^{-1} (y-x) = & - & \int d z \int d u \,
     \frac{\delta ^2 \, L^{\LLz} (\varphi )}
       {\delta \varphi (y) \, \delta \varphi (u)}\, C^{\LLz} (u-z)\, 
                \frac{\delta ^2 \, \Gf (\ufi)}{\delta \ufi (z) \,
     \delta \ufi (x)} \nonumber \\
 & + & \frac{\delta ^2 \, \Gf (\ufi)}{\delta \ufi (y) \,\delta \ufi (x)} \,.
\end{eqnarray} 
From this analogue of (\ref{s7}) follow the relations between
 the respective $n$-point functions of
$\Gf (\ufi) $ and $L^{\LLz}(\varphi )$ upon repeated functional
 derivation with respect to $\ufi$,
employing the chain rule together with the relation
  \begin{equation} \label{s11}
\varphi (x) = \int d y \, C^{\LLz} (x-y) \,
\frac{\delta \Gf (\ufi)}{\delta \ufi (y)} \, ,
 \end{equation} 
due to (\ref{s9}) and (\ref{s5}). With our conventions (\ref{f12}),
 (\ref{f16}) for the Fourier transformation,
the equations (\ref{s10} - \ref{s11}) appear in momentum space as
 \begin{eqnarray} 
(2\pi)^{-4}\, \frac{\delta (p+q)}{C^{\LLz}(p)} & = & - \, (2\pi)^8 \int_{k} 
\frac{\delta ^2 \, L^{\LLz}}{\delta \hat{\varphi }(p) \,
       \delta \hat{\varphi }(k)} \, C^{\LLz}(k)\,
   \frac{\delta ^2\, \Gf}{\delta \hat{\ufi}(-k) \, \delta \hat{\ufi}(q)}
          \nonumber \\
   &{}& + \, \frac{\delta ^2\, \Gf}{\delta \hat{\ufi}(p) \,
       \delta \hat{\ufi}(q)} \, ,\label{s12} \\
\hat{\varphi }(q) & = & (2\pi)^4 \, C^{\LLz} (q) \,
     \frac{\delta \Gf}{\delta \hat{\ufi}(-q)}\, .\label{s13}
 \end{eqnarray} 
In the tree order, $L^{\LLz}(\varphi )$ contains no
    2-point function, (\ref{p7}). 
Hence, setting in (\ref{s12}) $\varphi = \ufi = 0$ yields
 \begin{equation} \label{s14}
\frac{\delta ^2\, \Gf (\ufi)}{\delta \hat{\ufi}(p) \, \delta \hat{\ufi}(q)}
{\bigg |}_{\,\ufi \, \equiv \, 0}^{\,l\, = \,0} 
= (2\pi)^{-4} \, \frac{\delta (p+q)}{C^{\LLz}(p)} \, .
 \end{equation} 
To deduce the flow equation for the vertex functional,
  we derive equation (\ref{s4}) with respect
to the flow parameter $\Lambda$,
$$ (\partial _{\Lambda} \Gf ) (\ufi) + \int d y \,
     \frac{\delta \Gf}{\delta \ufi (y)}\, 
    \partial _{\Lambda} \ufi (y) = - \, \partial _{\Lambda} W^{\LLz} (J) +
              \int dy \, J(y) \,\partial _{\Lambda} \ufi (y) . $$
Hence, because of (\ref{s5}),
  \begin{equation} \label{s15}
( \partial _{\Lambda} \Gf ) (\ufi ) +
       \partial _{\Lambda} W^{\LLz} (J) \, = \, 0.
 \end{equation} 
Substituting $W^{\LLz}$ by $ L^{\LLz}$ according to (\ref{s8}) then yields 
  \begin{eqnarray}  \label{ss15}
( \partial _{\Lambda} \Gf ) (\ufi ) - 
      (\partial _{\Lambda} L^{\LLz}) (\varphi ) & - &
     \int dz \int dy \, \frac{\delta L^{\LLz}}{\delta \varphi (y)}
      {\dot C}^{\LLz} (y-z) J(z) \nonumber \\
        &+ & \frac{1}{2} \, \langle J,{\dot C}^{\LLz} J \,\rangle = \, 0 \,,
  \end{eqnarray} 
 ${\dot C}^{\LLz} $ denoting the derivative of the
 covariance $ C^{\LLz} $ with respect 
  to $\Lambda$. \\
There is an alternative way \cite{KKSch} to arrive at equation (\ref{ss15}),
starting from (\ref{s4}) but treating $\ufi$ as $\Lambda$-independent 
and thus $J$ to depend on $\Lambda$ according to (\ref{s2}).
Deriving (\ref{s4}) with respect to $\Lambda$ then reads
 \begin{eqnarray}
\partial_{\Lambda} \Gf (\ufi) & = & 
 \Big [ - (\partial_{\Lambda} W^{\LLz}) (J) 
  - \int dy \frac{\delta W^{\LLz}}{\delta J(y)}\,\partial_{\Lambda}J(y)
  \nonumber \\
  & {} & \qquad \qquad  + \int d y \,\partial_{\Lambda} J(y)\, \ufi(y)\,
    \Big]_{ J = \mathcal{J} (\ufi)}  \nonumber  \\
 & = &  - (\partial_{\Lambda} W^{\LLz})(J){\big |}_{J =\mathcal{J}(\ufi)}\,.
 \end{eqnarray} 
The substitution of $W^{\LLz}$ by $L^{\LLz}$ due to (\ref{s8} - \ref{s9})
again provides (\ref{ss15}).\\
Given (\ref{ss15}) the flow equation (\ref{f10}) with its vacuum part
 subtracted can be taken into account, leading to
  \begin{eqnarray}
&{}& (\partial _{\Lambda} \Gf ) (\ufi)  +
 \frac{1}{2} \langle \,\frac{\delta L^{\LLz}}{\delta \varphi } 
   - J \, ,{\dot C}^{\LLz} \, ( \frac{\delta L^{\LLz}}{\delta \varphi }
        - J \,) \rangle  \nonumber \\
  & = & \frac{\hbar}{2}   \left ( \langle \frac{\delta }{\delta \varphi }\,
                , {\dot C}^{\LLz}
              \frac{\delta }{\delta \varphi } \rangle L^{\LLz} (\varphi) 
             -  \langle \frac{\delta }{\delta \varphi } \, , {\dot C}^{\LLz}
              \frac{\delta }{\delta \varphi } \rangle
           L^{\LLz}(\varphi ) {\bigg |}_{\varphi \equiv 0} \right ) .
                      \nonumber
  \end{eqnarray} 
In the second term on the l.h.s. we use the relation
\begin{equation} \label{s16}
\int d x \, ( C^{\LLz} )^{- 1} \, (z - x)\, \ufi (x) =\,-
 \,\frac{\delta L^{\LLz}}{\delta \varphi (z)} + J(z) \, ,
\end{equation}    
resulting from (\ref{s2}) together with (\ref{s8}),
 and note $ (\partial _{\Lambda} C ) C^{-1}+
 C \partial _{\Lambda} C^{-1} = 0$. Hence, the flow equation
 for the vertex functional turns out as  
 \begin{equation} \label{s17}
 ( \partial _{\Lambda} \Gf ) (\ufi )\, -
 \, \frac{1}{2} \langle \, \ufi ,
 (\partial _{\Lambda} ( C^{\LLz})^{-1}) \ufi \,\rangle 
 = \frac{\hbar}{2}\, \langle \frac{\delta }{\delta \ufi} \,
 , (\partial _{\Lambda} C^{\LLz})
     \frac{\delta }{\delta \ufi}\, \rangle
 \,{\tilde \Gamma }^{\LLz} (\ufi)  \, ,
 \end{equation} 
with the r.h.s. defined by
 \begin{equation} \label{s18}
 \frac{\delta ^2 \, \Gtf (\ufi)}{\delta \ufi (x) \, \delta \ufi (y)} :=
\frac{\delta ^2 L^{\LLz} (\varphi )}{\delta \varphi (x) \,
 \delta \varphi (y)}{\bigg |}_{\varphi =\,
 C^{\LLz}  \mathcal{J} (\ufi)} \, - \,  
\frac{\delta ^2 L^{\LLz} (\varphi )}{\delta \varphi (x) \,
 \delta \varphi (y)}{\bigg |}_{\varphi =\, 0 } \, .
 \end{equation} 
Looking at this definition of the functional
 ${\tilde \Gamma}^{\LLz} (\ufi) $ we notice, that its 2-point
function vanishes, and furthermore, that its higher
 $n$-point functions, $n=4, 6, 8, \cdots ,$ emerge
from the first term on the r.h.s.. These latter are recursively
 determined by the functional equations
(\ref{s10}) or (\ref{s12}) (which could also be
 obtained via (\ref{s16}) and (\ref{s5})),
by performing successively two, four, six, $\cdots$  functional
 derivations with respect to $\ufi$.\\
The r.h.s. of the flow equation (\ref{s17}) can also be given
another form, expressing (\ref{s18}) first in terms of the functional
$ W^{\LLz}(J) $ by way of (\ref{s8}),(\ref{s9}) and then using the 
functional relation (\ref{s7}), 
\begin{eqnarray}\label{s19}
 \frac{\hbar}{2}\, \langle \frac{\delta }{\delta \ufi} \,
 , (\partial _{\Lambda} C^{\LLz})
     \frac{\delta }{\delta \ufi}\, \rangle
 \,{\tilde \Gamma }^{\LLz} (\ufi)  \qquad \qquad \qquad \qquad  \\
 = \frac{\hbar}{2} \int dx \int dy \, \partial_{\Lambda}
  ( C^{\LLz})^{-1}(y-x)\left(  \frac{\delta ^2\, W^{\LLz} (J)}
   {\delta J(x) \, \delta J(y)} -\frac{\delta ^2\, W^{\LLz} (J)}
   {\delta J(x) \, \delta J(y)} \bigg |_{J = 0} \right ) \nonumber \\
=   \frac{\hbar}{2} \int dx \int dy \, \partial_{\Lambda}
  ( C^{\LLz})^{-1}(y-x) \qquad \qquad \qquad \qquad
     \qquad \qquad \quad \quad \quad \, \, \, \nonumber \\
 \cdot \left( \bigg (\frac{\delta ^2 \, \Gf (\ufi)}
{\delta \ufi \, \delta \ufi}\bigg)^{-1}(x,y) - 
  \bigg (\frac{\delta ^2 \, \Gf (\ufi)}
{\delta \ufi \, \delta \ufi}\bigg)^{-1}(x,y)\bigg |_{\ufi = 0}
 \right )\, . \nonumber
\end{eqnarray}
This form is (also) met in the literature \cite{ReWe, D'AM},
and the flow equation (\ref{s17}) 
called there ``exact renormalization group''. \\
Similar to (\ref{f14}) 
 regarding the functional $L^{\LLz}(\varphi)$ we define
 the $n$-point functions, 
$n \in 2\mathbf{N}$, of the functional $\Gf (\ufi)$ in momentum space as
 \begin{equation} \label{s20}
(2\pi)^{4(n-1)} \frac {\delta}{\delta \hat {\ufi} (p_1)} \cdots  
 \frac{\delta}{\delta \hat {\ufi} (p_n)}
 \Gf (\ufi) {\Big |}_{\,\ufi \,\equiv \,0}
 = \delta ( p_1+ \cdots + p_n)\, \Gf _n (p_1, \cdots, p_n ),
 \end{equation}   
and analogously in the case of the functional $ \Gtf (\ufi) $.
 Performing in addition a respective loop
expansion, $n \in 2\mathbf{N}$,
 \begin{equation} \label{s21}
\Gf_n (p_1, \cdots, p_n) \, = \, \sum_{l=0}^{\infty} \hbar^l \,
\Gf_{l,\, n} (p_1, \cdots, p_n) ,
 \end{equation}   
and for the functions $ \Gtf _n (p_1, \cdots, p_n) $ alike,
 the flow equation (\ref{s17}) is finally converted into the system
 of flow equations, satisfied by the $n$-point functions, $n \in 2\mathbf{N},
l \in \mathbf{N}$,
 \begin{equation} \label{s22}
\partial _{\Lambda} \Gf_{l, \, n}(p_1, \cdots, p_n) \,
 = \, \frac{1}{2} \int_{k}
 \partial _{\Lambda} C^{\LLz}(k) \cdot
 \Gtf _{l-1, \, n+2} (k, - k, p_1, \cdots, p_n) \, .
 \end{equation}   
In contrast to the system of flow equations (\ref{f20}) satisfied
 by the amputated truncated Schwinger
functions, here the r.h.s. is in total of lower loop order,
 but there is no closed form for it. As explained
before, it has to be determined recursively via (\ref{s12}), treated
in a loop expansion and using (\ref{s14}). It then emerges in the form
\begin{eqnarray} \label{s23}
 \Gtf _{l, \, n+2} (k, - k, p_1, \cdots, p_n) \, = \,
\Gf_{l, \, n+2} (k, - k, p_1, \cdots, p_n) \qquad \nonumber \\ 
 - \, \sum_{r \geq 2} \, \sum_{\{n_i\}, \, \{l_i\} }'
   \sigma \, \Gf_{l_1, \, n_1+1}(k, \cdots \,)\, C^{\LLz} \,
 \Gf_{l_2, \, n_2+2} \, \cdots \, \qquad \qquad \nonumber \\
 \cdots \Gf_{l_{r-1}, n_{r-1}+2}\, C^{\LLz} 
 \Gf_{l_r, \, n_r+1} ( - k, \cdots \,) \, .                        
\end{eqnarray} 
The prime restricts  summation to $l_1+l_2+ \cdots + l_r =
 l$ and $n_1+n_2+ \cdots + n_r = n+2$, in addition,
 2-point functions in the tree order are excluded as factors. 
 The momentum assignment has been suppressed,
it goes without saying that the sum inherits from the l.h.s. 
the complete symmetry in the momenta $p_1, \cdots, p_n$. 
Moreover, there is a sign factor $\sigma$ depending on $ \{n_i \}$ 
and $ \{ l_i\}$. The form of (\ref{s23}) is easily understood
 when represented by Feynman diagrams: To the
first term (on the r.h.s.) correspond 1PI - diagrams, whereas to
 the sum correspond chains of 1PI -   diagrams,
minimally connected by single lines and thus not of 1PI - type.
 These chains are closed to 1PI - diagrams
by the contraction involved in the flow equation.

The system of flow equations (\ref{s22}) can alternatively be
 employed to prove the renormalizability
of the theory considered. In the tree order, only the
 2-point function (\ref{s14}) and the 4-point function
$\Gf_{0, 4} (p_1,\cdots, p_4) = g $ are different from zero.
The latter is easily obtained via (\ref{s12}-\ref{s14}) from (\ref{p9})
 observing $f=0$ there. In each loop order $l \geq 1$,
 the three counterterms
\begin{equation} \label{s24}
\Gamma^{\LLzz}_{l , 2}(p, - p) = {\underline a}_{\, l} (\Lambda_0) 
+ {\underline z}_{\,l} (\Lambda_0)\, p^2 ,
  \quad \Gamma^{\LLzz}_{l, 4}(p_1, \cdots, p_4) = 
 {\underline c}_{\,l} (\Lambda_0)
\end{equation}  
form the respective bare action, determined  in the end by
 the renormalization conditions, $l\geq 1$,
 \begin{equation} \label{s25}
\Gamma^{\Lz}_{l ,\, 2}(p, - p) = {\underline a}_{\, l}^R 
+ {\underline z}_{\, l}^R \, p^2 + \mathcal{O}((p^2)^2),
  \quad \Gamma^{\Lz}_{l,\, 4}(0, 0, 0, 0 ) =  {\underline c}_{\, l}^R \, .
 \end{equation}   
The renormalization constants ${\underline a}_{\, l}^R, 
 {\underline z}_{\, l}^R, {\underline c}_{\, l}^R 
 $ can be freely chosen. To prove renormalizability, we also have
 to make use of momentum derivatives
of the flow equations (\ref{s22}), i.e. acting on them
 with $\partial ^w$, (\ref{f19}). Then, the proof
by induction follows step-by-step the proof given in section 2.3
 considering amputated truncated
Schwinger functions. It will therefore not be repeated. As result,
 the analogue of the Propositions 2.1
 and 2.2 is established, where the function $\Lf_{l, n}$ 
appearing there is now replaced by the function $\Gf_{l, n}$.

\chapter{Spontaneously Broken  SU(2) Yang-Mills Theory}
Attempting to prove renormalizability of a non-Abelian gauge theory
via flow equations, following the path taken before in the case of a
scalar field theory, one finds oneself confronted with a serious
obstacle to be surmounted. By their definition the Schwinger functions
of a non-Abelian gauge theory are not gauge invariant individually,
the local gauge invarince of the theory, however, compels them to
satisfy the system of \emph{Slavnov-Taylor identities}
\cite{SlTa}. These identities are inevitably violated, if one employs
in the intermediate regularization procedure a momentum
cutoff. Moreover, the Slavnov-Taylor identities are generated by
nonlinear transformations of the fields -- the
\emph{BRS-transformations} \cite{BeRoSt} -- which, being composite
fields, have to be renormalized, too. 
 In the sequel we follow the
general line of \cite{nKM}, hereafter referred to as ``II''.
 For the sake of readability, however, a coherent detailed
argumentation is kept up.  As concerns a number of proofs
and technical derivations, we refer to the original article. 
 After presenting in Section 4.1 the classical action of
 the theory considered,
as a first step we disregard in Section 4.2 the Slavnov-Taylor identities
and establish for an arbitrary set of renormalization conditions at a
physical renormalization point a finite UV-behaviour of the Schwinger
functions without and with the insertion of one BRS-variation.
 The procedure is essentially the same as
in the case of the scalar theory.  Having thus established a family of
finite theories, in Section 4.3 the violation of the Slavnov-Taylor
identities of the amputated truncated Schwinger functions at the
physical value $\Lambda = 0 $ of the flow parameter and fixed $
\Lambda_0 $ is worked out, as well as the BRS-variation of the bare
action. Moreover, the corresponding violated Slavnov-Taylor identities
in terms of proper vertex functions are  also deduced, as this
formulation appears more accessible to analysis. Inspection of
the relevant part of these identities reveals, that in
this part irrelevant contributions from the vertex functions
with insertion of a BRS-variation appear. To overcome this obstruction, 
in Section 4.4 a particular mass expansion is developed. It results
from scaling the super-renormalizable $3$-point couplings appearing
at tree level, i.e. in the classical action, and traces
the effect of these couplings in the perturbative expansion.
On account of the corresponding adapted renormalization scheme
in the relevant part of the violated Slavnov-Taylor identities
now no longer irrelevant terms contribute, but solely 
renormalization constants. In Section 4.5 the equation of motion
of the antighost is considered. Finally, in Section 4.6 the restoration
of the Slavnov-Taylor identities is dealt with. Requiring the
relevant part of the violated Slavnov-Taylor identities to
vanish amounts to satisfy 53 equations containing the 37+7
relevant parameters of the respective generating functionals
of proper vertex functions without and with insertion of a
BRS-variation. Taking also into account the field equation
of the antighost, it is shown that this can be accomplished
extracting a particular set of 7 relevant parameters as
renormalization constants to be chosen freely, in terms of 
which the remaining ones are uniquely determined in satisfying
these 53 equations. Given a vanishing relevant part of the
violated Slavnov-Taylor identities then implies that their
irrelevant part has a vanishing UV-limit, thus achieving
the Slavnov-Taylor identities in this limit.
%%%%%%%%%%%%%%%%%%%%%%%%%%%%%%%%%%%%%%%%%%%%%%%%%%%%%%%
\section{The classical action}    
We begin collecting some basic properties of the classical Euclidean
SU(2) Yang-Mills-Higgs model on four-dimensional Euclidean space-time,
following closely the monograph of Faddeev and Slavnov \cite{FaSl}.
This model involves the real Yang-Mills field $\{A^a_{\mu}\}_{a =
1,2,3}$ and the complex scalar doublet $\{\phi_{\alpha}\}_{\alpha =
1,2}$ assumed to be smooth functions which fall-off rapidly. The
classical action has the form
\begin{equation}\label{y1}
S_{inv} = \int dx \left\{ \frac14 F^a_{\mu\nu} F^a_{\mu\nu} +
\frac12(\nabla_{\mu}\phi)^{\ast}
\nabla_{\mu}\phi + \lambda(\phi^{\ast}\phi-\rho^2)^2\right\},
\end{equation}
with the curvature tensor
\begin{equation}\label{y2}
F^a_{\mu\nu}(x) = \partial_{\mu}A^a_{\nu}(x) -
\partial_{\nu}A^a_{\mu}(x) + g\epsilon^{abc} A^b_{\mu}(x) A^c_{\nu}(x)
\end{equation}
and the covariant derivative
\begin{equation}\label{y3}
\nabla_{\mu} = \partial_{\mu} + g \frac{1}{2i} \sigma^aA^a_{\mu}(x)
\end{equation}
acting on the SU(2)-spinor $\phi$. The parameters $g, \lambda, \rho$
are real positive, $\epsilon^{abc}$ is totally skew symmetric,
$\epsilon^{123} = +1$, and $\{\sigma^a\}_{a = 1,2,3}$ are the standard
Pauli matrices. For simplicity the wave function normalizations of the
fields are chosen equal to one. The action (\ref{y1}) is invariant
under local gauge transformations of the fields
\begin{equation}
\label{y4}
\begin{split}
&\frac{1}{2i} \sigma^a A^a_{\mu}(x) \longrightarrow u(x) \frac{1}{2i}
\sigma^a A^a_{\mu}(x) u^{\ast}(x) + g^{-1} u(x)
\partial_{\mu}u^{\ast}(x), \\ & \qquad \quad \phi(x) \longrightarrow
u(x) \phi(x), \quad\quad
\end{split}
\end{equation}
with $u: \mathbf{R}^4 \to $ SU(2), smooth.  A stable ground state of
the action (\ref{y1}) implies spontaneous symmetry breaking, taken
into account by reparametrizing the complex scalar doublet as
\begin{equation}
\phi(x) = \left( \begin{array}{c}
B^2(x) + i B^1(x) \\ \rho + h(x) - i B^3(x) \end{array} \right)
\label{y5}
\end{equation}
where $\{B^a(x)\}_{a=1,2,3}$ is a real triplet and $h(x)$ the real
Higgs field. Moreover, in place of the parameters $\rho, \lambda$ the
masses
\begin{equation}
m = \frac12 g \rho, \quad M = (8 \lambda \rho^2)^{\frac12}
\label{y6}
\end{equation}
are used. Since we aim at a quantized theory pure gauge degrees of
freedom have to be eliminated. We choose the 't Hooft gauge fixing
\footnote{The general $\alpha$-gauge \cite{FaSl} would lead to mixed
propagators, in the Lorentz gauge the fields $\{ B^a\}$ would be massless}
\begin{equation}
S_{g.f.} = \frac{1}{2\alpha} \int dx (\partial_{\mu}A^a_{\mu} - \alpha
m B^a)^2,
\label{y7}
\end{equation}
with $\alpha \in \mathbf{R}_+$, implying complete spontaneous symmetry
breaking.  With regard to functional integration this condition is
implemented by introducing anticommuting Faddeev-Popov ghost and
antighost fields $\{c^a\}_{a=1,2,3}$ and $\{ \bar{c}^a\}_{a=1,2,3}$,
respectively, and forming with these six independent scalar fields the
additional interaction term \begin{eqnarray} S_{gh} &=& - \int dx
\bar{c}^a \big\{ (-\partial_{\mu}\partial_{\mu} + \alpha m^2)
\delta^{ab} + \frac12 \alpha gmh \delta^{ab}  \nonumber \\
& & \hspace{2cm} +\frac12 \alpha gm \epsilon^{acb} B^c - g
\partial_{\mu} \epsilon^{acb} A^c_{\mu}
\big\} c^b.
\label{y8}
\end{eqnarray} 
Hence, the total ``classical action" is
\begin{equation}
S_{\rm BRS} = S_{\rm inv} + S_{\rm g.f.} + S_{\rm gh} ,
\label{y9a}
\end{equation}
which we decompose as
\begin{equation}
S_{\rm BRS} = \int dx \left\{ {\cal L}_{\rm quad} (x) + {\cal L}_{\rm
int}(x) \right\}
\label{y9b}
\end{equation}    
into its quadratic part, with $\Delta \equiv \partial_{\mu}
\partial_{\mu}$,
\begin{eqnarray}
{\cal L}_{\rm quad} &=& \frac14 ( \partial_{\mu}A^a_{\nu} -
\partial_{\nu}A^a_{\mu} )^2 + \frac{1}{2\alpha}
(\partial_{\mu}A^a_{\mu})^2 + \frac12 m^2A^a_{\mu}A^a_{\mu} \nonumber
\\ & & + \frac12 h (- \Delta + M^2)h + \frac12 B^a (- \Delta + \alpha
m^2)B^a \nonumber \\ & & - \bar{c}^a (- \Delta + \alpha m^2) c^{a} \,
,
\label{y10}
\end{eqnarray}
and into its interaction part 
\begin{eqnarray}
{\cal L}_{\rm int} &=& g \epsilon^{abc}
(\partial_{\mu}A^a_{\nu})A^b_{\mu} A^c_{\nu} +
\frac14 g^2(\epsilon^{abc} A^b_{\mu}A^c_{\nu})^2 \nonumber \\
& & + \frac12 g \left\{ (\partial_{\mu}h)A^a_{\mu}B^a - h
A^a_{\mu}\partial_{\mu}B^a -
\epsilon^{abc}A^a_{\mu}(\partial_{\mu}B^b) B^c \right\} \nonumber \\
& & + \frac18 g A^a_{\mu}A^a_{\mu} \left\{ 4 mh + g(h^2+B^aB^a)
\right\} \nonumber \\ & & + \frac14 g \frac{M^2}{m} h (h^2 + B^aB^a) +
\frac{1}{32} g^2 \left( \frac Mm \right)^2 (h^2+B^aB^a)^2 \nonumber \\
& & - \frac12 \alpha gm \bar{c}^a \left\{ h \delta^{ab} +
\epsilon^{acb}B^c \right\} c^b
\nonumber \\
& & - g \epsilon^{acb} (\partial_{\mu} \bar{c}^a) A^c_{\mu}c^b.
\label{y11}
\end{eqnarray}
In (\ref{y10}) we recognize the important properties that all fields
are massive and that no coupling term $A^a_{\mu}\partial_{\mu} B^a$
appears. \\ The classical action $S_{BRS}$, (\ref{y9b}), shows the
following symmetries:
\begin{enumerate}
\item[i)] Euclidean invariance: $S_{BRS}$ is an O(4)-scalar.
\item[ii)] Rigid SO(3)-isosymmetry:
 The fields $\{A^a_{\mu}\}, \{B^a\}, \{c^a\}, \{ \bar{c}^a\}$
are isovectors and $h$ an isoscalar; $S_{BRS}$ is invariant under
spacetime independent SO(3)-transformations.
\item[iii)] BRS-invariance:
Introducing the classical composite fields
\begin{equation} \label{y12}
\begin{split}
\psi^a_{\mu}(x)& = \left\{ \partial_{\mu} \delta^{ab} +
 g \epsilon^{arb}A^r_{\mu}(x)\right \}
c^b(x), \\
\psi(x) & = - \frac12 g B^a(x) c^a(x),   \\
\psi^a(x) &= \left\{ (m + \frac12 gh(x)) \delta^{ab} +
 \frac12 g \epsilon^{arb} B^r(x) \right\}
c^b(x),  \\
\Omega^a(x) &= \frac12 g\epsilon^{apq}c^p(x)c^q(x), 
\end{split}
\end{equation}
the BRS-transformations of the basic fields are defined as       
\begin{eqnarray} \label{y13}
A^a_{\mu}(x) &\longrightarrow &A^a_{\mu}(x) - \psi^a_{\mu}(x)\epsilon,
\nonumber \\
h(x) &\longrightarrow &h(x) - \psi(x) \epsilon,
\nonumber \\
B^a(x) &\longrightarrow& B^a(x) - \psi^a(x) \epsilon,   \\
c^a(x)& \longrightarrow& c^a(x) - \Omega^a(x) \epsilon, \nonumber \\
\bar{c}^a(x)& \longrightarrow& \bar{c}^a(x) -
 \frac{1}{\alpha} (\partial_{\nu}A^a_{\nu}(x)-
 \alpha m B^a(x)) \epsilon.\nonumber
\end{eqnarray}
\end{enumerate}  
In these transformations $\epsilon$ is a spacetime independent
Grassmann element that commutes with the fields $\{A^a_{\mu}, h,
B^a\}$ but anticommutes with the (anti-)\linebreak ghosts $\{c^a,
\bar{c}^a\}$. To show the BRS-invariance of the total classical action
(\ref{y9a}) one first observes that the composite classical fields
(\ref{y12}) are themselves invariant under the BRS-transformations
(\ref{y13}). Moreover, we can write (\ref{y8}) in the form
\begin{equation}
S_{gh} = - \int dx \bar{c}^a \{-\partial_{\mu}\psi^a_{\mu} +
          \alpha m \psi^a \}.
\label{y14}
\end{equation}
Using these properties the BRS-invariance of the classical action
(\ref{y9a}) follows upon direct verification. \\ It is convenient to
add to the classical action (\ref{y9a}) source terms both for the
fields and the composite fields introduced, defining the extended
action
\begin{eqnarray}
S_c = S_{\rm BRS} + \int dx \{ \gamma^a_{\mu} \psi^a_{\mu} + \gamma
\psi + \gamma^a \psi^a + \omega^a\Omega^a \} \nonumber \\
\qquad - \int dx \{ j^a_{\mu} A^a_{\mu} + sh + b^aB^a +
 \bar{\eta}^ac^a + \bar{c}^a\eta^a \}.
\label{y15}
\end{eqnarray}
The sources $\gamma^a_{\mu}, \gamma, \gamma^a$ all have canonical
dimension 2, ghost number -1 and are Grassmann elements, whereas
$\omega^a$ has canonical dimension 2 and ghost number -2; the sources
$\eta^a$ and $\bar{\eta}^a$ have ghost number +1 and -1, respectively,
and are Grassmann elements.  Employing the BRS-operator ${\cal D}$,
defined by
\begin{equation}
{\cal D} = \int dx \left\{ j^a_{\mu} \frac{\delta}{\delta\gamma^a_{\mu}}
        + s \frac{\delta}
{\delta\gamma} + b^a \frac{\delta}{\delta\gamma^a} +
 \bar{\eta}^a \frac{\delta}{\delta\omega^a}
+ \eta^a \left(\frac{1}{\alpha} \partial_{\nu}
 \frac{\delta}{\delta j^a_{\nu}} -
       m \frac{\delta}{\delta b^a} \right) \right\} ,
\label{y16}
\end{equation}
the BRS-transformation  of the extended action $S_c$ , (\ref{y15}), 
can be written as  
\begin{equation}
S_c \longrightarrow S_c + {\cal D} S_c \, \epsilon
\label{y17}
\end{equation}    
Of course, $\epsilon$ also anticommutes with the sources of Grassmannian type. 
%%%%%%%%%%%%%%%%%%%%%%%%%%%%%%%%%%%%%%%%%%%%%%%%%%%%%%%%%%%%%%%
\section{Flow equations: renormalizability \\without Slavnov-Taylor identities}
In view of the various fields present, it is convenient to introduce
 a short collective notation for the
fields and sources. We denote:  \\
i) the bosonic fields and the corresponding sources, respectively, by
\begin{equation} \label{w1}
\varphi _{\tau } = ( A^{a}_{\mu }\, , \, h,\, B^{a}) \quad ,\quad J_{\tau } =
 ( j^{a}_{\mu }\, , \, s, \,b^{a}) ,
\end{equation}
ii) all fields and their respective sources by
 \begin{equation} \label{w2}
\Phi  = ( \varphi _{\tau}\, , \, c^{a}, \, {\bar c}^{a})  \quad , 
               \quad  K = ( J_{\tau}\, ,\, {\bar \eta}^{a},\, \eta ^{a}) ,
 \end{equation} 
iii) the insertions and their sources
 \begin{equation} \label{w3}
   \psi _{\tau} = (\psi ^{a}_{\mu}\, , \, \psi , \, \psi ^{a} ) \quad
 ,\ \gamma _{\tau} = ( \gamma ^{a}_{\mu}\, , \, \gamma , \, \gamma ^{a}) \, ,
                        \, \, \xi = ( \gamma _{\tau}\, ,\, \omega ^{a}) .
 \end{equation}
The quadratic part of $  S_{\rm BRS} $, (\ref{y9b}), defines the
 inverses of the various unregularized free propagators.
 We start from the theory defined on finite volume,
 as described in Section 2.1. With the notation introduced there, we have
\begin{eqnarray} \label{w4}
\int dx  {\mathcal L}_{\rm quad} (x) \equiv  Q(\Phi ) & = &
 \frac12 \langle A^{a}_\mu, \big( C^{-1}\big)
                _{\mu  \nu } A^{a}_\nu \rangle +
 \frac12 \langle h, C^{-1} h \rangle \nonumber \\
&  & +\frac12 \langle B^{a}, S^{-1} B^{a} \rangle -
 \langle {\bar c}^{a}, S^{-1} c^{a} \rangle ,
\end{eqnarray}  
where the Fourier transforms of these free propagators,
 (compare (\ref{g6}) with $ \Lambda=0, 
   \Lambda_0 =\infty $) turn out to be 
\begin{eqnarray} \label{w5}
  C(k) = \frac{1}{k^2 +M^2} \quad , \quad S(k) =
 \frac{1}{k^2 + \alpha m^2} \, , \nonumber \\ 
 C_{\mu  \nu}(k) = \frac{1}{k^2+m^2} \Big( \delta _{\mu \nu} - (1-\alpha )
                 \frac{k_\mu k_\nu}{k^2+\alpha m^2} \Big) .
\end{eqnarray}  
Again the notation has been abused omitting the ``hat".
 Furthermore, we shall use $ C(k) $ as a collective symbol
 for these propagators. A Gaussian product measure, the covariances
 of which are a regularized version
of the propagators (\ref{w5}), forms the basis to quantize
 the theory by functional integration. 
Although gauge symmetry is violated by any momentum cutoff
one should try to reduce the bothersome consequences as far
 as possible. Instead of the simple form
(\ref{g9}) we choose the cutoff function
 \begin{equation} \label{w6}
\sigma_{\Lambda}(k^2) \,=\,
 \exp \Big ( -  \frac{(k^2 + m^2)(k^2 + \alpha m^2)(k^2 + M^2) ( k^2)^2 }
{\Lambda^{10}} \,\Big )\ . 
\end{equation}
It is positive, invertible and analytic as the former, 
but satisfies in addition
 \begin{equation} \label{w7}
\frac{d}{d k^2}\sigma_{\Lambda}(k^2) |_{k^2 =0} \, = \, 0.
 \end{equation} 
 This property is the raison d'\^etre for the particular choice (\ref{w6}),
 turning out helpful in the later analysis of the relevant part of the STI.
 Employing this cutoff function we define the regularized propagators,
$ 0 \leq \Lambda \leq \Lambda_0 < \infty $,
 \begin{equation} \label{w8}
 C^{\Lambda,\Lambda_0}(k) \equiv  C(k) \sigma_{\Lambda,\Lambda_0}(k^2)
   : = C(k)\,(\, \sigma _{\Lambda_0}(k^2) - \sigma _{\Lambda}(k^2 )\,).
 \end{equation}
They satisfy the bounds
\begin{equation} \label{w9}
 % {\big|} \partial^w
{\bigg|}\Big( \prod_{i=1}^{|w|}\frac{\partial}{\partial k_{\mu _i}}\Big)
   \partial_{\Lambda}\, C^{\Lambda,\Lambda_0}(k) {\bigg |}
   \le \left\{ \begin{array}{ll} c_{|w|} \,
   \sigma_{2\Lambda}(k^2)
  & for \quad 0 \le \Lambda \le m\ , \\
 \Lambda ^{-3-|w|} P_{|w|}(\frac{|k|}{\Lambda}) \,\sigma _{\Lambda}(k^2) 
 & for \quad \Lambda > m\ . \end{array} \right\}
\end{equation}  
with polynomials $ P_{|w|} $ having positive coefficients. These
coefficients, as well as the constants $ c_{|w|}$, only depend on
$ \alpha, m, M, |w|$.
 Considering $ {\sigma_{\Lambda}}(k^2) $, (\ref{w6}),
 as a function of $( \Lambda , k^2 )$, it cannot be extended continuously 
to $ (0,0) $.
We set $ \sigma_{0}(0):= \lim_{k^2 \to 0} \sigma_{0}(k^2)= 0 $, and hence 
  $\sigma_{0,\Lambda_0}(0)= \sigma_{\Lambda_0}(0)= 1$.\\
Writing 
 \begin{equation} \label{w10}
\langle \Phi , K \rangle :\, =
 \int dx \Big( \sum_{\tau} \fitx \Jtx + \cbx \etx + \etbx \cax \Big) ,
 \end{equation}
the characteristic functional of the Gaussian product
 measure with covariances 
$\hbar C^{\Lambda, \Lambda_0} $, (\ref{w8}), (\ref{w5}), is given by
 \begin{equation} \label{w11}
\int d\mu_{\LLz} (\Phi ) \, e^{\, \frac{1}{\hbar} \langle \Phi , K \rangle} =
 e^{\, \frac{1}{\hbar} P(K)},
 \end{equation}
 \begin{eqnarray} \label{w12}
P(K) & = & \frac12 \langle \jm, C^{\LLz}_{\mu \nu} \jn \rangle +
     \frac12 \langle s, C^{\LLz}\,  s \rangle 
            \nonumber   \\
 & & + \frac12 \langle \ba , S^{\LLz} \,\ba \rangle - \langle \etb, S^{\LLz}
 \,\et \rangle \, .
 \end{eqnarray} 
The free propagators (\ref{w5}) reveal the mass dimensions of the
 corresponding quantum  fields: each
of the fields has a mass dimension equal to one, attributing equal
 values to the ghost and antighost field.  \\
To promote the classical model to a quantum field theory we consider
 the generating functional
$ L^{\LLz} (\Phi ) $ of the amputated truncated Schwinger functions.
 It unfolds according to the
integrated flow equation, cf. (\ref{f11}),
 \begin{equation} \label{w13}
e^{- \frac{1}{\hbar} \big ( L^{\LLz}(\Phi ) + I^{\LLz} \big ) } \, = \,
   e^{ \, \hbar \Delta _{\LLz} } \, e^{ - \frac{1}{\hbar}
      L^{\Lambda_0, \Lambda_0} (\Phi ) }
 \end{equation}
from the bare functional $L^{\Lambda_0, \Lambda_0} (\Phi ) $,
 which forms its initial value at
$ \Lambda = \Lambda_0 $. The functional Laplace operator
 appearing has the  form
 \begin{eqnarray} \label{w14}
\Delta _{\LLz} \, & = & \frac12 \langle
      \frac{\delta }{\delta \Am} , C^{\LLz}_{\mu \nu}
     \frac{\delta }{\delta \An} \rangle   
  + \frac12 \langle \frac{\delta }{\delta h} , C^{\LLz}
       \frac{\delta }{\delta h} \rangle\  \nonumber   \\
  & & + \frac12 \langle \frac{\delta }{\delta \B} ,
         S^{\LLz}\frac{\delta }{\delta \B} \rangle
    + \langle \frac{\delta }{\delta \ca} ,
         S^{\LLz}\frac{\delta }{\delta \cb} \rangle  \, .
 \end{eqnarray} 
Since the local gauge symmetry is violated by the regularization,
 the bare functional
 \begin{equation} \label{w15}
   L^{\LLzz} (\Phi ) = \int dx \, \mathcal{L}_{\rm int} (x) +
          L^{\LLzz}_{c.t.} (\Phi ) 
 \end{equation}
has at first to be chosen sufficiently general in order to allow
 the restoration of the Slavnov-Taylor
identities at the end. Therefore, we add as counterterms to the
 given interaction part (\ref{y11})
of classical origin all local terms of mass dimension $\leq 4 $, which
 are permitted by the unbroken
global symmetries, i.e. Euclidean $O(4)$- invariance and $
 SO(3) $- isosymmetry.
 There are 37 such 
terms, by definition all of order $\mathcal{O}(\hbar) $. The bare functional
 is presented in Appendix A.

The decomposition of the generating functional $ L^{\LLz}(\Phi ) $ is
 written employing a multiindex $n$,
 the components of which  denote the number of each source field
 species appearing:
 \begin{equation} \label{w16}
  n = ( n_A,\, n_h,\, n_B,\, n_{\bar c},\, n_c ) \quad ,
    \quad |n| = n_A + n_h+ n_B + n_{\bar c} + n_c \, .
 \end{equation}   
Moreover, we consider the functional within a formal loop expansion, hence
 \begin{equation} \label{w17}
L^{\LLz}(\Phi ) = \sum_{|n|=3}^{\infty} L^{\LLz}_{l=0, n}(\Phi)
       + \sum_{l=1}^{\infty} \hbar^l
\sum_{|n|=1}^{\infty} L^{\LLz}_{l,n}(\Phi) .
 \end{equation}
Disregarding the vacuum part, we can study the flow of the $n$-point
 functions in the infinite volume
limit $ \Omega \rightarrow \mathbf{R}^4, \Phi \in \mathcal{S}(\mathbf{R}^4) $.
 With our conventions
 (\ref{f12}) of the Fourier transformation, the momentum representation
 of the $n$-point function
with multiindex $n$, (\ref{w16}), at loop order $l$ is obtained as an
 $|n|$-fold functional derivative
 \begin{equation} \label{w18}
(2\pi )^{4(|n|-1)} \delta ^n_{\hat{\Phi }(p)} L^{\LLz}_l |_{\hat{\Phi}=0} \,
  = \, \delta (p_1 + \cdots + p_{|n|}) \Lf_{l,n} (p_1, \cdots , p_{|n|}) .
 \end{equation}
To avoid clumsiness, the notation does not reveal how the momenta are
 assigned to the multiindex $n$,
 and in addition, it suppresses the $ O(4) $- and $SO(3)$- tensor structure.
 From the definition 
(\ref{w18}) of the $n$-point function  follows that it is completely
 symmetric (antisymmetric) upon
permuting the variables belonging to each of the bosonic (fermionic) 
species occurring. Proceeding 
exactly as in the case of the scalar field, the flow equation (\ref{w13})
 is converted into a system of flow equations relating 
the $n$-point functions.
 It looks like (\ref{f20}), where $ n$ is now a multiindex
and the residual symmetrization has to be extended to a
 corresponding antisymmetrization in case of
the (anti)ghost fields.
The system is integrated in the familiar way. At first the tree
 order $l=0$ has to be gained, fully determined by the classical
 descendant (\ref{y11}) appearing in the
 initial condition (\ref{w15}) at $ \Lambda = \Lambda_0 $.
Given the tree order $l=0$, the inductive integration ascends
 in the loop order $l$, for fixed $l$ ascends in
$|n|$ , and for fixed $l,n$ descends in $w$ from $|w|=4$ to $w=0$,
 with initial conditions as follows: \\
$A_1)$ For $ |n|+|w|>4$ at $\Lambda=\Lambda_0$ ,
 \begin{equation} \label{w21} 
\partial ^w \Lb_{l,n}(p_1, \cdots , p_{|n|}) = 0 ,
 \end{equation} 
due to the choice of the bare functional (\ref{w15}). \\
$A_2)$ For the (relevant) cases $ |n|+|w|\leq 4$, renormalization
 conditions at the physical value
$\Lambda=0$ and a chosen renormalization point are freely
 prescribed order-by-order, subject only
to the unbroken $O(4)$- and $SO(3)$-symmetries. These
 conditions determine the 37 local counterterms
entering the bare functional (\ref{w15}). For simplicity we choose, as
 before, vanishing momenta as renormalization point.

Repeating exactly the steps that in the case of the scalar field
 led to the Propositions 2.1 and 2.2,
one establishes in the present case analogous
bounds, just reading now $ n$ as a multiindex. As a consequence
of these bounds a finite theory results in the limit
$\Lambda_0 \rightarrow \infty$, however, not yet the
 gauge theory looked for! The problem still to be
solved is to select renormalization conditions $A_2)$ such that the 
$n$-point functions in the limit
$\Lambda =0, \Lambda_0 \rightarrow \infty $ satisfy the Slavnov-Taylor 
identities. 

As worked out in the next section, to establish the 
Slavnov-Taylor identities necessitates
to consider Schwinger functions with a composite field inserted.
 There will appear two kinds of such insertions: the composite
 BRS-fields forming local insertions, and a space-time integrated insertion
describing the intermediate violation of the Slavnov-Taylor identities. \\
The classical composite BRS-fields (\ref{y12}) all have mass 
dimension 2 and transform
as vector-isovector, scalar-isoscalar, scalar-isovector 
and scalar-isovector, respectively. Moreover, the
first three have ghost number 1, whereas the last one has
 ghost number 2. Thus, adding counterterms
according to the rules formulated in Section 2.4, we 
introduce the bare composite fields 
\begin{equation} \label{w25}
\begin{split}
\psi^a_{\mu}(x) & =  R^0_1 \, \partial_{\mu} \cax + R^0_2 \,
 g \,\epsilon^{arb}A^r_{\mu}(x)c^b(x), \\
\psi(x) & =  - R^0_3 \, \frac{g}{2} B^a(x) c^a(x),  \\
\psi^a(x) & =  R^0_4 \, m \cax+ R^0_5 \,\frac{g}{2} h(x) \cax +
 R^0_6 \,\frac{g}{2}\, \epsilon^{arb} B^r(x) c^b(x),  \\
\Omega^a(x) & =  R^0_7 \,\frac{g}{2}\,\epsilon^{apq}c^p(x)c^q(x),
\end{split}
\end{equation}
keeping the notation introduced for the classical terms and using
 it henceforth exclusively according to (\ref{w25}). We set
 \begin{equation} \label{w26}
  R^0_{i} \, = \, 1+ \mathcal{O}(\hbar) , 
 \end{equation}
thus viewing the counterterms again as formal power series in $\hbar $;
 the tree order ${\hbar}^0 $ 
provides the classical terms (\ref{y12}). The reader notices
 that there is no insertion 
attributed to the \emph{linear} variation of the antighost field. It will be
 seen that the Slavnov-Taylor identities  can be established generating
 this variation  by functional derivation with respect to
the sources of the fields involved. We shall have to deal with
 Schwinger functions with \emph{ one}
insertion. Similarly as in 
Section 2.4, the bare interaction
(\ref{w15}) is modified adding the composite fields (\ref{w25}) coupled
 to corresponding sources, introduced in (\ref{y15}),
 \begin{equation} \label{w27}
\tilde {L}^{\LLzz} : = L^{\LLzz} + L^{\LLzz} (\xi ) ,
 \end{equation}
  \begin{equation} \label{w28}
 L^{\LLzz} (\xi ) = \int dx \{ \gamma^a_{\mu}(x) \psi^a_{\mu}(x) +
         \gamma(x) \psi(x)
 + \gamma^a (x)\psi^a(x)+ \omega^a(x) \Omega^a(x) \} . 
 \end{equation}
Then, from the corresponding
 generating functional of the regularized amputated truncated 
Schwinger functions with \emph{one} insertion $\psi(x) $,
 \begin{equation} \label{w30}
L^{\LLz}_{\gamma }(x \, ; \Phi ) := 
\frac{\delta }{\delta \gamma (x)} {\tilde L}^{\LLz}|_{\xi =0} ,\quad
 {\hat L}^{\LLz}_{\gamma }( q \, ; \Phi ) =
 \int dx \, e^{iqx} L^{\LLz}_{\gamma }(x \, ; \Phi )
  \end{equation}
with analogous expressions for the other insertions,
after a loop expansion, follows for the $n$-point
 functions with one insertion $\psi $,
 \begin{equation} \label{w32}
\delta ( q+p_1 + \cdots + p_{|n|}) \Lf_{\gamma ; l, n}
(q \, ;p_1, \cdots , p_{|n|}) :=
(2\pi)^{4(|n|-1)} \delta ^{n}_{\hat{\Phi }(p)} 
{\hat L}^{\LLz}_{\gamma ; \, l}(q \, ;\Phi )|_{\Phi =0} ,
 \end{equation}
a system of flow equations. From each of these systems
the renormalizability of the amputated truncated Schwinger functions with
 one insertion can be deduced inductively in the familiar way.
 We denote by $\xi $ any
of the labels $\gamma ^{a}_{\mu}\, , \gamma , \gamma ^{a}, \omega ^{a}$.
 First, the tree order $l=0$ is obtained from its initial 
condition at $\Lambda = \Lambda_0 $. For $l \geq 1$ the
 initial conditions are:  \\
$B_1)$ If $|n|+|w|>2 $ at $\Lambda = \Lambda_0$ ,
 \begin{equation} \label{w34}
\partial ^w \Lb_{\xi \, ; l, n}(q \, ;p_1, \cdots , p_{|n|}) = 0 \, .
 \end{equation}
$B_2)$ If $|n|+|w|\leq 2 $ at $\Lambda=0$ and at vanishing
 momenta (the renormalization point) the
initial condition can be fixed freely in each loop order,
 provided the Euclidean symmetry and the isosymmetry are respected.
 In total, there are 7 such renormalization conditions
 which then determine the 7 parameters $R^0_i$ entering
 the bare insertions (\ref{w25}).
 Given the bounds
 of the case without insertion one deduces 
 inductively the analogues of the Propositions 2.3 and 2.4,
with $n$ now a multiindex and $D=2$.
Hence, we have boundedness and convergence of the amputated truncated 
Schwinger functions with the insertion of one BRS-variation. 
 
The intermediate violation of the Slavnov-Taylor identities, as will
 be derived in the following section,
leads to a bare space-time integrated insertion of the form
\begin{eqnarray}
L^{\LLzz}_1(\Phi ) & = & \int dx N(x) ,   \label{w37a}  \\
 N(x) & = & Q(x) + Q '( x ; (\Lambda_0)^{-1}) .  \label{w37b}
\end{eqnarray}  
Here $Q(x)$ is a local polynomial in the fields and their derivatives,
 having  canonical mass dimension $ D=5$,
 whereas $ Q '( x ; (\Lambda_0)^{-1}) $ is 
 nonpolynomial in the field derivatives
 but with powers $(\Lambda_0)^{-1}$ as coefficients such that it becomes 
 irrelevant. The individual
 terms composing $N(x)$ involve at most five fields and
 have ghost number equal to one. We have to
control $L^{\LLz}_1(\Phi )$, the $L$-functional with one
 (bare) insertion (\ref{w37a}). Hence, in analogy
to the local case, cf. (\ref{i4}), a modified bare action
 \begin{equation} \label{w38}
L^{\LLzz} (\Phi ) + \chi L^{\LLzz}_1 (\Phi )
 \end{equation}
is introduced as initial condition in the
 (integrated form of the) flow equation
 \begin{equation} \label{w39}
e^{- \frac{1}{\hbar}(L^{\LLz}_{\chi } + I^{\LLz})}:\, =
 e^{\,\hbar \Delta _{\LLz}}
   \, e^{- \frac{1}{\hbar}(L^{\LLzz} +\chi  L^{\LLzz}_1)} \, .
 \end{equation}
Herefrom results the generating functional of the (regularized)
 amputated truncated Schwinger functions
 with one insertion (\ref{w37a}) as 
 \begin{equation} \label{w40}
L^{\LLz}_1 (\Phi ) \, = \, \frac{\partial }{\partial \chi }
    L^{\LLz}_{\chi} (\Phi ) |_{\chi = 0} \, .
 \end{equation}
It satisfies a linear differential flow equation which is
 easily obtained relating it to the case of a bare
local insertion, cf. (\ref{i4}-\ref{i12}),
   $$ \int dx \, \varrho (x) N(x) $$
and observing
$$ \frac{\partial }{\partial \chi } L^{\LLz}_{\chi} (\Phi ) |_{\chi = 0} =
\int dx \frac{\delta }{\delta \varrho (x)}
 {\tilde L}^{\LLz} (\varrho ; \Phi )|_{\varrho =0} =
  \int dx L^{\LLz}_{(1)} (x; \Phi ) = {\hat L}^{\LLz}_{(1)}(0; \Phi ) $$
Hence, the differential flow equation satisfied by the
 functional $ L^{\LLz}_1(\Phi )$ is the space-time
integrated analogue of (\ref{i10}). Performing a loop expansion,
 the amputated truncated $n$-point functions with one
 insertion (\ref{w37a}),
 $n$ a multiindex (\ref{w16}),
 \begin{equation} \label{w41}
\delta ( p_1 + \cdots + p_{|n|})
 {\mathcal L}^{\LLz}_{1; \, l, n}(p_1, \cdots , p_{|n|}) :=
(2 \pi)^{4(|n|-1)} \delta ^{\, n}_{{\hat \Phi }(p)}
 L^{\LLz}_{1; \, l}(\Phi ) |_{\Phi =0} 
 \end{equation}
then satisfy a system of flow equations similar 
to the case of the local BRS-insertions, 
 letting there the momentum take the value $q=0$. 
  As a consequence, we obtain analogous bounds, but 
 observing in the present case the dimension $D=5$. The irrelevant
 part appearing in the bare insertion (\ref{w37a} - \ref{w37b})
satisfies the required bounds to be admitted, cf. (\ref{p25}).
 %%%%%%%%%%%%%%%%%%%%%%%%%%%%%%%%%%%%%%%%%%%%%%%%%%%%%%%%%%%
 \section{Violated Slavnov-Taylor identities}
The Schwinger functions of the spontaneously broken Yang-Mills theory
should be uniquely determined by its free physical parameters $g,
\lambda , m $ and the gauge fixing parameter $\alpha $, once the
normalization of the fields has been fixed. This uniqueness - as well
as the physical gauge invariance - is accomplished by requiring the
Schwinger functions to satisfy the Slavnov-Taylor identities. These
identities, however, are inevitably violated by the intermediate
regularization in momentum space. Our ultimate goal is to show, that
by a proper choice of the renormalization conditions the
Slavnov-Taylor identities emerge upon removing the regularization. To
this end we first examine the violation of the Slavnov-Taylor
identities produced by the UV-cutoff $\Lambda_0$. Our starting point
is the generating functional of the regularized Schwinger functions,
here considered at the physical value $\Lambda = 0$ of the flow
parameter, $$ Z^{\Lz}(K) \, = \, \int d\mu_{\Lz}(\Phi)
\,e^{-\frac{1}{\hbar} L^{\LLzz} +\frac{1}{\hbar} \langle \Phi , K
\rangle } .$$ The Gaussian measure $ d\mu _{\Lz}(\Phi ) $ corresponds
to the quadratic form $ \frac{1}{\hbar} Q^{\Lz}(\Phi ) $,
cf. (\ref{w4}), (\ref{w8}), to wit:
\begin{eqnarray} \label{v1}
  Q^{\Lz}(\Phi ) & = & \frac12 \langle A^{a}_\mu, \big( C^{\Lz}\big)^{-1}
                _{\mu  \nu } A^{a}_\nu \rangle +
  \frac12 \langle h, (C^{\Lz})^{-1} h \rangle \nonumber \\
&  & +\frac12 \langle B^{a}, (S^{\Lz})^{-1} B^{a} \rangle -
  \langle {\bar c}^{a}, (S^{\Lz})^{-1} c^{a} \rangle .
\end{eqnarray}  
Defining \emph{regularized} BRS-variations (\ref{y13}),(\ref{w25})
 of the fields by
\begin{eqnarray} \label{v2}
\delta _{BRS} \,\fitx & = & - \, ( \si \psi _{\tau})(x) \,\epsilon 
      , \nonumber \\
\delta _{BRS} \,\ca (x) & = & - \, (\si \Omega ^{a}) (x)\, \epsilon ,  \\
\delta _{BRS} \,\cb (x) & = & - \, \big (\si (\, \frac{1}{\alpha }
 \,\partial _{\nu } \An -m \B)\big ) (x) \,\epsilon , \nonumber
 \end{eqnarray} 
 the BRS-variation of the Gaussian measure follows as
\begin{equation} \label{v3}
d\mu _{\Lz}(\Phi ) \mapsto  d\mu _{\Lz}(\Phi ) \Big( 1- \frac{1}{\hbar}\,
 \delta _{BRS}\, Q^{\Lz}(\Phi ) \Big) .
\end{equation}
Written more explicitly,  
 \begin{eqnarray} \label{v4}
\delta _{BRS} \, Q^{\Lz}(\Phi ) & =& \Big( - \sum_{\tau}
  \langle \fit, \big( C^{\Lz}_{\tau}\big)^{-1}
    \si \psi _{\tau} \rangle + \langle \cb ,
  \big( S^{\Lz}\big)^{-1} \si \Omega ^{a} \rangle \nonumber \\
 & & \quad - \,  \langle \, \frac{1}{\alpha } \,\partial _{\nu } \An
      -m \B , \si \big( S^{\Lz} \big)^{-1} 
             \ca \rangle \Big) \, \epsilon \, ,
 \end{eqnarray}  
it reveals that $\si$ just cancels its inverse appearing in the
 inverted propagators, and as a consequence, the BRS-variation of 
the Gaussian measure has mass dimension D=5. The essential reason 
for using  regularized BRS-variations (\ref{v2}) is to assure this property.
 From the requirement, that the 
regularized generating functional $Z^{\Lz}(K)$ be invariant 
under the  BRS-variations (\ref{v2}),
result the \emph{violated Slavnov-Taylor identities}
 \footnote{As long as the vacuum part is involved,
one has to stay in finite volume}
 \begin{equation} \label{v5}
    0 \stackrel{!}{=}
 \int d\mu_{\Lz}(\Phi) \,e^{-\frac{1}{\hbar} L^{\LLzz}
 +\frac{1}{\hbar} \langle \Phi , K \rangle }
 \Bigl( \delta _{BRS}\, \langle \Phi , K \rangle - \delta _{BRS}\,
        ( Q^{\Lz} + L^{\LLzz}) \Bigr) .
 \end{equation}  
This equation can be rewritten, introducing modified
  generating functionals:  \\
i) With the modified bare interaction (\ref{w27}) we define
 \begin{equation} \label{v6}
{\tilde Z}^{\Lz} (K,\xi ) := \int d\mu_{\Lz}(\Phi)
    \,e^{-\frac{1}{\hbar} {\tilde L}^{\LLzz} 
+\frac{1}{\hbar} \langle \Phi , K \rangle } ,
 \end{equation} 
in combination with a regularized version of the BRS-operator (\ref{y16}) ,
 \begin{equation} \label{v7}
{\mathcal D}_{\Lambda_0} = \sum_{\tau} 
\langle \Jt \, , \si \frac{\delta }{\delta {\gamma}_{\tau}} \rangle
 + \langle \etb, \si \frac{\delta }{\delta \omega ^{a}} \rangle  
 +\langle \, \frac{1}{\alpha } \partial _{\nu} \frac{\delta }{\delta \jn}
    - m \frac{\delta}{\delta \ba}\, ,\si \et \rangle  .
 \end{equation} 
ii) In addition, we treat the BRS-variation of the bare action,
 \begin{equation} \label{v8}
L^{\LLzz}_1 \epsilon :\, = \,
      - \delta _{BRS} \Big( Q^{\Lz} + L^{\LLzz} \Bigr) ,
 \end{equation} 
as a space-time integrated insertion with ghost number 1.
 Because of the regularizing factor $ \si$ ,
cf. (\ref{v2}), the integrand is not a polynomial in the fields
 and their derivatives. 
With $ \chi \in \mathbf{R} $, we then define
 \begin{equation} \label{v9}
Z^{\Lz}_{\chi} (K ) := \int d\mu_{\Lz}(\Phi) \,e^{-\frac{1}{\hbar}
        ( L^{\LLzz}+\, \chi L^{\LLzz}_1 ) 
+\frac{1}{\hbar} \langle \Phi , K \rangle } .
  \end{equation} 
Due to these definitions, the violated Slavnov-Taylor identities
 (\ref{v5}) can be written in the form
 \begin{equation} \label{v10}
\mathcal{D}_{\Lambda_0} \, \tilde{Z}^{\Lz} (K, \xi ) |_{\xi =0} \, = \, 
    \frac{d}{d \chi } Z^{\Lz}_{\chi } ( K) |_{\chi  =0} .
 \end{equation} 
From the modified functionals (\ref{v6}) and (\ref{v9}) follow, cf.
 (\ref{i8}), the generating functionals
of the corresponding amputated truncated Schwinger functions
\begin{eqnarray} 
{\tilde Z}^{\Lz} (K,\xi ) & = & e^{\frac{1}{\hbar} P(K) } \,
 e^{-\frac{1}{\hbar} ( {\tilde L}^{\Lz}
( \fit , \, c, \,\bar c \,;\, \xi ) + I^{\Lz}) } , \label{v11}  \\
 Z^{\Lz}_{\chi} (K ) & = & e^{\frac{1}{\hbar} P(K) }\,
 e^{-\frac{1}{\hbar} ( L^{\Lz}_{\chi }
    ( \fit ,\, c,\, \bar c ) + I^{\Lz} )} , \label{v12}
\end{eqnarray}
where the variables of the $Z$- and the $L$-functional are related as
\begin{eqnarray}  \label{v13}
 \fitx & = & \int dy \, C^{\Lz}_{\tau} (x-y)\, \Jt (y) ,  \nonumber \\
\cax & = & - \int dy \,S^{\Lz} (x-y)\, \et (y) ,   \\
\cbx & = & - \int dy\, S^{\Lz} (x-y)\, \etb (y) . \nonumber
 \end{eqnarray} 
Furthermore, $P(K)$, (\ref{w12}), has to be taken here at $\Lambda =0$.
 We observe, that the vacuum
part $I^{\Lz}$ present without insertions appears, since both 
 insertions have positive ghost number.
To have a less cumbersome notation in the rest
 of this section, we abbreviate
 \begin{eqnarray}  \label{v14}
L & \equiv  & L^{\Lz} \,\, \Bigl(\, = \tilde{L}^{\Lz} |_{\xi = 0} \,
 = \, L^{\Lz}_{\chi } |_{\chi =0} \, \Bigr) \, ,
  \qquad L^0 \equiv L^{\LLzz} \, ,\nonumber  \\
L_1 & \equiv  & L^{\Lz}_1 \    =
 \frac{d}{d \chi} \, L^{\Lz}_{\chi} |_{\chi = 0} \, , \qquad  L^0_1
                    \equiv L^{\LLzz}_1 ,  \\
L_{\gamma } & \equiv  & L^{\Lz}_{\gamma } (x \, ; \Phi ) \, ,
 \qquad L^0_{\gamma } \equiv 
                  L^{\LLzz}_{\gamma }(x \, ; \Phi ) \, \,
 \Bigl( = \, \frac{\delta }{\delta \gamma (x)}
                     L^{\LLzz}(\xi ) |_{\xi =0} \Bigr) , \nonumber 
 \end{eqnarray}   
see (\ref{w27}-\ref{w30}). Moreover, we denote the inverted
 unregularized propagators by
\begin{equation} \label{v15}
D_{\tau} \equiv \Bigl( (- \Delta +m^2 ) \delta_{\mu , \nu} -
 \frac{1-\alpha }{\alpha } \partial _{\mu}
  \partial _{\nu}\, ,\,\, - \Delta + M^2 ,\,\, -\Delta +
 \alpha  m^2 \equiv D \Bigr) .
\end{equation} 
From (\ref{v10}) we derive via (\ref{v11} - \ref{v13}), employing the
previous abbreviations, the \emph{ violated Slavnov-Taylor identities
of the amputated truncated Schwinger functions}:
 \begin{eqnarray}\label{v16}
 \langle \ca , D\, (\, \frac{1}{\alpha }\, \partial _{\nu}
\An - m\B) \rangle - \langle \ca ,\si (\, \partial _{\nu} \frac{\delta
L}{\delta \An} - m\frac{\delta L} {\delta \B} ) \rangle \nonumber \\ +
\sum_{\tau} \langle \fit \, , D_{\tau} L_{\gamma _{\tau}} \rangle -
\langle \cb , D L_{\om} \rangle \, = \, L_1 .\qquad \qquad
\end{eqnarray}
 As will turn out, we also need the explicit form of $
L^0_1$, (\ref{v8}), i.e. the BRS-variation of the bare action. From
its definition (\ref{v8}) follows directly, using (\ref{w27} -
\ref{w28}),
 \begin{eqnarray} \label{v17}
 L^0_1 = \langle \ca , D\, (\,
\frac{1}{\alpha }\, \partial _{\nu} \An - m\B) \rangle - \langle
\frac{\delta L^0}{\delta \cb}, \si (\, \frac{1}{\alpha }\, \partial
_{\nu} \An - m\B) \rangle + \qquad \, \\ \sum_{\tau} \langle \fit \, ,
D_{\tau} L^0_{\gamma _{\tau}} \rangle - \langle \cb , D L^0_{\om}
\rangle + \sum_{\tau} \langle \frac{\delta L^0}{\delta \fit} , \si
L^0_{\gamma _{\tau}} \rangle - \langle \frac{\delta L^0}{\delta \ca} ,
\si L^0_{\om} \rangle .\nonumber
 \end{eqnarray}
Moreover, to restore the Slavnov-Taylor identities  we shall rely 
on proper vertex functions, too.
Therefore, the violated form in terms of these
 functions is derived here, too. In the following, all
functionals appearing should carry the superscript $ \Lz$ which 
is omitted, cf. (\ref{v14}).
Considering the generating functional of the truncated Schwinger functions
\begin{equation} \label{v19}
e^{\frac{1}{\hbar} \tilde W (K,\xi )} \, =
 \, \frac{\tilde Z (K, \xi )}{\tilde Z (0, 0)} ,
\end{equation}
it follows from (\ref{v10}), together with (\ref{v11} - \ref{v12}) 
and using notation defined in (\ref{v14}), that
 \begin{equation} \label{v20}
\mathcal{D}_{\Lambda_0} \, \tilde W(K,\xi ) |_{\xi =0} \, = \,
         - L_1 (\fit \, , \ca, \cb ) ,
 \end{equation}
with arguments according to (\ref{v13}). Because of the inherent
 symmetries, the functional $L$, and
hence also $W$, contain only one $1$-point function, which we
 force to vanish by the renormalization
condition
 \begin{equation} \label{v21}
\frac{\delta L}{\delta h(x)} {\bigg |}_{\Phi =0} \, \stackrel{!}{=} \, 0 \,
 , \quad \rightarrow 
 \frac{\delta {\tilde L}}{\delta h(x)} {\bigg|}_{\Phi =0} = 0 \, .
 \end{equation}
A Legendre transformation yields the (modified) generating
 functional of the proper vertex functions,
 \begin{equation} \label{v22}
\tilde \Gamma ( \ufit, \uca, \ucb; \xi ) + \tilde W (\Jt, \et, \etb; \xi )
 = \int dx \Bigl( \sum_{\tau}\ufit \Jt 
  + \etb \uca + \ucb \et \Bigr) ,
 \end{equation}  
with variables related by
\begin{eqnarray} \label{v23}
\ufit (x)  =  \frac{\delta \tilde W}{ \delta \Jtx} \, , \qquad \qquad
 \,\Jtx & = & \frac{\delta \tilde \Gamma }{\delta \ufit (x)} \, , 
            \nonumber \   \\
\uca (x)  =  \frac{\delta \tilde W}{ \delta \etbx} \, , \,
         \qquad \qquad \, \etbx & = &
                   -\frac{\delta \tilde \Gamma }{\delta \uca (x)} \, ,  \\
\ucb (x)  =  - \frac{\delta \tilde W}{ \delta \etx} \, ,\, \,
 \qquad \quad \, \etx & = &
                    \frac{\delta \tilde \Gamma }{\delta \ucb (x)} \, .
  \nonumber 
 \end{eqnarray}
Since $ \tilde W $ does not contain $1$-point functions,
 because of (\ref{v21}), but begins with
$2$-point functions, the equations on the left in (\ref{v23}) imply,
 that the variables $ \ufit , \uca ,
 \ucb $ of $ \tilde \Gamma $ vanish, if the variables
 $ \Jt , \et , \etb $ are 
equal to zero. Inverting these equations provides $ \Jt , \et , \etb $
 as respective functions of  $ \ufit , \uca , \ucb $, to be used
in the definition (\ref{v22}) of $ \tilde \Gamma  $. It follows,
 that there is no $1$-point proper
vertex function, i.e.
\begin{equation} \label{v24}
\frac{\delta \Gamma }{\delta \uh (x)} {\bigg |}_{\,\uFi \, = \, 0} \,
             = \, 0 .
\end{equation}
From the functional derivation of (\ref{v22}) with respect to the
 source $\gamma (x) $ at fixed $\uFi $ ,
$$ \frac{\delta {\tilde \Gamma}}{\delta \gamma (x)}{\bigg |}_{\uFi} 
 + \frac{\delta {\tilde W}}{\delta \gamma (x)} {\bigg |}_K \qquad \qquad
     \qquad \qquad \qquad \qquad \qquad \qquad \qquad $$
 $$ + \int dy \left ( \sum_{\tau} 
      \frac{\delta {\tilde W}}{\delta J_{\tau}(y)} 
         \frac{\delta J_{\tau}(y)}{\delta \gamma (x)}  
+  \frac{\delta {\bar \eta}^{a}(y)} {\delta \gamma (x)} 
 \frac{\delta {\tilde W}}{\delta {\bar \eta}^{a}(y)}   
+   \frac{\delta { \eta}^{a}(y)} {\delta \gamma (x)} 
 \frac{\delta {\tilde W}}{\delta{ \eta}^{a}(y)}  \right )  $$
$$ = \int dy \left( \sum_{\tau} \ufit(y)\frac{\delta J_{\tau}(y)}
      {\delta \gamma (x)} 
 + \frac{\delta {\bar \eta}^{a}(y)} {\delta \gamma (x)}\, \uca (y)
 - \frac{\delta { \eta}^{a}(y)}{\delta \gamma (x)} \,\ucb(y) \right),\quad $$  
 we infer, because of (\ref{v23}),
 \begin{equation} \label{v25}
   \frac{\delta \tilde \Gamma }{\delta \gamma (x)}{\bigg |}_{\uFi}
  =\, -\, \frac{\delta \tilde W}{\delta \gamma (x)}{\bigg |}_K ,
 \end{equation}
and  similar relations for the derivatives with respect to the sources
 $\gm , \g $ and $ \om $. These relations are employed at $\xi = 0 $.
 Using a notation in accord with (\ref{v14})\,, 
  \begin{equation} \label{v26}
\Gamma \equiv \tilde \Gamma ^{\Lz} |_{ \, \xi \, = \,0} \, ,
  \qquad \Gamma _{\gt} (x) \equiv \frac{\delta {\tilde \Gamma}^{\Lz}
  }{\delta \gt (x)} {\bigg|}_{\, \xi\, =\, 0} \, , \end{equation} the
  \emph{ violated Slavnov-Taylor identities for proper vertex
  functions} emerge from (\ref{v20}) via (\ref{v23}),(\ref{v25}) as
\begin{eqnarray} \label{v27}
\sum_{\tau} \langle \frac{\delta \Gamma }{\delta \ufit} \, ,\si
 \Gamma _{\gt} \rangle 
           - \langle \frac{\delta \Gamma }{\delta \uca} ,\si
 \Gamma _{\om} \rangle  
  - \langle \frac{1}{\alpha } \partial _{\nu} \uAn - m\uB ,
 \si \frac{\delta \Gamma }{\delta \ucb}\rangle
    \nonumber  \\ = \, \Gamma _1 ( \ufit \, , \uca, \ucb ) \, ,
   \qquad  \qquad  \qquad \qquad \qquad 
\end{eqnarray} 
 with
\begin{equation} \label{v28}
\Gamma _1 (\ufit , \uca, \ucb ) \, = \, L_1 ( \fit , \ca, \cb ) \, .
 \end{equation}
In (\ref{v28}) the variables are related, suppressing the
 supersript $ \Lz$ of the propagators, as
 \begin{eqnarray} \label{v29}
\fit (x) \, = \int dy\, C_{\tau} (x-y) \frac{\delta \Gamma }
  {\delta \ufit (y)} \, , \qquad \qquad \qquad \, \\
\cax \, = - \int dy S (x-y) \frac{\delta \Gamma }{\delta \ucb (y)} \, ,\quad
 \cbx \, = \int dy \frac{\delta \Gamma }{\delta \uca (y)} S(y-x) \, .
 \nonumber \end{eqnarray} 
Comparing (\ref{v17}) with (\ref{v27}) we
 observe, that $L^0_1$ and $\Gamma _1$ have the same form! The
 apparently additional terms in $L^0_1$ result from the quadratic part
 of the classical action, which by definition is excluded from the
 bare interaction $L^0$, but is contained in $\Gamma $.
%%%%%%%%%%%%%%%%%%%%%%%%%%%%%%%%%%%%%%%%%%%%%%%%%%%%%%%%%%%
\section{Mass scaling of super-renormalizable\\ couplings}
The systems of amputated truncated Schwinger functions and of proper
vertex functions are equivalent formulations of the theory. To restore
the Slavnov-Taylor identities, however, still to be accomplished, 
analysing their relevant part in terms of the proper vertex functions
turns out to be simpler. 
A  necessary condition to achieve the Slavnov-Taylor identities
then is a vanishing  relevant part of the violating functional
$\, \Gamma^{\,0,\,\Lambda_0}_1$ ,(\ref{v27}). 
The means available to generate this property is the 
freedom in choosing the relevant terms appearing in the
functionals $\, \Gamma^{\,0,\,\Lambda_0}_{l,n}\,$
and $\, \Gamma^{\,0,\,\Lambda_0}_{\ga ;\, l,n}\,,
\ga = \ga_{\tau}, \om $, i.e. the renormalization conditions
of the functionals $ \Ga^{\Lz}$ and $\Ga^{\Lz}_{\ga}$.
Inspecting (\ref{v27}), however, reveals an obstacle in
bringing this freedom of choice to bear. In order to exhaust
all relevant terms of (\ref{v27}), up to 5 field- and momentum
derivatives have to be applied, according to the dimension 5
of the insertion defining $\, \Gamma^{\,0,\,\Lambda_0}_1$.
Due to the property (\ref{w7}) we observe that the
 momentum derivatives of the cutoff function
$ \,\si( k^2)  = {\sigma}_{\Lao}(k^2)  \,$ 
do not contribute to these terms. Thus, the field- and momentum
derivatives in question generate from (\ref{v27}) terms
with $ d_1 $ such derivatives applied to the factors of
the form  $\de \Ga/ \de \vp\,$  and $ d_2 $ such
derivatives applied to factors of the form 
 $\,\Ga_{\ga}\,$,  $\pa A^a\,\,$, or $m B^a\,$, 
with $\,d_1+d_2\,\le 5$. Applying $\,d_2 \ge 3\,$ derivatives 
on the functionals 
$\, \Gamma^{\,0,\,\Lambda_0}_{\ga ;\,l,n}\, $, however,
generates irrelevant contributions, since the respective
insertions of these functionals are of dimension 2. It appears,
that the elimination of these unknown irrelevant terms can only
be accomplished making  the relevant terms from
 $\, \Gamma^{\,0,\,\Lambda_0}_{l,n}\,$, which form products
with them, disappear. One notices, however, that already
in the tree order  $\, \Gamma^{\,0,\,\Lambda_0}_{0,n}\,$ 
there are terms preventing this procedure, to wit,
the nonvanishing super-renormalizable three-point couplings
and the mass terms of the two-point functions (see Appendix A).

To tackle this problem requires to trace in the perturbative
expansion the effects of the super-renormalizable three-point
couplings appearing. To this end, in the tree-level part 
 (\ref{y11}) of the interaction (\ref{w15}) the mass parameters 
entering the three-point couplings, as well as in the
BRS-insertions (\ref{w25}) the mass parameter appearing there
in $ \psi^a(x)$, are scaled by a common factor  $\la > 0\,$:  
\eq
m \to \la m\ ,\quad
M \to \la M\  .
\label{scale}
\eqe
{\it It is important to notice that the mass parameters present
in the regularized propagators appearing in the flow equations
are not scaled.}
All amputated truncated Schwinger functions, considered first,
 will  then depend smoothly on  $\,\la\,$, and are expanded as
\eq
{\mathcal L}^{\La,\Lao}_{l,n}(\la;\vec{p}) \,=\,
\sum_{\nu=0}^{\infty} (m\, \la)^{\nu}\
{\mathcal L}^{(\nu),\La,\Lao}_{l,n}(\vec{p}) \ ,\quad
 \vec{p}= (p_1,\cdots,p_{|n|})\, ,
\label{nl1}
\eqe
\eq
{\mathcal L}^{\La,\Lao}_{\ga ;\, l,n}(\la;q ;\vec{p}) \,=\,
\sum_{\nu=0}^{\infty} (m\,\la)^{\nu}\
{\mathcal L}^{(\nu),\La,\Lao}_{\ga;\,l,n}(q ;\vec{p}) \ ,
\label{nl2}
\eqe
with finite sums in suitable renormalization schemes.\\
In the \emph{Renormalization scheme} adopted here
 \emph{relevant terms} satisfy \\
i) $|n|+|w|+\nu \le 4\,$ in case of the functional $\,L^{\LLz}\,$,\\
ii)  $|n|+|w|+\nu \le 2\,$ in case of the functionals
 $\,L^{\LLz}_{\ga}\,, \ga = \ga_{\tau}, \om $,\\
in accord with the bounds stated below.\\ 
 At tree level holds ( there are no functions with $ |n| \leq 2 $ at $ l=0 $)
\eq
(\partial^w {\mathcal L}^{(\nu),\La,\Lao}_{0,n})(\vec{0})\,=\,0\,,
\quad \mbox{if }\quad |n|+|w|+\nu\, < \, 4\ .
\label{nl3}
\eqe
For $ l \geq 1$  renormalization
 conditions on the relevant terms are imposed as follows:
\eq
(\partial^w {\mathcal L}^{(\nu),0,\Lao}_{l,n})(\vec 0) 
\stackrel{!}{=} \,0\,,
\quad \mbox{if }\quad |n|+|w|+\nu \, < \, 4  \ ,
\label{nl4}
\eqe
whereas if $\, |n|+|w|+\nu \, = \, 4 \,$, on the r.h.s. a free
 constant $ r_{(\nu),\, l,\, n}$  can be  chosen.\\
Correspondingly, in the case of a BRS-insertion, at the tree level holds
\eq \label{nl5}
 (\partial^w {\cal L}^{(\nu), \La,\Lao}_{\ga ;\,0, n})(0; \vec{0}\,) = 0\, ,
\quad \mbox{if}\quad |n| + |w|+ \nu < 2 \,,
\eqe
and renormalization conditions are imposed as
\eq \label{nl6}
 (\partial^w {\cal L}^{(\nu), 0 ,\Lao}_{\ga ;\,l,n})(0 ; \vec{0}\,)
   \stackrel{!}{=} 0 \, ,
 \quad \mbox{if}\quad  |n| + |w|+ \nu < 2 \,,
\eqe 
but if $\, |n|+|w|+\nu \, = \, 2 \,$, on the r.h.s. again
 a free constant can be  chosen.\\   
According to the expansions (\ref{nl1}) and (\ref{nl2}) the flow equations
of the form (\ref{f20}) and (\ref{i15}) have to be adjusted attributing a
 superscript $ (\nu) $ to the n-point functions
 $\, {\cal L}^{\LLz}_{l,n}\,$ and  $\, {\cal L}^{\LLz}_{\ga; l,n}\,$ 
 and in the quadratic term
 to sum $ \nu_1 + \nu_2 = \nu $,
 in complete analogy to the loop index $l$. Employing these extended
 flow equations the following bounds can be deduced,\\
\textbf{ Proposition 1}, II \\
\emph{Let $ l \in \mathbf{ N}_0 $ and $\, 0 \leq \La \leq \Lao $, then}
\eq
|\,\partial^w {\cal L}^{(\nu) , \La,\Lao}_{l,n}(\vec{p}\,)|
\,\leq\,
 (\La+m)^{4-|n|-|w|- \nu}\,{\cal P}_1(\log{\La+m \over m})\,
{\cal P}_2(\frac{|\vec{p}|}{\La+m})\ ,
\label{b1}
\eqe
\eq    
|\,\partial^w {\cal L}^{(\nu), \La,\Lao}_{\ga ;\,l,n}(q ; \vec{p}\,)| 
\,\leq\,
(\La+m)^{2-|n|-|w|- \nu}\,{\cal P}_1(\log{\La+m \over m})\,
{\cal P}_2(\frac{|q,\vec{p}|}{\La+m})\ .
\label{b2}
\eqe
\emph{ In these bounds ${\cal P}_i\,,\,i=1,2, $ denote (each time they
 appear possibly new) polynomials with nonnegative coefficients
 independent of $\La,\Lao,\, \vec p,\, q,\, m\,$. The
 coefficients may depend on $n,\,l,\, w,\,
$ and the other free parameters of the theory $\al,\,M/m\, ,\, g $.}\\
 These bounds are uniform in $\, \Lao $.
 The proof in II follows closely the line of proof presented 
 before in the case of a scalar field.
 
The system of flow equations satisfied by the proper
 vertex functions $\, \Ga^{\LLz}_{l, n}\,$ is a direct extension
of (\ref{s22}) to various types of fields now present, as before
in the case of the Schwinger functions. The functions 
$\,{\tilde \Ga}^{\LLz}_{l, n}\,$ appearing on the r.h.s. of
the flow equation (\ref{s22}), denoted in II, eq.(84), by
 $\, L^{\LLz}_{l, n}\,$, have to be determined recursively
according to their definition (\ref{s18}) from the crucial
 relation (\ref{s12}) extended to various types of fields,
providing finally the functions $\,{\tilde \Ga}^{\LLz}_{l, n}\,$
in the form (\ref{s23}). In addition, as in the case of Schwinger 
 functions, flow equations for proper vertex functions with a 
BRS-insertion,$\,\Ga^{\LLz}_{\ga; l, n}\,$, have to be 
considered, too, see II, eq.(86). The r.h.s. of these flow
equations has again to be determined recursively as
described above regarding proper vertex functions without 
insertion. Having obtained this way the functions entering 
the r.h.s. of the flow equations in the form (\ref{s23})
the mass scaling (\ref{scale}) in the tree-level
interaction and the insertion can be performed, leading
to the expansions    
\eq
{\Ga}^{\La,\Lao}_{l,n}(\la;\vec{p}\,) \,=\,
\sum_{\nu=0}^{\infty}(m\la) ^{\nu}\ 
{\Ga}^{(\nu),\La,\Lao}_{l,n}(\vec{p}\,) \ ,\quad
 \vec{p}= (p_1,\cdots,p_{|n|})\, ,
\label{nv1}
\eqe
\eq
{\Ga}^{\La,\Lao}_{\ga ;\, l,n}(\la;q ; \vec{p}\,) \,=\,
\sum_{\nu=0}^{\infty}(m\la) ^{\nu}\
{\Ga}^{(\nu),\La,\Lao}_{\ga;\,l,n}(q ; \vec{p}\,) \ .
% \vec{p}= (p_1,\cdots,p_{|n|})\, ,
\label{nv2}
\eqe
Considering the tree level $ l=0$ first, the scaling (\ref{scale}) 
leads to \\
i) in case of (\ref{nv1}),
\eq \label{nv3}
(\pa^{w}\,{\Ga}^{(\nu),0,\Lao}_{0, n})(\vec{0}\,) = 0 \,,
\quad |n| = 3,\quad |w| + \nu\, {\not =}\, 1\ ,     
\eqe
ii) the masses of the two-point functions,
 fixed by the regularized propagators, are not scaled,
 and there is no $ |n| = 1 $ content,\\  
iii) in case of (\ref{nv2}),     
 \eq \label{nv4}
(\pa^{ w}\,{\Ga}^{(\nu),0,\Lao}_{\ga ;\,0, n})(0 ; \vec{0}\,) = 0 \,,
\quad |n|+ |w| + \nu <  2\,.     
\eqe
Implementing the expansions (\ref{nv1}) and (\ref{nv2}) in the
flow equations of $ \Ga^{\LLz}_{l, n}$ and  $ \Ga^{\LLz}_{\ga; l, n}$,
respectively, a superscript $ (\nu) $ with corresponding values
has to be attached to the various $n$-point functions involved there.
Employing these flow equations and proceeding inductively
as in the case of the Schwinger functions,
renormalizability of the proper vertex functions may 
be deduced. The relevant terms are subject to the renormalization
conditions as follows, where $ l\geq 1\,$:
\eq \label{nv5}
(\pa^{\, w}\,{\Ga}^{(\nu),0,\Lao}_{l ,\, n})( \vec{0}\,)   
 \stackrel{!}{=} 0 \,,
 \quad \mbox {if} \quad |n|+ |w| + \nu < 4\,,     
\eqe
whereas if $ |n|+ |w| + \nu = 4\,$, on the r.h.s. a nonvanishing 
 constant can be chosen.\\
 In  the case of a BRS-insertion
 \eq \label{nv6}
(\pa^{\, w}\,{\Ga}^{(\nu),0,\Lao}_{\ga ;\, l ,\, n})(0 ;\, \vec{0}\,)
 \stackrel{!}{=} 0 \,,
 \quad \mbox {if} \quad |n|+ |w| + \nu < 2\,,     
\eqe
whereas if $ |n|+ |w| + \nu = 2\,$, again a nonvanishing 
 constant on the r.h.s. may be imposed. \\
Proceeding inductively as indicated provides the bounds:\\
\textbf{ Proposition 2}, II
\begin{eqnarray}\label{bv1}
|\,\partial^w\, \Ga^{(\nu) , \La,\Lao}_{l,n}(\vec{p})|
\,\leq\,
 (\La+m)^{4-|n|-|w|-\nu }\,{\cal P}_1(\log{\La+m \over m})\,
     {\cal P}_2(\frac{|\vec{p}|}{\La+m})\ ,\\
 (l, |n|)\, {\not =}\, (0, 2)\qquad \qquad \qquad \qquad \nonumber
\end{eqnarray}
\eq
|\,\partial^w \Ga^{(\nu), \La,\Lao}_{\ga ;\,l,n}(q ; \vec{p})| 
\,\leq\,
(\La+m)^{2-|n|-|w|- \nu}\,{\cal P}_1(\log{\La+m \over m})\,
{\cal P}_2(\frac{|q,\vec{p}|}{\La+m})\ ,
\label{bv2}
\eqe
\emph{The notations are those from (\ref{b1}),(\ref{b2}).}\\

\noindent
Using the expansions (\ref{nv1}) and (\ref{nv2}) of the 
$n$-point functions of $ \Ga^{\LLz}_{l, n}$ and 
 $ \Ga^{\LLz}_{\ga; l, n}$, respectively, which record
 the number of inherent super-renormalizable vertices,
allows now to proceed as envisaged at the beginning of this
section in making vanish the relevant part of $ \Ga^{\Lz}_1 $.
The bounds (\ref{bv1}) and (\ref{bv2}) show that the degree
 of divergence decreases with this number.\\
The expansions of the $n$-point functions of the functionals
$ \Ga^{\Lz}_1 $ and $ L^{\Lz}_1 $ resulting from the mass
 scaling (\ref{scale}),
 \eq
{\mathcal L}^{\La,\Lao}_{1;\,l, n}(\la;\vec{p}\,) \,=\,
\sum_{\nu=0}^{\infty}(m \la)^{\nu}\
{\mathcal L}^{(\nu),\La,\Lao}_{1;\,l, n}(\vec{p}\,) \ ,\quad
 \vec{p}= (p_1,\cdots,p_{|n|})\, ,
\label{ma1}
\eqe
 \eq
{\Ga}^{\La,\Lao}_{1;\,l, n}(\la;\vec{p}\,) \,=\,
\sum_{\nu=0}^{\infty}(m \la)^{\nu}\
{\Ga}^{(\nu),\La,\Lao}_{1;\,l, n}(\vec{p}\,) \ ,
\label{ma2}
\eqe
form the starting point of the demonstration.
Moreover, in a consistent mass expansion of the violated
Slavnov-Taylor identities (\ref{v27}), conform with the
treatment of the BRS-insertions, the mass scaling (\ref{scale})
has to be performed in the BRS-variation
$\,\frac{1}{\alpha} (\partial_{\nu}A^a_{\nu}(x) - \alpha m B^a(x)) \,$
 of the antighost, too. The aim envisaged is to determine the
 relevant part of the functional $\, \Ga^{\Lz}_1 $, given by
 the values 
 $\,( \pa^{w} \Ga^{(\nu),\, 0, \,\Lao}_{1;\,l, n}) (\vec 0) \, ,
 \,\, |n|+|w|+\nu \leq 5 \,$, via (\ref{v27}).
It is important  to observe that  irrelevant  contributions only emerge
from the functionals containing a BRS-insertion.
Upon requiring the vertex functions entering (\ref{v27}) to satisfy 
 the boundary conditions, $\, l \in {\mathbf N}_0 $,
 \eq \label{ma3}
 ( \pa^{w} \Ga^{(\nu),\, 0, \,\Lao}_{ l,\, n}) (\vec 0) \,
  \stackrel{!}{=}\, 0 \,, \quad
  \mbox{if} \quad |n|+|w|+\nu <  4 \,,
 \eqe
then  annihilates the irrelevant contributions generated 
 from the functionals\, $ {\Ga }^{(\nu), \Lz}_{\ga_{\tau}} ,
 {\Ga }^{(\nu), \Lz}_{\om}\, $ by multiplication 
 and only contributions of these functionals with 
 $ |n_2|+|w_2|+ \nu_2 \leq 2\,$ field-, momentum- and 
mass derivatives, i.e. relevant terms, do  appear. 
The boundary conditions (\ref{ma3}) are satisfied for
 $ \, l\geq 1 \,$ due to the renormalization
 conditions (\ref{nv5}), and in the tree order, 
if $ |n|=3\,$, (\ref{nv3}). \\
At this stage one has to remember that the mass scaling
is only performed with regard to the boundary terms
appearing in the the flow equations, but \emph{not}
touching the free propagators. In (\ref{v27}), however, 
the inverted free propagators  
$\Ga ^{0,\Lao}_{0,\,n},\, |n| = 2 ,\,$ appear 
as boundary terms at $\La=0\,$ for the functions 
$\Ga ^{\La,\Lao}_{1;l,n}\,$, and their masses then are
scaled (\ref{scale}), satisfying (\ref{ma3}) in this case, too.
One should notice, that in the approach followed
the flow equations and the violated Slavnov-Taylor identities
have been derived \emph{before} the mass expansion,
and only afterwards the mass expansion is applied
to all boundary terms, and the bounds on the 
vertex functions have been verified inductively.

Because of the renormalization conditions (\ref{nv5}) imposed
 on (a subset of) the relevant terms of the vertex functions,
in all two-point functions the leading terms of the mass
expansion are fixed to zero, i.e. with the notation of Appendix A,
\eq
\de m^2_{(\nu)} = 0\ ,\quad
 \Sigma^{\bar{c}c(\nu)}(0)=0\ ,\quad \Sigma^{BB(\nu)}(0)=0\ ,\quad 
 \Sigma^{hh(\nu)}(0)=0\  \mbox{ for }\ \nu \le 1\ ,
\label{2pt}
\eqe
and also
\eq
\Sigma^{AB(\nu)}(0)=0  \    \mbox{ for }\  \nu = 0~;\quad
\ka^{(\nu)} =0\    \mbox{ for }\  \nu \le 2\ .
\label{21pt}
\eqe
The respective relevant parts of vertex functions with a
BRS-insertion are listed in Appendix B. The corresponding
restricted set of renormalization conditions (\ref{nv6}) is 
automatically fulfilled.

 The functionals  $\, L^{\LLz}_1, \,\Ga^{\LLz}_1 \,$ serve to
 control the violation of the Slavnov-Taylor identities.
In contrast to the functionals $\, \Ga^{\LLz}\,$ and 
$\,\Ga^{\LLz}_{\ga}\,$, these functionals contain
 irrelevant boundary terms at $\La =\Lao\,$, resulting from
 the presence of the factors $\,\si $.
\footnote{Due to the property (\ref{w7}) these factors do not
 affect the relevant part.} 
 The bare functional $ \Ga^{\LLzz}_1 $ satisfies
 the bound, $ l \in {\mathbf N}_0 $,
 \eq \label{ma5}
 |\,\partial^w \Ga^{(\nu),\, \LLzz}_{1;\,l, n}( \vec p \,)| \,\leq
  \,(\Lao +m)^{\,5 -|n|-|w|- \nu} \,
   \Big ( \log{\Lao\over m}\Big)^{r} \,
    {\cal P}(\frac{|\vec p \,|}{\Lao})\ ,
\eqe
which holds trivially, unless $ 2 \leq |n| \leq 5 $.
Using the relation
 \eq \label{clv3}
   \Ga^{\Lao,\Lao}_{1;\,l,n}(p_1,\cdots,p_{|n|}) =
     {\mathcal L}^{\Lao,\Lao}_{1;\,l,n}(p_1,\cdots,p_{|n|}) \, ,
 \eqe 
this assertion can be established from (\ref{v17}) employing there
the previous bounds on $ \pa^{\, w} {\mathcal L}^{(\nu), \,\LLz}_{l, n} $,
 (\ref{b1}), and on
 $ \pa^{\, w} {\mathcal L}^{(\nu),\, \LLz}_{1;\,l, n} \,$,
 (\ref{b2}) , at the value $ \La = \Lao $, together
with  a bound on $ \pa^{\, w} \si (k^2)\,$ easily 
obtained from (\ref{w6}).
The bound on the functional  $ \Ga^{\LLz}_1 $ as stated
 below in (\ref{bdf}) 
 does not follow from the choice of standard renormalization
 conditions for insertions.
 Rather  it is \emph{assumed} here that the relevant part
 of this functional at the physical
 value $ \La = 0 $ of the flow parameter does vanish, 
 $\, l \in {\mathbf N}_0 \,$ ,
  \eq \label{ma6}
 ( \pa^{\, w} \Ga^{(\nu),\, \Lz}_{1; \,l, n})( \vec{0}\, )  = 0\, ,
      \qquad |n | + | w | + \nu \leq 5 \, .
 \eqe  
In Section  4.6 it will be shown that this assumption can
be accomplished via the violated Slavnov-Taylor identities
(\ref{v27}) choosing for the functionals
 entering the l.h.s. suitable renormalization conditions within
 the class (\ref{nv5}), (\ref{nv6}) considered.\\
Taking this assumption for granted the irrelevant part
of the functional $\, \Ga^{\Lz}_1 \,$ then vanishes upon
 shifting the UV- cutoff to infinity, owing to the \\
 \textbf{Proposition 3},\, II  \\
 \emph{ Given (\ref{ma6}), then for $ l \in {\mathbf N}_0 , |n| \geq 2 $
  and $ 0 \leq \La \leq \Lao $ ,}
 \eq \label{bdf}
 |\,\partial^w \Ga^{(\nu),\La,\Lao}_{1;\,l, n}( \vec p\, )| \,\leq
  \, \frac{1}{\Lao} \,(\La+m)^{\,5+1-|n|-|w|- \nu}
   \Big ( \log{\Lao\over m}\Big)^{r} \,
    {\cal P}(\frac{|\vec p\, |}{\La+m})\ .
\eqe
\emph{ with a positive integer $r$ depending on $ n, l, w\, $, and a
  polynomial ${\mathcal P} $ as in (\ref{b1}),(\ref{b2}). } \\
In the proof given in II the system of flow equations for the
vertex functions $\,\Ga^{(\nu),\, \Lz}_{1; \,l, n}\,$ is 
integrated inductively. The flow equations in the case
of an integrated insertion occurring here coincide with those
in the case of a local insertion at insertion momentum zero,
cf.(\ref{w40})-(\ref{w41}). 
Given the condition (\ref{ma6}), from
the bound (\ref{bdf}) then follows that  
 {\it the  Slavnov-Taylor identities are restored
in the limit  $\Lao \to \infty\,$.}  
%%%%%%%%%%%%%%%%%%%%%%%%%%%%%%%%%%%%%%%%%%%%%%%%
\section{Equation of motion of the antighost}
The renormalization of a non-Abelian gauge theory using 
a gauge invariant renormalization scheme is generally  
based on the Slavnov-Taylor identities, complemented by
the equation of motion of the antighost \cite{FaSl},\cite{ZJ}.
Approaching renormalization via flow equations 
this equation is derived from the regularized functional
integral representation of the generating functional
and its renormalization is considered in conjunction
with the restoration of the Slavnov -Taylor identities.
In Section 3.1 the equation of motion in connection with
renormalization within the theory of a scalar field
has been treated, in the present theory the field equation
of the antighost is obtained as the analogue of (\ref{a5}),
\eq \label{vs13}
 \frac{\de L^{\LLz} (\Phi)}{\de \cbx} \,=\,
 L^{\LLz}_{\zeta^a }( x; \Phi)\ ,
 \eqe
 which emerges from extending the original bare interaction
 $ L^{\LLzz}(\Phi) $  by the insertion, cf. (\ref{a3}),
  \eq \label{vs14}
      L^{\LLzz} (\zeta ; \Phi) \, = 
          \int dx \ \zeta^a (x)\  \frac{\de L^{\LLzz} (\Phi)}{\de \cbx}\ ,
 \eqe
where the source $ \zeta^a(x) $ is now a Grassmann element
 carrying ghost number $ -1$. The BRS-invariant classical
   action (\ref{y9a}) satisfies the classical field equation
   $ \de / \de \cbx S_{BRS} = 
    \partial_{\mu}\psi^a_{\mu}(x) - \alpha m \psi^a (x) $,
  where the form (\ref{y14}) of $\, S_{\rm gh}\,$ has been used.
The aim is to show that at the physical value $ \La = 0 $
 of the flow parameter the field equation 
\eq \label{vs15}
   \frac{\de L^{\Lz} (\Phi)}{\de \cbx} \,=\,
    \pa_{\mu} L^{\Lz}_{\g_{\mu}}(x\, ; \Phi) |_{mod}
     - \alpha m L^{\Lz}_{\g}(x ; \Phi) |_{mod} \,,
    \eqe
still holds in the  renormalized theory, following
 at the tree level from the classical action. 
 The label "mod" prescribes in the bare insertions
 (\ref{w25}) to replace
  $ R^0_i \rightarrow { \tilde R}^0_i = O(\hbar) $
 for $ i=1,4 \,$  since the respective tree order 
does not appear on the l.h.s.\\ 
The field equation (\ref{vs15}) can be rewritten in terms of
proper vertex functions using the conventions introduced,
\begin{eqnarray} \label{vs28}
   (2\pi)^4 \, \frac{\de {\Ga}^{\Lz} (\uFi)}{\de \ucb(q)} \,=\,
   & - &\, \frac{q^2 + \al m^2}{\si(q^2)}\, \uca(-q) \\
   & - & i q_{\mu} {\Ga}^{\Lz}_{\g_{\mu}}(q ; \uFi) |_{mod}
     -\, \alpha m {\Ga}^{\Lz}_{\g}(q ; \uFi) |_{mod} \,.\nonumber
    \end{eqnarray} 
  The first term on the r.h.s. is the tree level
 $2$-point function. Restricting
  (\ref{vs28}) to its relevant part, $\si(q^2) $  is
  replaced by $ \si (0) = 1 $ due to (\ref{w7}), 
the first term then provides the 
  tree order of $R_1$ and $R_4$ excluded in the insertions as
  indicated by the  label \emph{mod}, cf. (\ref{vs15}).\\
Taken for granted in the present case the outcome of 
Section 3.1, the field equation (\ref{vs28}) holds in the limit
$ \Lambda_0 \rightarrow \infty $, if its relevant part is fulfilled.
%%%%%%%%%%%%%%%%%%%%%%%%%%%%%%%%%%%  
\section{Restoration of the Slavnov-Taylor\\ identities}
To achieve the Slavnov-Taylor identities it is necessary,
and due to Proposition 3 also sufficient, that the relevant part
of the functional $\,\Ga^{\Lz}_1 \,$ vanishes. Enforcing this
 behaviour in accord with the violated Slavnov-Taylor identities
(\ref{v27}), the conditions (\ref{ma6}) amount to satisfy the
53 equations listed in the Appendix C. These equations
are satisfied in the tree order. Since the analysis to be
performed for $\, l > 0\, $, although rather technical, is 
a crucial step in the method developed, it appears meaningful
to present it verbatim as given in II.\\
Noticing that the normalization constants of the BRS-insertions behave as  
$\, R_i = 1+ {\mathcal O}(\hbar) , i = 1,\cdots 7$, we first analyse the
 equations $ IX $ to $ XXIX $, but take already into account
the equations $ VII_d \,, \, VIII_c $ \,, the latter ones providing
\begin{equation} \label {c1}
    r^{hBA}_2 \, = \, r^{\bar{c} c A}_2 \, \stackrel{!}{=} 0 \, .
\end{equation} 
In proceeding we use conditions determined before, if needed. \\ 
From $XIV_b \,, \,XIV_e \,, XV_{1b}\,, XXIII \, $ directly follow
 \begin{equation}\label{c2}
  r^{AA{\bar c}c}_1 \, = \, r^{AA\bar{c} c }_2 \, =   
  r^{BB{\bar c}c}_1 \, = \, r^{AABB}_2 \, \stackrel{!}{=} 0 \, ,
 \end{equation} 
and then, from $ XIV_{a+c}\,, XVII_b \,, XVIII_c \,, XXVIII, XXIX $ ,
 \begin{equation} \label{c3}
  r ^{AAAA}_2 \, = \,r^{hh{\bar c}c} \, = \, r^{{\bar c}c \bar{c}c} \, =   
  r^{hB{\bar c}c} \,= \,   r^{BB{\bar c}c}_2\,\stackrel{!}{=} 0 \, .
 \end{equation}     
 $ XVI_a \,, XVIII_a \,$, and $XV_{2a}$ combined with $\,XVI_b \,$,
 respectively, require
 \begin{equation}\label{c4}
 R_2 \, \stackrel{!}{=} \,R_6 \, \stackrel{!}{=} \,R_7 \,, \qquad \quad
   R_3 \, R_5 \, \stackrel{!}{=} \,( R_2 )^2 \,.
   \end{equation}
   \begin{equation}\label{c5}
     XIV_c : \qquad  2 F^{AAAA}_1 \, R_1 \,
            \stackrel{!}{=}\,- F^{AAA} g R_2 \,
   \end{equation}  
   \begin{equation}\label{c6}
     XI  : \qquad \quad F^{{\bar c}c B\, (1)} \, R_5 \,
                    \stackrel{!}{=}\,- F^{{\bar c}c h \,(1)} R_2 \,.
   \end{equation} 
 From $ X \,, \,XX\,, \, XIX\, , \, IX \,$ follow for the
                self-coupling of the scalar field 
 \begin{eqnarray}
    8\, F^{BBBB} \, R_4 & \stackrel{!}{=} & F^{BBh (1)} \,g R_3 \,,
                           \label{c7} \\
     4\, F^{BBhh} \, R_4 & \stackrel{!}{=} & F^{BBh (1)}\, g R_5\,,
                              \label{c8} \\
     8\, F^{hhhh} \, R_4 R_3 & \stackrel{!}{=} & F^{BBh (1)}\, g (R_5 )^2 \,,
                               \label{c9} \\
     F^{hhh \,(1)} \, R_3 & \stackrel{!}{=} & F^{BBh (1)} \,R_5  \,,
                                 \label{c10}
     \end{eqnarray}
  and from $XVI_b\, ,\, XVII_a \,,\, XXI\,,\, XIII_2 \,$ for the
                         scalar-vector coupling
 \begin{eqnarray}
 2\, F^{BBA} \,R_5  & \stackrel{!}{=} & - \,F^{hBA}_1\, R_2 \,,
                            \label{c11} \\
 4\, F^{AAhh} \,R_1  & \stackrel{!}{=} &  \,F^{hBA}_1 \,g R_5 \,,
                                \label {c12} \\
 4\, F^{AABB}_1 \,R_1  & \stackrel{!}{=} &  \,F^{hBA}_1\, g R_3 \,,
                                 \label{c13}\\
      F^{AAh (1)} \,R_1  & \stackrel{!}{=} &  \,F^{hBA}_1\,  R_4 \,.
                                 \label{c14} 
 \end{eqnarray} 
One easily verifies that the remaining equations of $ IX $ to $ XXIX $
   are satisfied due
to these conditions (\ref{c1})-(\ref{c14}).\\
At this stage, all those relevant couplings with $\, |n| = 3,4\,$ not 
  appearing already
 in the tree order are required to vanish: (\ref{c1})-(\ref{c3}).
   All other couplings
 involving four fields are determined by particular couplings
  with $ |n|=3\, $: (\ref{c5}),
 (\ref{c7})-(\ref{c9}), (\ref{c12}),(\ref{c13}). In addition, there
  are $4$ conditions
 relating couplings with $ |n|=3 $ : (\ref{c6}), (\ref{c10}), (\ref{c11})
       and (\ref{c14}).
 Moreover, the normalization constants of the BRS-insertions 
 are required to satisfy the three conditions (\ref{c4}).\\
 \noindent
 There are still $ 18-2 $ equations among $ I $ to $ VIII $ to be considered.
 They contain the relevant parameters of $\,\Gamma^{\,0, \Lambda_0}\,$
 with $ |\,n| = 1,2,3 \,$, except $ F^{hhh} $,
 together with the normalization constants of the BRS-insertions.
 Since two of these parameters have been fixed before, (\ref{c1}),
  there remain
  $26 $ to be dealt with.\,($ F^{hhh} $ will then be determined
  by (\ref{c10}).)
These parameters in addition have to obey the conditions derived before: 
 We first observe that the condition (\ref{c14}) is identical to
 equation $ VI_b \, $. 
 There remain the $5$ conditions  to be satisfied:
$3$ conditions (\ref{c4}), together
 with (\ref{c6}), (\ref{c11}). 
All these conditions generate $4$ linear relations among the 
  equations still to be considered: denoting by $ \{X\} $ the content
 of the bracket
 $ \{ \cdots \} $ appearing in equation $ X $, we find  
  \begin{eqnarray}
 0 &=& \alpha^{-1} \{VIII_b \} +gR_2 \{I_b\} + 
        R_1 \big( \{III_a\} + \{III_b\}\big )\, , \label{c15} \\
 0 &=& gR_2 \{II_b\} - \{ VIII_b\} + R_1 \{IV_b\} - 2 R_4 \{V\} \, ,
  \label{c16}\\
 0 &=& R_2 \{ IV_a\} - R_3 \big ( \{VI_a\} - \{ VI_b\} \big ) \, ,
 \label{c17}\\
 0 &=& R_2 \{V \} - R_3 \{ VII_c \} \, .\label{c18}
 \end{eqnarray} 
 Hence, the $ 26 $ parameters in question are constrained by
 $ 16+5-4 = 17 $ equations.
 As \emph{renormalization conditions} we then fix
 $ \kappa^{(3)} = 0 \,$ and let
 \begin{equation}\label{c19}
 \Sigma_{\,\rm trans}\, , \Sigma_{\,\rm long}\, , \Sigma^{AB(1)} ,
 {\dot \Sigma}^{{\bar c} c},
    {\dot \Sigma}^{BB} , F^{AAA} , F^{BBh(1)} , R_3
 \end{equation}
 be chosen freely. These parameters correspond  to the number
of wave function renormalizations (including one for the BRS sector)
and coupling constant renormalizations of the theory.  
Thus, there are $ 26 - 9 $ parameters left,
 together with $ 17 $
 equations. These parameters are now determined successively in
 terms of (\ref{c19})
 and possibly parameters determined before in proceeding. We list
 them in this order,
 writing in bracket the particular equation fulfilled:
$$ R_1(I_b)\, , R_4(II_b)\,, R_2(III_b) \rightarrow R_6 , R_7 , R_5 
 \quad { \rm due \,to}\quad (\ref{c4}) \,, $$
  $$\,F^{{\bar c} c A}_1(III_a)\, , 
  F^{BBA} (V) \rightarrow  F^{hBA}_1 \quad {\rm due \,to}\quad
 (\ref{c11}) \, , $$
$$  F^{AAh(1)}( VI_b ) \, , \,F^{{\bar c} c B(1)}(IV_a) \rightarrow
    F^{{\bar c} c h(1)}\quad {\rm due \, to} \quad (\ref{c6}) \, ,$$
  \begin{equation} \label{c20}
  \Sigma^{{\bar c} c (2) }(VIII_a)\,, \,\Sigma^{BB(2)}(II_a) \, , \,
 \de m^2_{\,(2)}(I_a)
   \, , \, \Sigma^{hh(2)}(VII_a) \, ,\, {\dot \Sigma}^{hh}(VII_{b+c}) \,.
\end{equation}
Now all parameters are determined, without using the
 equations $ IV_b \,, 
  VI_a\,$, $ VII_c \, , VIII_b \,.$ These equations, however,
 are satisfied because of 
 the relations (\ref{c15})-(\ref{c18}). Finally, the relevant
 couplings with $ |n| = 4 $
 as well as $ F^{hhh(1)} $ then are explicitly given
 by (\ref{c5}), (\ref{c7})-(\ref{c10}),
 (\ref{c12}) and (\ref{c13}). 
 
 We have not yet implemented the field equation of
 the antighost (\ref{vs28}).
Performing the mass scaling as before and then extracting the local content 
 $ |n|+ |w| + \nu  \leq 4 \,$ leads to the relations   
\begin{eqnarray}
 1 + {\dot \Sigma}^{{\bar c} c} & = & R_1 \, , \label{c22} \\
     \alpha + \Sigma^{{\bar c} c (2)} & = & \alpha R_4 \, , \label{c23} \\
            F^{{\bar c} c A}_1 & = & g R_2 \, , \label{c24} \\
   F^{{\bar c} c B(1)}\ & = & \frac{\alpha}{2} \,g R_6 \,, \label{c25}\\
  F^{{\bar c} c h(1)}\,  & = & - \,\frac{\alpha}{2}\, g R_5 \,. \label{c26} 
  \end{eqnarray}
   Fixing now the hitherto free renormalization constant
 $ \Sigma_{\rm long} \,$ at 
   the particular value  $ \Sigma_{\rm long} = 0 \,$, we claim these
 relations to be satisfied: 
   (\ref{c22}) and (\ref{c24}) follow at once from $ I_b $
   and $ III_{a+b} $, respectively;
   (\ref{c25}) follows from $ 2\{IV_a \} - \{IV_b \} $, due
 to (\ref{c24}) and (\ref{c4}); and
   herefrom follow (\ref{c26}) due to (\ref{c6}), and (\ref{c23})
 because of $ VIII_a \,$, thus establishing the claim. \\
 Given these additional relations (\ref{c22})-(\ref{c26}) we can adjust
 the procedure (\ref{c20}) choosing now a \emph{reduced set of free
  renormalization conditions (\ref{c19}) in which} $\Sigma_{\rm long}$
  \emph{is excluded.} 
 Proceeding similarly as before we find
  \begin{equation} \label{c27}
  I_b : \quad \Sigma_{\rm long} \, = \, 0 \,,\qquad
         II_a : \quad \Sigma^{BB(2)} \,= \, 0 \, ,
  \end{equation}  
  \begin{equation} \label{c28}
     III_b  : \qquad g R_2 \, = \,  - \, 2 F^{AAA}\,
        \frac{ 1 + {\dot \Sigma}^{{\bar c} c}}{ 1+ \Sigma_{\rm trans}}  
 \longrightarrow \, R_6,\, R_7,\, R_5
 \end{equation}
 $$ {\rm due \, to} \quad (\ref{c4}), $$
 \begin{equation} \label{c29}
  II_b  : \qquad   R_4 \,= \,   
          \frac{ 1 + {\dot \Sigma}^{{\bar c} c}}{ 1 + {\dot \Sigma}^{BB}}
                        \Big ( 1+ \Sigma^{AB(1)} \Big) \, ,
  \end{equation}   
  \begin{equation} \label{c30}
  I_a  : \qquad   1+ \de m^2_{\,(2)} \,= \,   
          \frac{ 1}{ 1 + {\dot \Sigma}^{BB}}
                        \Big ( 1+ \Sigma^{AB(1)} \Big)^2 \,, 
  \end{equation}   
  \begin{equation} \label{c31}
     V  : \quad  2\, F^{BBA} \,= \, \, F^{AAA}    
    \frac{ 1 + {\dot \Sigma}^{BB}}{ 1+ \Sigma_{\rm trans}} \,\, \,
        \longrightarrow  F^{hBA}_1\, \longrightarrow \, F^{AAh(1)}
 \end{equation}
  $$  {\rm due \,to}\quad (\ref{c11}),(\ref{c14}),$$ 
  \begin{equation} \label{c32}
  VII_a  :  \qquad \Big (\frac{M}{m} \Big )^2 + \Sigma^{hh(2)} 
          \, = \, \frac{4}{g} \, F^{BBh(1)}\, \frac{R_4}{R_3}  \,,    
    \end{equation}
     \begin{equation} \label{c33}
   VII_{b+c}  :  \qquad  1 + {\dot \Sigma}^{hh} \, = \,
               ( 1 + {\dot \Sigma}^{BB}) \,\frac{ R_5}{R_3} \,.
  \end{equation}
This list closes the analysis of (\ref{ma6}) quoted from II.

In retrospect the method followed first dealt with the
renormalization of the functional $\,\Ga^{\Lz} \,$ and of
the accompanying functionals with a BRS-insertion
 $\, \Ga^{\Lz}_{\ga_{\tau}} , \Ga^{\Lz}_{\om}\,$,
 disregarding the Slavnov-Taylor identities. In these functionals
 there appear $ 37 + 7 $ relevant parameters.  
Requiring the absence of tadpoles $ (\, \kappa = 0\,)$ and 
 fixing $ \,\Sigma_{\rm long} = 0 \,$ because of the field equation 
 of the antighost, it is then shown that  the set (\ref{c19})
\emph{without} $ \Sigma_{\rm long} \,$ can be used as
 \emph{ renormalization constants to be chosen freely}:
 given these renormalization constants the remaining relevant
 parameters are determined uniquely upon requiring
 that the relevant part (\ref{ma6}) of the functional
 $\, \Ga^{\Lz}_1 \,$ does vanish via the violated
Slavnov-Taylor identities (\ref{v27}). According 
 to Proposition 3, the irrelevant part of the functional
 $\, \Ga^{\Lz}_1 \,$ then vanishes in the limit 
  $\, \Lambda_0 \to \infty \,$, too. Thus, within
 perturbation theory, the functionals $\, \Ga^{ 0, \, \infty} \,$ 
 and $\, \Ga^{0,\, \infty}_{\ga_{\tau}} , \Ga^{0,\, \infty}_{\om}\,$
 are finite and satisfy the Slavnov-Taylor identities,
 i.e. equation (\ref{v27}) for
 $\, \Lambda_0 \to \infty \, $ with the r.h.s. vanishing.
 \\[.5cm]  

%%%%%%%%%%%%%%%%%%%%%%%%%%%%%%%%%%%%%%%%%%%%%%%%

%%%%%%%%%%%%%%%%%%%%%%%%%%%%%%%%%%%%%%%%%%%%%%%%
\begin{appendix}
\renewcommand{\thesection}{\Alph{section}}
\section{The relevant part of $ \Gamma $}
The bare functional $ L^{\Lambda_0, \Lambda_0} $ and the relevant
 part of the generating functional $\Gamma ^{0, \Lambda_0} $ for
 the proper vertex functions have the same general form. We present the
 latter and give the tree order of both explicitly. The cutoff symbols
   $\,0, \Lambda_0$ are suppressed. We write

$$ \Gamma(\underline{A},\underline{h},\underline{B},\underline{\bar{c}},
\underline{c})   =\sum^4_{|n|=1} \Gamma_{|n|} + \Gamma_{(|n| > 4)}, $$
 $|n|$ counting the number of fields, and extract its relevant part,
 i.e. its local field
content with mass dimension not greater than four. Generally we
 will not underline the field variable
symbols in the Appendices, though of course all arguments in the
 $\Gamma $- functional should
appear underlined. The modification to obtain the bare functional
 $ L^{\Lambda_0, \Lambda_0} $ is stated at the end.

\vspace{0.3cm}
\noindent
1) One-point function:\\
$$ \Gamma_1 = \kappa \hat{h}(0). $$

\noindent
2) Two-point functions: \\
\begin{eqnarray*}
\Gamma_2 =  \int_p \Bigg\{ \frac12 A^a_{\mu}(p) A^a_{\nu}(-p)
     \Gamma^{AA}_{\mu\nu}(p) + \frac12 h(p) h(-p) \Gamma^{hh}(p)\quad\\
       + \frac12\, B^a(p)B^a(-p) \Gamma^{BB}(p) 
        -\, \bar{c}^a(p) c^a(-p) \Gamma^{\bar{c}c}(p)\\
       +\, A^a_{\mu}(p) B^a(-p) \Gamma^{AB}_{\mu}(p) \Bigg\},
 \end{eqnarray*}
\begin{eqnarray*}
\Gamma^{AA}_{\mu\nu}(p) &=& \delta_{\mu\nu}(m^2+\delta m^2) + 
(p^2\delta_{\mu\nu}-p_{\mu}p_{\nu}) (1 + \Sigma_{\rm trans} (p^2)) \\ 
& & + \frac{1}{\alpha} p_{\mu}p_{\nu} (1 + \Sigma_{\rm long} (p^2)), \\ 
\Gamma^{hh}(p) &=& p^2 + M^2 + \Sigma^{hh}(p^2), \quad 
\Gamma^{BB}(p) = p^2 + \alpha m^2 + \Sigma^{BB}(p^2), \\
\Gamma^{\bar{c}c}(p) &=& p^2 + \alpha m^2 + \Sigma^{\bar{c}c}(p^2), \quad
\Gamma^{AB}_{\mu}(p) = ip_{\mu} \Sigma^{AB}(p^2).  
 \end{eqnarray*} 
Besides the unregularized tree order explicitly stated, 
there emerge 10 relevant parameters from the various self-energies:
$$\delta m^2, \Sigma_{\rm trans}(0), \Sigma_{\rm long}(0), \Sigma^{hh}(0), 
\dot{\Sigma}^{hh}(0),  \Sigma^{BB}(0), \dot{\Sigma}^{BB}(0),
 \Sigma^{\bar{c}c}(0), \dot{\Sigma}
^{\bar{c}c}(0), \Sigma^{AB}(0)$$ where the notation 
$\dot{\sum}(0) \equiv (\partial_{ p^2 }\sum)(0)$ has been used.
 We note, that in transforming the regularized $L$-functional
 into the corresponding $\Gamma $-functional the inverse of the
 regularized propagators (\ref{w8}) become the $2$-point functions
 of the latter in the tree order $l=0$. The factor
$(\sigma _{\Lambda, \Lambda_0}(p^2) )^{-1}$ thus appearing,
 however, does not contribute to the relevant part due
 to the property (\ref{w7}).
\noindent
3) Three-point functions: \\
Only the relevant part is given explicitly: $r \in \mathcal{O}(\hbar)$
 denotes a relevant parameter
which vanishes in the tree order, otherwise a relevant parameter
 is denoted by $F$. Moreover, we
indicate an irrelevant part by a symbol
 ${\mathcal O}_n, \; n \in \mathbf{N}$, indicating that this
part vanishes as an $n$-th power of the momentum in the limit
 when all momenta tend to zero homogeneously.
\begin{eqnarray*}
\Gamma_3 &=& \int_p\int_q
  \Big\{ \epsilon^{rst} A^r_{\mu}(p) A^s_{\nu}(q) A^t_{\lambda}(-p-q)
       \Gamma^{AAA}_{\mu\nu\lambda}(p,q)   \\
&+ &  A^r_{\mu}(p) A^r_{\nu}(q) h(-p-q) \Gamma^{AAh}_{\mu\nu}(p,q) \\
&+ & \epsilon^{rst} B^r(p) B^s(q) A^t_{\mu}(-p-q) \Gamma^{BBA}_{\mu}(p,q) \\
& +&  h(p) B^r(q) A^r_{\mu}(-p-q) \Gamma^{hBA}_{\mu}(p,q) 
       + \epsilon^{rst} \bar{c}^r(p) c^s(q) A^t_{\mu}(-p-q)
         \Gamma^{\bar{c}cA}_{\mu}(p,q) \\
& +&  B^r(p) B^r(q) h(-p-q) \Gamma^{BBh}(p,q) + h(p) h(q) h(-p-q)
             \Gamma^{hhh}(p,q) \\
& +&  \bar{c}^r(p) c^r(q) h(-p-q) \Gamma^{\bar{c}ch}(p,q) 
     + \epsilon^{rst} \bar{c}^r(p) c^s(q) B^t(-p-q)
         \Gamma^{\bar{c}cB}(p,q) \Big \},
\end{eqnarray*}
\begin{eqnarray*} \begin{array}{llllll}
\Gamma^{AAA}_{\mu\nu\lambda}(p,q) &=& \delta_{\mu\nu} i(p-q)_{\lambda} 
F^{AAA} + {\cal O}_3, \quad &  F^{AAA} &=& - \frac12 g + r^{AAA}, \\
\Gamma^{AAh}_{\mu\nu}(p,q) &=& \delta_{\mu\nu} F^{AAh} + {\cal O}_2, &
        F^{AAh} &=&  \frac12 mg + r^{AAh}, \\
\Gamma^{BBA}_{\mu}(p,q) &=& i(p-q)_{\mu} F^{BBA} + {\cal O}_3, &
F^{BBA} &=&  - \frac14 g + r^{BBA}, \\
\Gamma^{hBA}_{\mu}(p,q) &=& i(p-q)_{\mu} F_1^{hBA} &
       F_1^{hBA} &=&  \frac12 g + r_1^{hBA}, \\
& & + i(p+q)_{\mu} r^{hBA}_2 + {\mathcal O}_3,\\
     \Gamma^{\bar{c}cA}_{\mu}(p,q) &=& ip_{\mu} F_1^{\bar{c}cA}
      + iq_\mu r^{\bar{c}cA}_2 
       + {\cal O}_3, &
F_1^{\bar{c}cA} &=& g + r_1^{\bar{c}cA}, \\
\Gamma^{BBh}(p,q) &=& F^{BBh} + {\mathcal O}_2, &
F^{BBh} &=&  \frac14 g \frac{M^2}{m} + r^{BBh}, \\
\Gamma^{hhh}(p,q) &=& F^{hhh} + {\mathcal O}_2, &
F^{hhh} &=&  \frac14 g \frac{M^2}{m} + r^{hhh}, \\
\Gamma^{\bar{c}ch}(p,q) &=& F^{\bar{c}ch} + {\mathcal O}_2, &
F^{\bar{c}ch} &=&  - \frac12 \alpha gm + r^{\bar{c}ch}, \\
\Gamma^{\bar{c}cB}(p,q) &=& F^{\bar{c}cB} + {\mathcal O}_2, &
F^{\bar{c}cB} &=&  \frac12 \alpha gm + r^{\bar{c}cB}. 
\end{array}
\end{eqnarray*}
The 3-point functions $AAB$ and $BBB$ have no relevant local content. \\
\noindent
4) Four-point functions: \\
With parameters $r$ and $F$ defined as before 
\begin{eqnarray*}
\Gamma_4|_{\rm rel} &=& \int_k \int_p \int_q
     \big\{ \epsilon^{abc} \epsilon^{ars} A^b_{\mu}(k)
     A^c_\nu(p)A^r_\mu(q) A^s_\nu(-k-p-q) F_1^{AAAA} \\
& & + A^r_\mu(k) A^r_\mu(p) A^s_\nu(q) A^s_\nu(-k-p-q) r_2^{AAAA} \\
& & + A^a_\mu(k)A^b_\mu(p) \bar{c}^r(q) c^s(-k-p-q)
       (\delta^{ab}\delta^{rs}r_1^{AA\bar{c}c}
+ \delta^{ar}\delta^{bs}r_2^{AA\bar{c}c}) \\
& & + A^a_\mu(k)A^b_\mu(p) B^r(q) B^s(-k-p-q)
       (\delta^{ab}\delta^{rs}F_1^{AABB}
+ \delta^{ar}\delta^{bs}r_2^{AABB}) \\
& & + B^a(k)B^b(p) \bar{c}^r(q) c^s(-k-p-q)
    (\delta^{ab}\delta^{rs}r_1^{BB\bar{c}c}
+ \delta^{ar}\delta^{bs}r_2^{BB\bar{c}c}) \\
& & + h(k)h(p)h(q)h(-k-p-q) F^{hhhh} \\
& & + B^r(k)B^r(p)h(q)h(-k-p-q) F^{BBhh} \\
& & + B^r(k)B^r(p)B^s(q)B^s(-k-p-q) F^{BBBB} \\
& & + A^r_\mu(k)A^r_\mu(p)h(q)h(-k-p-q) F^{AAhh} \\
& & + h(k)h(p)\bar{c}^r(q)c^r(-k-p-q) r^{hh\bar{c}c} \\
& & + \bar{c}^a(k)c^a(p)\bar{c}^r(q)c^r(-k-p-q) r^{\bar{c}c\bar{c}c} \\
& & + \epsilon^{rst}h(k)B^r(p)\bar{c}^s(q)c^t(-k-p-q) r^{hB\bar{c}c} \big\},
\end{eqnarray*} 
\begin{eqnarray*}\begin{array}{llllll}
F_1^{AAAA} &=& \frac14 g^2 + r_1^{AAAA}, \quad &
F_1^{AABB} &=& \frac18 g^2 + r_1^{AABB}, \\
F^{hhhh} &=& \frac{1}{32} g^2 \left( \frac Mm \right)^2 + r^{hhhh}, &
F^{BBhh} &=& \frac{1}{16} g^2 \left( \frac Mm \right)^2 + r^{BBhh}, \\
F^{BBBB} &=& \frac{1}{32} g^2 \left( \frac Mm \right)^2 + r^{BBBB}, &
F^{AAhh} &=& \frac18 g^2 + r^{AAhh}.
\end{array} 
\end{eqnarray*}
Hence, in total $\Gamma$ involves $1 + 10 + 11 + 15 = 37$
 relevant parameters. \\
After deleting in the two-point functions the contributions of the 
order $l=0$, i.e. keeping only the
10 parameters which appear in the various self-energies, we have
 the form of the bare functional
$ L^{\Lambda_0, \Lambda_0}$, and its order $l=0$ also given explicitly. 
%%%%%%%%%%%%%%%%%%%%%%%%%%%%%%%%%%%%%%%%%%%%%%%%%%%
\section{The relevant part of the BRS-insertions}
We also have to consider the vertex functions (\ref{v25}-\ref{v26})
 with one operator insertion, 
generated by the BRS-variations. These insertions have mass
 dimension $D=2$. Performing the Fourier-transform
$$ \hat{\Gamma }^{0, \Lambda_0}_{\gamma }(q) \, = \,
    \int dx\,  e^{iqx} \, {\Gamma }^{0, \Lambda_0}_{\gamma }(x) $$
and similarly in the other cases, we list the respective relevant
 part of these four vertex functions with
 one insertion, suppressing  the superscript $0, \Lambda_0$:
\begin{eqnarray*}
\hat{\Gamma}_{\gamma^a_\mu}(q)|_{\rm rel} &=& -iq_\mu c^a(-q)R_1 + 
\epsilon^{arb} \int_k A^r_\mu(k) c^b(-q-k) gR_2 \, ,\\
 \hat{\Gamma}_{\gamma}(q)|_{\rm rel} &=&
      \int_k B^r(k)c^r(-q-k) (- \frac12 gR_3)\,, \\
\hat{\Gamma}_{\gamma^a}(q)|_{\rm rel} &=& m c^a(-q)R_4 \\ 
  & &+ \int_k h(k)c^a(-q-k)\frac12 gR_5 +
    \epsilon^{arb} \int_k B^r(k) c^b(-q-k) \frac12 gR_6 \, , \\
\hat{\Gamma}_{\omega^a}(q)|_{\rm rel} &=&
       \epsilon^{ars} \int_k c^r(k) c^s(-q-k) \frac12 gR_7.
\end{eqnarray*} 
There appear 7 relevant parameters 
$$R_i = 1 + r_i \, , \qquad r_i = {\cal O}(\hbar), \qquad i = 1,...,7. $$
All the other 2-point functions, and the higher ones, of course,
 are of irrelevant type.
%%%%%%%%%%%%%%%%%%%%%%%%%%%%%%%%%%%%%%%%%%%%%%%%%%%%
 \section{The relevant part of $\,\Gamma_1 $}
As a consequence of the expansion in the mass parameters
the conditions following from the fact that the 
relevant part of the functional $\Ga_1\,$ should vanish 
\[ 
\Gamma_1(\uA,\uh,\uB,\ubc,\uc)|_{\dim \le 5} \begin{array}{c}
 ! \\ = \\ \\
\end{array} 0 .
\]
can be reordered according to the value of
$\nu\,$ which appears. We get contributions for $0 \le \nu \le 3\,$. 
The value of $\, \nu\,$ in the various relevant couplings is indicated
 as a superscript in parentheses {\it if $\nu >0\,$}.  
We explicitly indicate the momentum and the power of $m\,$ in front of
each STI. The power of $m\,$ indicates the value of $\nu\,$
in the corresponding contribution to $\Ga_1\,$.\\[0.2cm]
\noindent
Two fields
\begin{enumerate}
\item[I)] $\; \delta_{A^a_\mu(q)} \delta_{c^r(k)} \Gamma_1|_0$
\begin{enumerate}
\item[a) $0 \, \stackrel {!} {= } \,$]$   m^2\, q_\mu \left\{ -
    (1+\de m^2_{\,(2)} )R_1 + \sum^{AB(1)}  R_4 + 
1 + \frac{1}{\alpha} \sum^{\bar{c}c(2)} \right\} $,
\item[b) $0  \, \stackrel {!} {= } \,$]$  q^2 q_\mu \Big\{ - {1\over
      \al}(1 + \sum_{\rm long} ) R_1 
+{1\over  \al} (1 + \dot{\sum}^{\bar{c}c}) 
\Big\} $.
\end{enumerate}
\end{enumerate}  
\begin{enumerate}
\item[II)] $\; \delta_{B^a(q)} \delta_{c^r(k)} \Gamma_1|_0$
\begin{enumerate}
\item[a) $0  \, \stackrel {!} {= }$]$ m^3 \left\{(\alpha +\sum^{BB(2)})R_4
- (\alpha + \sum^{\bar{c}c(2)}) 
          - \frac{g}{2}\, \ka^{(3)} R_3 \, \right\}$.
\item[b) $0   \stackrel {!} {=}$]$m\,q^2  \Big\{ - \sum^{AB(1)} R_1
+ (1 +\dot{\sum}^{BB})R_4- (1 + \dot{\sum}^{\bar{c}c}) \Big\} $.
\end{enumerate}
\end{enumerate}
\vspace{0.2cm}
\noindent
Three fields
\begin{enumerate}
\item[III)] $\; \delta_{A^r_\mu(p)} \delta_{A^s_\nu(q)} \delta_{c^t(k)}
\Gamma_1|_0$
\begin{enumerate}
\item[a) $ 0    \stackrel {!} {= }($]$\!\!\!p_\mu p_\nu - q_\mu q_\nu)\!
\Big\{\!\! -\! 2 F^{AAA}R_1\! - \! \frac{1}{\alpha}
(F_1^{\bar{c}cA}\! - r_2^{\bar{c}cA})\! +\!
\left[\frac{1}{\alpha} ( 1\! +
\sum_{\rm long})\!-\!
(1\!+\sum_{\rm trans})\right]gR_2\Big\},$
\item[b) $0   \stackrel {!} {= }    ($]$\!\! p^2-q^2)\delta_{\mu\nu}
\left\{ 2 F^{AAA} R_1 +  (1 + \sum_{\rm trans} ) g R_2
\right\} $,
\end{enumerate}
\item[IV)] $\; \delta_{A^r_\mu(p)} \delta_{B^s(q)} \delta_{c^t(k)}
\Gamma_1|_0$
\begin{enumerate}
\item[a) $0   \, \stackrel {!} {= } \,$]$   m\, p_\mu \left\{ 2F^{BBA} R_4 +
\frac12 g \sum^{AB(1)} R_6 + \frac{1}{\alpha}
F^{\bar{c}cB, (1)} -  r_2^{\bar{c}cA}\right\}$,
\item[b) $0   \, \stackrel {!} {= } \,$]$ m\, q_\mu 
\left\{ g \sum^{AB(1)} R_2
+ 4F^{BBA} R_4 +  (F_1^{\bar{c}cA} - r_2^{\bar{c}cA}) \right\}$,
\end{enumerate}
\item[V)] $ \delta_{B^r(p)} \delta_{B^s(q)} \delta_{c^t(k)} \Gamma_1|_0
$
\begin{enumerate}
\item[] $\!\!\!\!\!\!\!\!\!\!\!\!\!\! \!\!
0 \! \stackrel {!} {= }\! (p^2\!-q^2)\!\left\{ 2R_1 F^{BBA} +
(1 + \dot{\sum}^{BB}) {g \over 2} R_6 
\right\} $,
\end{enumerate}
\item[VI)] $\; \delta_{A^r_\mu(p)} \delta_{h(q)} \delta_{c^t(k)}
\Gamma_1|_0$
\begin{enumerate}
\item[a) $0 \, \stackrel {!} {= }\, $]$   m\, p_\mu \Big\{ -2R_1F^{AAh(1)}+
R_4 (F_1^{hBA} - r_2^{hBA}) 
+ \sum^{AB(1)} \frac12 gR_5 - \frac{1}{\alpha} F^{\bar{c}ch(1)} \Big\}$,
\item[b) $0   \, \stackrel {!} {= }
 \,$]$ m\, q_\mu \left\{ -2 R_1F^{AAh(1)} +
2R_4  F_1^{hBA}\right\}$,
\end{enumerate}
\item[VII)] $\; \delta_{h(p)} \delta_{B^s(q)} \delta_{c^t(k)}
\Gamma_1|_0$
\begin{enumerate}
\item[a) $0   \, \stackrel {!} {= } \,$]$   m^2\left\{
(\frac{M^2}{m^2} + \sum^{hh(2)}) (-
\frac12 g R_3) + 2 F^{BBh(1)} R_4 + F^{\bar{c}ch(1)}
 + (\alpha+ \sum^{BB(2)}) \frac12 g R_5\right\} $,
\item[b) $0   \, \stackrel {!} {= } \,$]$ p^2 \Big\{ F_1^{hBA}R_1 - (1 +
   \dot{\sum}^{hh}) \frac12 gR_3 
\Big\}$,
\item[c) $0   \, \stackrel {!} {= } \,$]$ q^2 \Big\{- F^{hBA}_1R_1 + (1 +
  \dot{\sum}^{BB}) \frac12 gR_5 
\Big\}$,
\item[d) $0   \, \stackrel {!} {= } \,$]$ k^2 \left\{ r_2^{hBA} R_1
\right\}$,
\end{enumerate}   
\item[VIII)] $\; \delta_{c^t(q)} \delta_{c^s(p)} \delta_{\bar{c}^r(k)}
\Gamma_1|_0$
\begin{enumerate}
\item[a) $0   \, \stackrel {!} {= } \,$]$   m^2 \left\{2
    F^{\bar{c}cB(1)}R_4 - (\alpha  +  \sum^{\bar{c}c (2)})  g R_7\right\}$,
\item[b) $0   \, \stackrel {!} {= } \,$]$ k^2 \Big\{ F_1^{\bar{c}cA}R_1 -
r_2^{\bar{c}cA}R_1 - (1 + \dot{\sum}^{\bar{c}c}) gR_7 \Big\}$,
\item[c) $0   \, \stackrel {!} {= } \,$]$ (p^2 + q^2) \Big\{
r_2^{\bar{c}cA}R_1 \Big\}$.
\end{enumerate}
\end{enumerate}
\noindent 
Four fields
\begin{enumerate}
\item[IX)] $\; \delta_{h(p)} \delta_{h(q)} \delta_{B^1(k)}
\delta_{c^1(l)} \Gamma_1|_0$ \\
\\
$0   \, \stackrel {!} {= } \, m\,\left\{
6 F^{hhh, (1)} (- \frac12 gR_3) + 4 F^{BBhh} 
R_4 + 2 F^{BBh, (1)} g R_5 + 2 r^{hh\bar{c}c}\right\} $.
\item[X)] $\; \delta_{B^1(k)} \delta_{B^1(p)} \delta_{B^2(q)}
\delta_{c^2(l)} \Gamma_1|_0$ \\
\\
$0   \, \stackrel {!} {= } \, m\left\{- F^{BBh, (1)} gR_3 + 8F^{BBBB}  R_4
 +  \left( 2r_1^{BB\bar{c}c} +r^{BB\bar{c}c}_2 \right)\right\} $.
\item[XI)] $\; \delta_{h(l)} \delta_{\bar{c}^3(k)} \delta_{c^1(p)}
\delta_{c^2(q)} \Gamma_1|_0$ \\
\\
$0   \, \stackrel {!} {= } \,  m\left\{2r^{hB\bar{c}c}  R_4 
+ F^{\bar{c}cB (1)} gR_5 + F^{\bar{c}ch, (1)} gR_7 \right\} $.
\item[XII)] $\; \delta_{c^2(k)} \delta_{\bar{c}^2(l)} \delta_{c^1(p)}
\delta_{B^1(q)} \Gamma_1|_0$ \\
\\
$0   \, \stackrel {!} {= } \,  m\left\{ F^{\bar{c}ch(1)} (- \frac12 g R_3) +
(2r_1^{BB\bar{c}c}- r_2^{BB\bar{c}c}) R_4 
 + F^{\bar{c}cB(1)} (\frac12 gR_6 - gR_7) + 2
 r^{\bar{c}c\bar{c}c}\right\}   $.
\item[XIII)$_1$] $\; \delta_{A^1_\mu(k)} \delta_{A^2_\nu(p)}
\delta_{B^1(q)} \delta_{c^2(l)} \Gamma_1|_0$ \\
\\
$0   \, \stackrel {!} {= } \,  2r_2^{AABB} R_4 + r_2^{AA\bar{c}c} $.
\item[XIII)$_2$] $\; \delta_{A^1_\mu(k)} \delta_{A^1_\nu(p)}
\delta_{B^2(q)} \delta_{c^2(l)} \Gamma_1|_0$ \\
\\
$0   \, \stackrel {!} {= } \, 
\,  m\left\{ -F^{AAh(1)}  g R_3 +
4F_1^{AABB} R_4 
 + 2  r_1^{AA\bar{c}c}\right\}   $.
\item[XIV)] $\; \delta_{A^1_\mu(p)} \delta_{A^1_\nu(q)}
\delta_{A^2_\rho(k)} \delta_{c^2(l)} \Gamma_1|_0$
\begin{enumerate}
\item[a)] $0   \, \stackrel {!} {= } \, 2 \delta_{\mu\nu} l_\rho \Big\{
4 (F^{AAAA}_1 + r_2^{AAAA}) R_1 
+ 2F^{AAA} gR_2 + \frac{1}{\alpha} r_1^{AA\bar{c}c}
\Big\}$,
\item[b)] $0   \, \stackrel {!} {= } \, \delta_{\mu\nu} (p_\rho +
q_\rho)  \left\{ \frac{2}{\alpha} r_1^{AA\bar{c}c}
\right\}$,
\item[c)] $0   \, \stackrel {!} {= } \, (\delta_{\mu\rho}l_\nu +
\delta_{\nu\rho}l_\mu) \left\{ - 4 F_1^{AAAA}R_1 - 2 F^{AAA} gR_2 
\right\}$,
\item[d)] $0   \, \stackrel {!} {= } \, (\delta_{\mu\rho}p_\nu +
\delta_{\nu\rho}q_\mu) \left\{ 0 \right\}$,
\item[e)] $0   \, \stackrel {!} {= } \, (\delta_{\mu\rho}q_\nu +
\delta_{\nu\rho}p_\mu)  \left\{ - \frac{1}{\alpha} r_2^{AA\bar{c}c}\right\}$.
\end{enumerate}   
\item[XV)$_1$] $\; \delta_{B^1(p)} \delta_{B^1(q)} \delta_{A^2_\mu(k)}
\delta_{c^2(l)} \Gamma_1|_0$
\begin{enumerate}
\item[a)] $0   \, \stackrel {!} {= } \, l_\mu  \Big\{ 4 F_1^{AABB}R_1 +
2F^{BBA} gR_6 \Big\}$,
\item[b)] $0   \, \stackrel {!} {= } \, k_\mu 
\left\{ r_1^{BB\bar{c}c}\right\}$,
\end{enumerate}
\item[XV)$_2$] $\; \delta_{B^1(p)} \delta_{B^2(q)} \delta_{A^1_\mu(k)}
\delta_{c^2(l)} \Gamma_1|_0$
\begin{enumerate}
\item[a)] $0   \, \stackrel {!} {= } \, p_\mu  \left\{ -2r_2^{AABB}R_1 +
2F^{BBA} gR_2 + F_1^{hBA} gR_3 \right\}$,
\item[b)] $0   \, \stackrel {!} {= } \, q_\mu \left\{-2 r_2^{AABB}R_1 -
2F^{BBA} gR_2 + 2F^{BBA}gR_6 \right\}$,
\item[c)] $0   \, \stackrel {!} {= } \, k_\mu \Big\{-2 r_2^{AABB}R_1 +
F_1^{hBA} \frac12 gR_3 + r_2^{hBA}\frac12 gR_3
 + F^{BBA} gR_6 - \frac{1}{\alpha} r_2^{BB\bar{c}c}
\Big\}$,
\end{enumerate}
\item[XVI)] $\; \delta_{h(p)} \delta_{A_{\mu}^1(k)} \delta_{B^2(q)}
\delta_{c^3(l)} \Gamma_1|_0$
\begin{enumerate}
\item[a)] $0   \, \stackrel {!} {= } \, p_\mu  \left\{ F_1^{hBA}g(R_6
-R_2) - r_2^{hBA} gR_2 \right\}$,
\item[b)] $0   \, \stackrel {!} {= } \, q_\mu \left\{F_1^{hBA}gR_2 -
r_2^{hBA} gR_2 + 2F^{BBA}gR_5 \right\}$,
\item[c)] $0   \, \stackrel {!} {= } \, k_\mu \Big\{F_1^{hBA} \frac12
gR_6 - r_2^{hBA} \frac12 gR_6 + F^{BBA} gR_5 
 - \frac{1}{\alpha} r^{hB\bar{c}c} \Big\}$,
\end{enumerate}  
\item[XVII)] $\; \delta_{h(p)} \delta_{h(q)} \delta_{A^1_\mu(k)}
\delta_{c^1(l)} \Gamma_1|_0$
\begin{enumerate}
\item[a)] $0   \, \stackrel {!} {= } \, l_\mu  \left\{ 4F^{AAhh}R_1 -
F_1^{hBA} gR_5 \right\}$,
\item[b)] $0   \, \stackrel {!} {= } \, k_\mu \left\{r_2^{hBA}gR_5 +
\frac{2}{\alpha} r^{hh\bar{c}c}  \right\}$.
\end{enumerate}
\item[XVIII)] $\; \delta_{A^2_\mu(k)} \delta_{c^2(p)} \delta_{c^1(q)}
\delta_{\bar{c}^1(l)} \Gamma_1|_0$
\begin{enumerate}
\item[a)] $0   \, \stackrel {!} {= } \, l_\mu  \left\{
F_1^{\bar{c}cA}g(R_2 -R_7) + \frac{2}{\alpha} r^{\bar{c}c\bar{c}c}\right\}$,
\item[b)] $0   \, \stackrel {!} {= } \, p_\mu \Big\{
2r_1^{AA\bar{c}c}R_1 + r_2^{\bar{c}cA} g(R_2 - R_7) $
$ + \frac{2}{\alpha} r^{\bar{c}c\bar{c}c} \Big\} $,
\item[c)] $0   \, \stackrel {!} {= } \, q_\mu \Big\{ -r_2^{AA\bar{c}c}
R_1 - r_2^{\bar{c}cA} gR_7  + \frac{2}{\alpha} r^{\bar{c}c\bar{c}c}\Big\}$.
\end{enumerate}
\end{enumerate}
\vspace{0.2cm}
\noindent
Five fields
\begin{enumerate}
\item[XIX)] $\; \delta_{h(p)} \delta_{h(q)} \delta_{h(k)}
\delta_{B^1(l)} \delta_{c^1(l^\prime)} \Gamma_1|_0$ \\
\\
$0   \, \stackrel {!} {= } \, -2F^{hhhh}R_3 + F^{hhBB} R_5
$.
\item[XX)] $\; \delta_{h(p)} \delta_{B^1(q)} \delta_{B^1(k)}
\delta_{B^2(l)} \delta_{c^2(l^\prime)} \Gamma_1|_0$ \\
\\
$0   \, \stackrel {!} {= } \, -F^{BBhh} R_3 + 2F^{BBBB} R_5
$.
\item[XXI)] $\; \delta_{A^1_\mu(k)} \delta_{A^1_\nu(p)} \delta_{h(k)}
\delta_{B^2(l)} \delta_{c^2(l^\prime)} \Gamma_1|_0$ \\
\\
$0   \, \stackrel {!} {= } \, -F^{AAhh} R_3 + F_1^{AABB} R_5
$.
\item[XXII)] $\; \delta_{A^1_\mu(k)} \delta_{B^1(p)}
\delta_{c^1(l^\prime)} \delta_{A^2_\nu(q)} \delta_{B^3(l)} \Gamma_1|_0$
\\
\\
$0   \, \stackrel {!} {= } \, r_2^{AABB} (R_6- 2R_2) $.
\item[XXIII)] $\; \delta_{A^1_\mu(k)} \delta_{B^1(q)}
\delta_{A^2_\nu(p)} \delta_{c^2(l^\prime)} \delta_{h(l)} \Gamma_1|_0$ 
\\
\\
$0   \, \stackrel {!} {= } \, r_2^{AABB} R_5  $.
\item[XXIV)] $\; \delta_{A^3_\mu(k)} \delta_{A^3_\nu(p)}
\delta_{\bar{c}^2(q)} \delta_{c^3(l)}  \delta_{c^1(l^\prime)}
\Gamma_1|_0$ \\
\\
$0   \, \stackrel {!} {= } \, r_2^{AA\bar{c}c} R_2  + r_1^{AA\bar{c}c}
R_7 $.   
\item[XXV)] $\; \delta_{A^3_\mu(k)} \delta_{\bar{c}^3(q)}
\delta_{A^2_\nu(p)}  \delta_{c^3(l)} \delta_{c^1(l^\prime)}
\Gamma_1|_0$ \\
\\
$0   \, \stackrel {!} {= } \, r_2^{AA\bar{c}c} (3R_2 - R_7)
$.
\item[XXVI)] $\; \delta_{B^1(p)} \delta_{B^1(q)} \delta_{\bar{c}^1(k)}
\delta_{c^2(l)}  \delta_{c^3(l^\prime)}  \Gamma_1|_0$ \\
\\
$0   \, \stackrel {!} {= } \, r_2^{BB\bar{c}c} (R_6 - R_7)  -
r_1^{BB\bar{c}c} R_7 $.
\item[XXVII)] $\; \delta_{B^1(p)}  \delta_{\bar{c}^1(k)} \delta_{B^2(q)}
\delta_{c^3(l)}  \delta_{c^1(l^\prime)}  \Gamma_1|_0$ \\
\\
$0   \, \stackrel {!} {= } \, -r^{hB\bar{c}c} R_3  + r_2^{BB\bar{c}c}
(3R_6 - 2R_7) $.
\item[XXVIII)] $\; \delta_{h(p)}  \delta_{h(q)} \delta_{\bar{c}^1(k)}
\delta_{c^2(l)} \delta_{c^3(l^\prime)}   \Gamma_1|_0$ \\
\\
$0   \, \stackrel {!} {= } \, r^{hB\bar{c}c} R_5  + r^{hh\bar{c}c} R_7$.
\item[XXIX)] $\; \delta_{h(p)}  \delta_{B^1(q)} \delta_{c^1(l)}
\delta_{\bar{c}^2(k)} \delta_{c^2(l^\prime)}   \Gamma_1|_0$ \\
\\
$0   \, \stackrel {!} {= } \, 2r^{hh\bar{c}c} R_3  - 2r_1^{BB\bar{c}c} R_5
+ r_2^{BB\bar{c}c} R_5 + r^{hB\bar{c}c} (-R_6 + 2R_7) $.
\end{enumerate}

\end{appendix}

\renewcommand{\thesection}{}
\section{Acknowledgements}
The author is much indebted to Christoph Kopper for his
encouragement and gratefully acknowledges his careful
reading of the manuscript and his
 numerous valuable suggestions.


\begin{thebibliography}{99}
\bibitem{Dy} F.J. Dyson: ``The Radiation Theories of Tomonaga, Schwinger,
        and Feynman'',
       \emph{ Phys. Rev.} \textbf{75} (1949) 486 - 502,
       ``The S Matrix in Quantum Electrodynamics'',
         \emph{ Phys. Rev.} \textbf{75} (1949) 1736 - 1755. 
\bibitem{BPHZ} N.N. Bogoliubov, O.S. Parasiuk:``\"Uber die Multiplikation
      der Kausalfunktionen
       in der Quantentheorie der Felder'',\emph{ Acta Math.}
       \textbf{97} (1957)
        227 - 266,  
       K. Hepp: \emph{Th\'eorie de la renormalisation}, 
              Lecture  Notes in Physics, Springer, Berlin 1969, \\
      W. Zimmermann:``Local Operator Products and Renormalization in
       Quantum Field Theory'', in:
       \emph{Lectures on Elementary Particles and Quantum Field Theory},
        1970 Brandeis University Summer
       Institute in Theoretical Physics, Vol.1, M.I.T. Press, Cambridge 1970.
\bibitem{PaVi} W. Pauli, F. Villars:``On Invariant Regularization 
      in Relativitic Quantum Theory'',
                   \emph{Rev. Mod. Phys.} \textbf{21} (1949) 434 - 444.
\bibitem{Spe} E. Speer:\emph{ Generalized Feynman Amplitudes},
           Princeton University, Princeton 1969.
\bibitem{dim} G.'t Hooft, M.Veltman: ``Regularization and Renormalization
        of Gauge Fields'', \emph{Nucl. Phys.} \textbf{B44} (1972) 189-213,\\
         C.G.Bollini, J.J.Giambiagi: ``Lowest order ``divergent'' Graphs in
 $\nu$-dimensional space'', \emph{Phys. Lett.} \textbf{40B} (1972) 566-568,\\
        P.Breitenlohner, D.Maison: ``Dimensional Renormalization and
  the Action Principle'', \emph{Commun. Math. Phys.} \textbf{52} (1977) 11-38,
        ``Dimensionally Renormalized Green's Functions for Theories with
 Massless Particles. I,II '',\emph{Commun.Math.Phys.} \textbf{52} (1977) 39-54,
                           \textbf{52} (1977) 55-75.
\bibitem{VW} \emph{Renormalization Theory, Erice 1975}, A.S. Wightman,
             G. Velo (Eds.), Reidel, 1976.
\bibitem{Zi} W. Zimmermann: ``Convergence of Bogoliubov's Method
 of Renormalization in Momentum Space'',
\emph{ Commun. Math. Phys.} \textbf{15} (1969) 208 - 234. 
\bibitem{BPHZL}J.H.Lowenstein: ``Convergence Theorems
 for Renormalized Feynman Integrals with Zero - Mass Propagators'',
\emph{ Commun. Math. Phys.} \textbf{47} (1976) 53 - 68.
\bibitem{HV}G. 't Hooft: ``Renormalization of Massless Yang - Mills Fields'', 
               \emph{Nucl. Phys.} \textbf{B33} (1971) 173 - 199,
      ``Renormalizable Lagrangians for Massive Yang - Mills Fields'', 
                  \emph{Nucl. Phys.} \textbf{B35} (1971) 167 - 188,  \\
             G. 't Hooft, M. Veltman: ``Combinatorics of Gauge Fields'',\\
                     \emph{ Nucl. Phys.} \textbf{B50} (1972) 318 - 353.
\bibitem{Wil} K.G.Wilson: ``Renormalization Group and Critical Phenomena. I.
               Renormalization Group and the Kadanoff Scaling Picture'',\\
                    \emph{ Phys. Rev.} \textbf{B4} (1971) 3174 - 3183,\\
           ``Renormalization Group and Critical Phenomena. II.
         Phase - Space - Cell  Analysis  of Critical Behaviour'',
                 \emph{ Phys. Rev.} \textbf{B4} (1971) 3184 - 3205,\\
     K.G. Wilson, J. Kogut: ``The Renormalization Group and the $\varepsilon$ 
               Expansion'',\emph{ Phys. Rep.} \textbf{12C} (1974) 75 - 199.
\bibitem{GaKu} K. Gawedzki, A. Kupiainen: ``Gross-Neveu Model
         Through Convergent Perturbation Expansions ", 
          \emph{ Commun. Math. Phys.} \textbf{102} (1985) 1-30.
\bibitem{FMRS} J. Feldman, J. Magnen, V. Rivasseau, R. S\'en\'eor:
    ``A renormalizable
   field theory: the massive Gross-Neveu model in two dimensions ", \\
       \emph{Commun. Math. Phys.} \textbf{103} (1986) 67-103.
\bibitem{Bry} D.C. Brydges: \emph{ Functional Integrals and their
          Applications},\\ Lausanne lectures 1992, mp-arc 93-24.
\bibitem{Riv1} V.Rivasseau: \emph{Constructive Renormalization Theory},\\
              arXiv: math-ph/9902023. 
\bibitem{Riv2} V. Rivasseau: \emph{ From Perturbative to Constructive
         Renormalization}, Princeton University Press, 1991.
\bibitem{GN} G. Gallavotti, F. Nicolo: ``Renormalization Theory in
      Four-Dimensional Scalar Fields  I, II ",\\
       \emph{ Commun. Math. Phys.} \textbf{100} (1985) 545-590, 
           \textbf{101} (1986) 247-282.  \\
      G. Gallavotti: ``Renormalization theory and ultraviolet 
       stability for scalar fields via renormalization group methods " ,\\
        \emph{ Rev. Mod. Phys.} \textbf{57} (1985) 471-562. 
\bibitem {FHRW} J. S. Feldman, T. R. Hurd, L. Rosen, J. D. Wright:
     ``QED: A Proof of Renormalizability ",
            Springer, Lecture Notes in Physics 312, 1988.
 \bibitem{Pol} J. Polchinski: ``Renormalization And Effective Lagrangians'',\\
                       \emph{Nucl. Phys.} \textbf{B231} (1984) 269 - 295.
\bibitem{MiRa} P. K. Mitter, T. R. Ramadas: ``The Two-Dimensional
   $O(N)$ Nonlinear $\sigma$-Model: Renormalisation and Effective Actions ",\\
       \emph{Commun. Math. Phys.} \textbf{122} (1989) 575-596.
\bibitem{KKS} G. Keller, Ch. Kopper, M. Salmhofer:
 ``Perturbative Renormalization  and Effective Lagrangians in $\Phi ^4_4 $\,'',
                  \emph{ Helv. Phys. Acta} \textbf{65} (1992) 32 - 52.
\bibitem{KK3} G. Keller, Ch. Kopper: ``Perturbative Renormalization
                      of Composite Operators via Flow Equations I\,'',\\
                \emph{Commun. Math. Phys.} \textbf{148} (1992) 445 - 467.  
\bibitem{KK4} G. Keller, Ch. Kopper: 
             ``Perturbative Renormalization of Composite
              Operators via Flow Equations II: Short distance expansion'',\\
                \emph{ Commun. Math. Phys.} \textbf{153} (1993) 245 - 276.
\bibitem{Wiec} Ch. Wieczerkowski: ``Symanzik's Improved actions
           from the viewpoint of the Renormalization Group'',\\
          \emph{Commun. Math. Phys.} \textbf{120} (1988) 148 - 176,
\bibitem{Impr} G. Keller: ``The Perturbative Construction of Symanzik's
           improved Action for $\Phi ^4_4$ and $QED_4$ ``,
                 \emph{Helv. Phys. Acta} \textbf{66} (1993) 453 - 470.
\bibitem{Ke2} G. Keller: ``Local Borel summability of Euclidean
    $\Phi _4^4$: A simple proof via Differential Flow Equations'',\\
                 \emph{ Commun. Math. Phys.} \textbf{161} (1994) 311 - 323.  
\bibitem{KK1} G. Keller, Ch. Kopper: ``Perturbative Renormalization 
          of Massless $\Phi ^4_4$ with Flow Equations'',
                  \emph{Commun. Math. Phys.} \textbf{161} (1994) 515 - 532.
\bibitem{QED} G. Keller, Ch. Kopper: ``Perturbative renormalization
       of QED via flow equations'',
           \emph{Phys. Lett.} \textbf{B273} (1991) 323-332,\\
          ``Renormalizability Proof for QED Based on Flow Equations'',\\
          \emph{Commun. Math. Phys.} \textbf{176} (1996) 193 - 226.
\bibitem{Kim} C. Kim: ``A Renormalization Group Flow Approach to
              Decoupling and Irrelevant Operators'',
          \emph{Ann. Phys.(N.Y.}) \textbf{243} (1995) 117 - 143.
\bibitem{KKSch} G. Keller, Ch. Kopper, C. Schophaus:
             ``Perturbative Renormalization with Flow Equations in
                    Minkowski Space'',\\
         \emph{ Helv. Phys. Acta} \textbf{70} (1997) 247 - 274.
\bibitem{KM} Ch. Kopper, V.F. M\"uller: ``Renormalization Proof for
          Spontaneously Broken Yang - Mills Theory with Flow Equations'',\\
                \emph{ Commun. Math. Phys.} \textbf{209} (2000) 477 - 516.
\bibitem{KMR} Ch. Kopper, V.F. M\"uller, Th. Reisz: ``Temperature Independent
                   Renormalization of Finite Temperature Field Theory'',\\
            \emph{Ann. Henri Poincar{\'e}} \textbf{2} (2001) 387 - 402.
\bibitem{Salm} M. Salmhofer: \emph{Renormalization, An Introduction},
              Springer, 1999.
\bibitem{Vert} M. Bonini, M. D' Attanasio, G. Marchesini:
                   ``Perturbative renormalization and infrared
  finiteness in the Wilson renormalization group: the massless scalar case'',
 \emph{ Nucl. Phys.} \textbf{B409} (1993) 441 - 464,\\
   Ch. Wetterich: ``Exact evolution equation for the effective potential'',
                \emph{Phys. Lett.} \textbf{B301} (1993) 90 - 94.
\bibitem{Bon}M. Bonini, M. D'Attanasio, G. Marchesini:
         ``Ward Identities and Wilson Renormalization Group in QED'',
            \emph{ Nucl. Phys.} \textbf{B418} (1994) 81 - 112, 
    ``Renormalization Group flow for $SU(2)$ Yang - Mills Theory and 
    Gauge Invariance'', \emph{Nucl. Phys.} \textbf{B421} (1994) 429 - 455, 
  ``BRS - Symmetry for Yang - Mills Theory and Exact Renormalization Group'',\\
               \emph{Nucl. Phys.} \textbf{B437} (1994) 163 - 186,  \\
 L. Girardello, A. Zaffaroni: ``Exact Renormalization Group Equation and
      Decoupling in Quantum Field Theory'',\\ \emph{ Nucl. Phys.}
      \textbf{B424} (1994) 219 - 238. \\
 C. Becchi: ``On the Construction of Renormalized Gauge Theories Using
          Renormalization Group Techniques'', arXiv: hep-th/ 9607188,  \\
 M. Bonini, F. Vian: ``Chiral Gauge Theories and Anomalies in the
     Wilson Renormalization  Group Approach'',\\
    \emph{Nucl. Phys.} \textbf{B511} (1998) 479 - 494,\\
        ``Wilson Renormalization Group for Supersymmetric Gauge Theories
    and Gauge Anomalies'', \emph{Nucl. Phys.} \textbf{B532} (1998) 473 - 497.
\bibitem{D'AM} M. D' Attanasio, T.R. Morris: ``Gauge Invariance, the
    Quantum Action Principle, and the Renormalization Group''\,\\
        \emph{Phys. Lett.} \textbf{B378} (1996) 213 - 221.
\bibitem{AKMT} S. Arnone, Y.A. Kubyshin, T.R. Morris, J.F.Tighe:
     ``A gauge invariant regulator for the ERG '', 
            \emph{Int.J.Mod.Phys.} \textbf{A16} (2001) 1989
\bibitem{EHW} U. Ellwanger: ``Flow equations and BRS invariance for
   Yang-Mills theories ``, \emph{Phys. Lett.} \textbf{B335} (1994) 364 - 370,
 U. Ellwanger, M. Hirsch, A. Weber: ``Flow equations for the relevant part
  of the pure Yang-Mills action ``, \emph{Z. Phys.} \textbf{C69}
                 (1996) 687 - 697,
   U. Ellwanger: ``Confinement, Monopoles
      and Wilsonian effective Action ``,\\ \emph{Nucl. Phys.}
                   \textbf{ B531} (1998) 593 - 612. 
\bibitem{ReWe} M. Reuter, C. Wetterich: ``Effective average action for
     gauge theories and exact evolution equations ``, \emph{Nucl. Phys.}
           \textbf{B417} (1994) 181 - 214, ``Exact evolution equation
   for scalar electrodynamics ``, \emph{Nucl. Phys.} \textbf{B427}
    (1994) 291 - 324, ``Gluon Condensation in Non-Perturbative Flow 
       Equations ``, \emph{Phys. Rev.} \textbf{D56} (1997) 7893 - 7916,\\
  H. Gies: ``Running coupling in Yang-Mills theory - a flow equation \\
       study - '', \emph{Phys. Rev.} \textbf{D66} (2002) 025006.
\bibitem{LaRe} O. Lauscher, M. Reuter: ``Towards Nonperturbative
   Renormalizability of Quantum Einstein Gravity ``,
   \emph{Int.J.Mod.Phys.} \textbf{A17} (2002) 993 - 1002, ``Is Quantum
  Einstein Gravity Nonperturbatively Renormalizable? ``,
   \emph{Class. Quant. Grav.} \textbf{19} (2002) 483 - 492. 
 \bibitem{BB} C. Bagnuls, C. Bervillier: ``Exact Renormalizstion Group
                  Equations. An Introductory Review'', 
            \emph{Phys. Rep.} \textbf{348} (2001) 91 - 157.  
\bibitem{Kopp} Ch. Kopper: \emph{Renormierungstheorie mit Flussgleichungen},\\
                      Shaker Verlag, Aachen 1998.       
\bibitem{GJ}J. Glimm, A. Jaffe: \emph{Quantum Physics},\\ Sec. Ed., Springer,
                        New York 1987, chapt. 9.1.
\bibitem{HiYa} T. Hida: \emph{Stationary Stochastic Processes},
        Princeton 1970, Theor. 4.2. \\
      Y.Yamasaki: \emph{Measures on Infinite Dimensional Spaces} World 
     Scientific, Singapore 1985,  Part A, Theor. 17.1 .
\bibitem{Ko+Me} Ch. Kopper, F.Meunier: 
              ``Large Momentum bounds from Flow Equations'', 
            \emph{Ann. Henri Poincar{\'e}} \textbf{3} (2002) 435 - 449.      
\bibitem{Wei} S.Weinberg: ``High Energy behaviour in Quantum Field theory'',\\
                     \emph{Phys. Rev.} \textbf{118} (1960) 838-849.
\bibitem{LvW} N.P.Landsman, C. van Weert: ``Real- and Imaginary-Time 
       Field Theory  at Finite Temperature and Density'',
      \emph{ Phys.Rep.} \textbf{145} (1987) 141- 249.  
\bibitem{Bour} N.Bourbaki, \emph{Fonctions d' une variable r{\'e}elle}, \\
                           Editions Hermann, Paris, 1976, chap. 6
\bibitem{Lam} Y.M.P. Lam: ``Perturbation Lagrangian Theory for Scalar
            Fields - Ward-Takahashi Identity and Current Algebra'',\\
                 \emph{ Phys. Rev.} \textbf{D6} (1972) 2145 - 2161,
 ``Equivalence Theorem on\\ Bogoliubov-Parasiuk-Hepp-Zimmermann-Renormalized
 Lagrangian\\ Field Theories'',
           \emph{Phys. Rev.} \textbf{D7} (1973) 2943 - 2949.
\bibitem{Low} J.H. Lowenstein: ``Differential vertex operations
          in Lagrangian field\\ theory'',
        \emph{Commun. Math. Phys.} \textbf{24} (1971) 1- 21.
\bibitem{PiSo} O. Piguet, S.P. Sorella: \emph{Algebraic Renormalization},\\
                          Springer, Berlin 1995. 
\bibitem{ZJ} J. Zinn - Justin:\emph{ Quantum Field Theory and Critical
              Phenomena},\\
          ch. 21, Clarendon Press,Ox\-ford, 3rd ed. 1997,\\ and 
   J. Zinn-Justin in: \emph{Trends in Elementary Particle Theory},\\
      Lecture Notes in Physics {\bf 37}, 2 - 40, Springer-Verlag 1975. 
\bibitem{SlTa} A.A. Slavnov, \emph{Math. Theor. Phys.} \textbf{10} (1972) 99 \\
       J.C. Taylor: ``Ward Identities and Charge Renormalization of the\\
     Yang - Mills Field'', \emph{ Nucl. Phys.} \textbf{B33} (1971) 436 - 444.
\bibitem{BeRoSt} C. Becchi, A. Rouet, R. Stora: ``Renormalization of
                     the Abelian Higgs - Kibble Model'', 
                 \emph{ Commun. Math. Phys.} \textbf{42} (1975) 127 - 162,
       ``Renormalization of Gauge Theories'',
                       \emph{Ann. Phys.(N.Y.)} \textbf{98} (1976) 287 - 321. 
\bibitem{FaSl} L.D. Faddeev, A.A. Slavnov: \emph{Gauge Fields: Introduction
             to\\ Quantum Theory}, Benjamin, Reading MA 1980.
\bibitem{nKM} Ch. Kopper, V.F. M\"uller: ``Renormalization of
   Spontaneously Broken $SU(2)$ Yang - Mills Theory with Flow Equations'',\\
    \emph{Rev. Math. Phys.} \textbf{21} (2009) 781 - 820. (arXiv: 0902.2486),
\end{thebibliography}
\end{document}